\documentstyle[12pt,epsf]{report}

\newcounter{sub}
\newcounter{subeqn}[sub]
\renewcommand{\thesubeqn}{\alph{subeqn}}
\renewcommand{\theequation}{\thesub\thesubeqn}

\title{Normal modes of relativistic systems in post-Newtonian
approximation\\and\\the stability curve of r-modes}
\vspace{2cm}
\author{Vahid Rezania}
\vspace{4cm}

\def\be{\begin{equation}}
\def\ee{\end{equation}}
\def\bt{\begin{tabular}}
\def\et{\end{tabular}}
\def\lp{\left(}
\def\rp{\right)}
\def\ls{\left[}
\def\rs{\right]}

\def\st{\stepcounter{sub}}
\def\stq{\stepcounter{subeqn}}
\def\bea{\begin{eqnarray}}
\def\eea{\end{eqnarray}}
\def\beas{\begin{eqnarray*}}
\def\eeas{\end{eqnarray*}}
\def\L{{\cal L}}

\def\H{{\cal H}}
\def\Lcl{{\cal L}^{cl}}
\def\Lpn{{\cal L}^{pn}}
\def\x{{\bf x}}
\def\u{{\bf u}}
\def\v{{\bf v}}
\def\p{{\bf p}}
\def\xp{{\bf x'}}
\def\up{{\bf u'}}
\def\vp{{\bf v'}}
\def\xpp{{\bf x''}}
\def\upp{{\bf u''}}
\def\vpp{{\bf v''}}
\def\Y{{\bf Y}}
\def\xxi{\xi\hspace{-.18cm}\xi\hspace{-.18cm}\xi}
\def\nab{\nabla\hspace{-.35cm}\nabla\hspace{-.35cm}\nabla}
\def\varp{\varpi\hspace{-.34cm}\varpi\hspace{-.34cm}\varpi}

\def\P{{\cal P}}

\def\d{\delta}
\def\T{{\cal T}}
\def\Pm{{\cal P}_m}
\begin{document}

\maketitle
\newpage

\pagenumbering{roman}

%************************** Abstract *****************************************

\begin{abstract}
This thesis consist of two parts, post-Newtonian modes of relativistic systems
and the stability curve of $r$-modes of neutron stars.
In part I,
we use the post-Newtonian ($pn$) order of Liouville's equation to
study the normal modes of oscillation of a spherically symmetric
relativistic system.   Perturbations that are neutral in Newtonian
approximation develop into a new sequence of normal modes. In the
first $pn$ order; a) their frequency is an order $q$ smaller than
the classical frequencies, where $q$ is a $pn$ expansion
parameter; b) they are not damped, for there is no gravitational
wave radiation in this order; c) they are not coupled with the
classical modes in $q$ order; d) because of the spherical symmetry
of the underlying equilibrium configuration they are designated by
a pair of angular momentum eigennumbers, ($j,m$), of a pair of
phase space angular momentum operators ($J^2,J_z$).    The
eigenmodes are, however, $m$-independent. Hydrodynamics of these
modes is also investigated; a) they generate oscillating
macroscopic toroidal motions that are neutral in classical case ;
and b) they give rise to an oscillatory $g_{0i}$ component of the
metric tensor that otherwise is zero in the unperturbed system.
The conventional classical modes, which in their hydrodynamic
behavior emerge as $p$ and $g$ modes are, of course, perturbed to
order $q$.   These, however, have not been of concern in this
work.

In part II, stability curve of $r$-modes of neutron stars are
calculated by considering vorticity-shear viscosity coupling. The
coupling is predicted by kinetic theory, a causal theory of fluid
rather than the Navier-Stocks theory. We calculate this coupling
and show that it can in principle significantly modify the
stability diagram at lower temperatures. As a result, colder stars
can remain stable at higher spin rates.

As an application, the loss of angular momentum
through gravitational radiation, driven by the excitation of
r-modes, is considered in neutron stars having rotation
frequencies smaller than the associated critical frequency. We
find that for reasonable values of the initial amplitudes of such
pulsation modes of the star, being excited at the event of a
glitch in a pulsar, the total post-glitch losses correspond to a
negligible fraction of the initial rise of the spin frequency in
the case of Vela and older pulsars. However, for the Crab
pulsar the same effect would result, within a few months, in a
decrease in its spin frequency by an amount larger than its
glitch-induced frequency increase. This could provide an
explanation for the peculiar behavior observed in the post-glitch
relaxations of the Crab.

\end{abstract}

%*********************** Acknowledgements ***********************************

\chapter*{Acknowledgements}
\typeout{Acknowledgements}
\addcontentsline{toc}{chapter}{Acknowledgements}
I wish to appreciate Prof. Yousef Sobouti deeply for his great
scientific guideness and warmfull advices.  I also thank Dr.
Mehdi Jahan-Miri who introduced me neutron stars
instabilities firstly.  Also would like to appreciate Prof. Roy Maartens who opened
a new vision of life to me.

I thank Prof. Y. Sobouti, director, and Dr. M. Khajehpoor, deputy director, of
Institute for Advanced Studies in Basic Sciences (IASBS), Zanjan- Iran,
for great hospitality during my Ph. D. study.
I would like to thank Prof. R. Maartens, director of
Relativity and Cosmology Group in Portsmouth University, UK, for great kindness
during my visit from UK, where part of this work was completed.

I thank N. Andersson, S. Morsink, M. Bruni, M. T. Mirtorabi, M. Saadatfar, H. G. Khosroshahi,
M. Mahmoudi, and A. Dianat for their helpful discussions.
I would like to give my warmful appreciation to Sharareh, my wife, for every thing.

%****************************************************************************

\tableofcontents
\listoftables
\listoffigures

%*****************************************************************************

\newpage

\noindent In this thesis I present the evolution and dynamics of
compact objects through a number of different approaches which
will be described in the following parts.   In part one using relativistic
Liuoville's equation, I studied
normal modes of a relativistic system in post-Newtonain approximation. This
part is supervised by Prof. Yousef Sobouti (IASBS) as the main part of my Ph. D. thesis
\cite{RSo00, SRe00}. Beside this,
I became in the recently discovered instability in
newly borne neutron stars.  In this respect, I've done some research under the
supervisons of Dr. M. Jahan-Miri (IASBS) \cite{RJa00} and Prof. Roy Maartens
(Portsmouth,
UK) \cite{RMa00}.  Second part of my thesis is devoted to the $r$-mode
instability.

%**************************   Part I    ************************************

\newpage
\pagenumbering{arabic}
\part{Post-Newtonian modes of relativistic systems}

\newpage

\chapter{Introduction}

Chandrasekhar's \cite{Cha65} formulation of
post-Newtonian ($pn$) hydrodynamics is among the pioneering ones.
He
generalized Eulerian equations of Newtonian hydrodynamics to
$pn$ order consistent with Einstein's field equations,
and applied them to obtain the $pn$ corrections to
the equilibrium and stability of uniformly rotating homogeneous masses.
Blanchet,
Damour and Sch$\ddot{\rm a}$fer \cite{DSc90} studied the
gravitational wave generation of a self gravitating
fluid
by adding an appropriate term to $pn$ equation of
hydrodynamics. Cutler \cite{Cut91} employed the $pn$
hydrodynamics
and a perturbation technique to derive an expression
for the $pn$ correction to Newtonian eigenfrequencies.
Cutler and Lindblom \cite{CLi92} adopted Cutler's method
to calculate numerically the oscillation frequencies of the $l=m$
$f$-modes of rapidly rotating polytropic
neutron stars.

In this work we study normal modes of a non-rotating
relativistic system in $pn$
approximation
through the relativistic Liouville's
equation rather than the relativistic hydrodynamics.    The reason for doing
so is to avoid thermodynamic concepts being incorporated into hydrodynamics.
Liouville's equation is a purely dynamical theory and free from such
complexities.
Furthermore
in many cosmological and astrophysical
situations, an idealized fluid model of matter is inappropriate,
and a self-consistent microscopic model based on relativistic
kinetic theory \cite{Ehl71} gives a more detailed physical description.
A well-known example is the case of collisionless particles, as
for cosmological neutrinos and photons \cite{TMa67}, or stellar clusters in
equilibrium \cite{RFa91}. Also there are other examples among
non-equilibrium evolving systems, such as stellar clusters with
collisions \cite{Pod70}, the early evolution of a FRW universe into an
anisotropic Bianchi universe or an inhomogeneous universe via a
disturbance of the equilibrium collisional balance \cite{MEl89}.   The
relativistic transport of photons \cite{LDa66} and cosmic rays \cite{Web85}, or
mixture of cosmic elementary particles \cite{Ber88} are other
non-equilibrium situations suited to a kinetic approach.

The kinetic theory offers an alternative approach to describe the
matter, rather than the phenomenological fluid dynamics and its
associated thermodynamics. For example, the standard
thermodynamics of fluids violates causality and is unstable \cite{HLi83}. A
casual and stable generalization emerges from the kinetic theory,
as developed by Israel and Stewart \cite{ISt79}.  In some cases, a fluid
model leads to the loss of information and is unable to account
for certain effects.    For example, Landau-type damping of
gravitational perturbations by a kinetic gas is not present in the
fluid models \cite{Ste72}.    Another example is the rotational
perturbations coexisting with initial singularity, which is
impossible in the fluid case \cite{Reb91}.

In compiling this work we have relied heavily on the
following studies dealing with various aspects of Liouville's,
Liouville-Poisson's and Antonov's equations.

O(3) symmetry and mode classification of classical Liouville's
equation for spherically symmetric potentials was studied by
Sobouti \cite{Sob89a}. Simple harmonic potentials in one, two, and
three dimensions were discussed by Sobouti \cite{Sob89b}.      He
obtained exact and complete eigensolutions by means of raising and
lowering ladders for Liouville operator.    Furthermore, he
investigated potentials of self gravitating spheres, oblate or
prolate spheroids, and ellipsoids in details. A systematic method
to elaborate the symmetries of Liouville's equation for an
arbitrary potential were introduced by Sobouti and Dehghani
\cite{SDe92}. They showed that the symmetry group of $r^2$ potential
is GL(3, c) and classified eigenmodes of Liouville's equation for
quadratic potential. O(4) symmetry of $r^{-1}$ potential were
obtained by Dehgahni and Sobouti \cite{DSo93}.     Dynamical
symmetry of Liouville's equation for $r^2$ potential was worked
out by Dehghani and Sobouti \cite{DSo95}. Dynamical symmetry group
of general relativistic Liouville's equation was discussed by
Dehghani and Rezania \cite{DRe96}.    In particular they found that
in de Sitter's space-time the group is SO(4,1) $\bigotimes$
SO(4,1).

In applications to self-gravitating systems the pioneering work
was done by Antonov \cite{Ant62}.      He reduced the linearized
Liouville-Poisson equations to a self adjoint operator in phase
space.     Further elaborations on Antonov's equation were made by
Lynden-Bell \cite{Lyn66}, Milder \cite{Mid67}, Lynden-Bell and Sanitt
\cite{LSa69},
Ipser and Thorne \cite{ITo68}.     These authors were concerned with the
stability of a given isotropic distribution function.
Stabilities of anisotropic distribution functions were
investigated by Doremus et. al. \cite{DFB70, DFB71}, Doremus and Feix
\cite{DFe73}, Gillon et. al. \cite{GCDB76}, Kandrup and Sygnet \cite{KSy85}.
Attempts to solve the linearized Liouville-Poisson equation for
eigenfrequencies and eigenmodes of oscillations were made by
Sobouti \cite{Sob84,Sob85,Sob86}.      Further and more transparent
exposition of mode classification and mode calculations were given
by Sobouti and Samimi \cite{SSa89} and Samimi and Sobouti \cite{SSo95}.

Here, using the standard $pn$ expansion of the metric components
\cite{Wei72}, we derive the $pn$ approximation of Liouville's
equation ($pnl$).      In the time-independent case, we show that a
generalization of classical integrals (energy and angular
momentum, say) are the static solutions of $pnl$.
In time-dependent regimes,
the effect of the $pn$ corrections on the known solutions of the
classical equation can be analyzed by the usual perturbation techniques.
Whatever the procedure, the first order corrections on the known
modes will be small and will not change their nature.    We will not pursue
such issues here.    The main interest of this work is to study a new
class of solutions of $pnl$ that originate solely from the  $pn$  terms
and have no precedence in classical theories.    It is not difficult to
anticipate the existence of such modes.     Perturbations on an equilibrium
state, that are functions of classical integrals
do not disturb the equilibrium of the system at classical level.    That
is
they do not induce restoring forces in the system.    They, however, do so in
the $pn$ regime, and make the system oscillate about the $pn$ equilibrium
state.    Such perturbations may be considered as a class of infinitely
degenerate zero frequency modes of the classical system.      The $pn$ forces
unfold this degeneracy and turn them into a sequence of non zero frequency
modes distinct and uncoupled from the other classical modes.     We have termed
them as $pn$ modes.

A hydrodynamic analog of $pn$ modes is the following.     In
spherically symmetric fluids, toroidal motions are neutral.
Sliding one spherical shell over the other is not opposed by a
restoring force.     A small magnetic field or a slow rotation
(mainly through Coriolis forces) gives rigidity to the system.
The fluid resists against such displacements and a sequence of
well defined toroidal modes of oscillation develop.    See Sobouti
\cite{Sob80}, Hasan and Sobouti \cite{HSo87}, Nasiri and Sobouti \cite{NSo89},
and Nasiri \cite{Nas92} for examples and typical calculations.

The plan of this part is as follow.   In chapter 2 we
briefly review Liouville's equation both in classical and
relativistic regimes.
We introduce a distribution function and its equation of
evolution. Macroscopic quantities associated to the distribution
function are discussed.

In chapter 3 we adopt the post-Newtonian ($pn$) approximation to study a
self gravitating system imbedded in an otherwise flat space-time.
We obtain the $pn$ approximation of Liouville equation ($pnl$). We find
two integrals of $pnl$ that are the $pn$ generalizations of the
energy and angular momentum integrals of the classical Liouville's
equation. Post-Newtonian polytropes, as simultaneous solutions of
$pnl$ and Einstein's equations, are discussed and calculated

In chapter 4 we give the $pn$ order
of the linearized
Liouville equation that governs the evolution of small perturbations from an
equilibrium state.       We extract the
equation for a sequence of new modes that are generated solely by $pn$ force
but are absent in classical regime.
We explore the O(3) symmetry of the modes and classify them on
basis of this symmetry.
We study hydrodynamics of these modes.
We seek a variational approach to the calculation of $pn$ modes and
give numerical values for polytropes.

Post-Newtonian approximation is
reviwed in appendix A.     In appendix B coordinates transformation that we
need to extract Liouville's equation in
$pn$ approximation, is discussed.   In appendix C post-Newtonian hydrodynamics
are recovered by integration of $pnl$ over ${\bf v}$-space.
Simultaneous eigensolutions of $J^2$ and $J_z$
operators are constructed and elaborated in appendix D.\\

\setcounter{sub}{0}
\setcounter{subeqn}{0}
\renewcommand{\theequation}{2.\thesub\thesubeqn}

\chapter{Liouville's equation \\ Classical and Relativistic}

Kinetic theory has expanded in classical, quantum, and
relativistic directions \cite{Liboff90}.  Classical kinetic theory
is the foundation of fluid dynamics and thus is important to
aerospace, mechanical, and chemical engineering.  It is also
relevant to many problems in astrophysics, for example the
stability and evolution of stellar systems. Quantum kinetic theory is
applicable to problems in particles transportation, radiation
through material media, etc., which are important in solid
state and laser physics. Relativistic kinetic theory has became
important in certain plasma physics. It is also used to study the
evolution of relativistic stellar systems and the dynamics of
cosmological fluids.

In its most elementary version the kinetic theory of a simple gas
relies on the concept of $N$ pointlike particles which may
interact with each other. Collisions are assumed to establish a
local or global equilibrium of the system.  Between the collisions
the particles move on geodesics of a given spacetime.

Technically, the gas particles are described by an invariant
one-particle distribution function governed by Boltzmann or
Liouville's equation. The latter is best applicable to dilute
gases where collisions may be neglected. The macroscopic fluid
dynamics for such a system may be obtained in terms of the first
and second moments of the distribution function. A gas, however,
is the only system for which the correspondence between
microscopic variables, governed by a distribution function, and
phenomenological fluid quantities is sufficiently well understood.

The goal of this chapter is to introduce the Liouville's equation
both in classical and relativistic cases. Liouville's  equation
gives the time evolution of probability distribution function.  It provides
the dynamical basis of statistical mechanics , both at and away
from equilibrium . Its solution enables one to calculate the
ensemble average of any dynamical quantity .

In section 2.1 classical Liouville's equation is discussed. For
the application in astrophysical problems, the linearized
Liouville-Poisson equation is introduced. Relativistic Liouville's
equation is considered in section 2.2.
In section 2.3, macroscopic quantities are discussed.

\section{Classical Liouville's equation}
For a system of $N$ degrees of freedom, phase space is a $2N$
dimensional space whose axes are the $(x_i, p_i)$ variables. Thus
the state of the system at any given instant $(\x, \p)$ is a
single point, which is usually called system point, in $2N$
dimensional phase space.  Time is exhibited explicitly in $2N+1$
dimensional phase space, a $2N$ phase space with an additional
orthogonal time axis.  As time evolves, the system point moves on
a system trajectory, $[\x (t), \p (t)]$, which is a curve in
$2N+1$ space. The system trajectory is determined by solving the
equations of motion:

\st\bea\label{hamilton} \stq &&{\dot \x}={\partial H\over\partial
\p}\,,\label{hamilton-a}\\\stq && {\dot \p}=-{\partial
H\over\partial \x}\,,\label{hamilton-b}\eea where $H$ is the
Hamiltonian of the system.  It is clear that the state of the
system will be specified uniquely by $2N$ initial constants,
$(\x(0), \p(0))$.

An abstract collection of a large number of independently
identical system points is called an ensemble.  An important
property of the ensemble is that trajectories of the ensemble can
never cross in phase space. This follows from the fact that for a
system with $N$ degrees of freedom the system of trajectory, Eq.
(\ref{hamilton}), is uniqely specified by $2N$ initial values,
$[\x(0), \p(0)]$.

%The swarm of system points moving through the phase space be have
%much like a fluid in a multidimensional space, and there are
%numerous similarities between the discussion of the ensemble and
%the well-known notion of fluid dynamics.

Consider an infinitesimal volume in phase space surrounding a
given system point at time $t=0$. In the course of time the system
points defining a volume element move about in phase space and
the volume contained by them will take on different shape as time
progresses.  %The dotted curve in Fig. 2.1 schematically indicates
%the evolution of the volume element in time.
The Liouville theorem, however, states that the size of a volume element in
the phase space remains constant under canonical transformations induced
by the Hamiltonian,
i.e. the Jacobian of a canonical transformation is unity
\cite{Liboff90}.

%The size of volume element $d\Gamma$, is one of Poincare's
%integral invariant.  It does not vary with time. For, a time
%evolution of system points is a canonical transformation, and the
%volume elements remain invariant under such transformation . This
%is sometimes referred to as liouville's theorem \cite{Liboff90}.

Let $d{\cal N}$ denotes the number of system points in a phase
space volume element. It remains constant. For, a system point
initially inside can never get out, and one outside can never
enter the volume. Indeed, if some system point were to cross the
border, its trajectory would intersect a trajectory of a system
point defining the boundary surface.  But this is not possible.
For, if two trajectories were to coincide at one time, they would
coincide at all the times. Hence the number of the system points,
$d{\cal N }$, within a volume element of phase space, $d\Gamma$,
remains constant. In other words the probability density,
$f(\x,\p,t) = {\cal N}^{-1}d{\cal N}/d\Gamma$, should remain
constant in time. That is

\st\be\label{liouv1}\stq {df\over dt}=0\,. \label{liouv1-a}\ee
This is the Liouville's equation.  Assuming $f$ is differentiable,
we obtain

\stq\be {df\over dt }={\partial f\over \partial t}+ {\partial
f\over \partial x_i}{\dot x_i}+{\partial f\over \partial p_i}{\dot
p_i}=0. \label{2.2a}\ee
Taking equations of motion, Eq. (\ref{hamilton}), into account,
Eq. (\ref{2.2a}) becomes \st\be {df\over dt }={\partial f\over
\partial t}+ {\partial H \over \partial p_i}{\partial f\over
\partial x_i}-{\partial H \over
\partial x_i}{\partial f\over
\partial p_i}=0\,. \label{2.3}\ee
It is convenient to write it as

\st\be\label{2.4}\stq i{\partial f\over \partial t}=\Lcl f\,,
\label{2.4a}\ee where classical Liouville's operator, $\Lcl$, is
the linear operator

\stq\be \Lcl=-i({\partial H \over \partial p_i}{\partial\over
\partial x_i}-{\partial H \over \partial x_i}{\partial\over
\partial p_i})\,,\label{2.4b} \ee
As it will be shown later, the reason for including $i$ is to render
$\Lcl$ Hermitian.  In terms of $\Lcl$, the formal solution of Eqs.
(\ref{2.4}) is

\st\be f(\x,\p,t)=e^{-it \Lcl}f(\x,\p,0)\,.\label{2.5} \ee It is
easy to show that if the initial $f(\x,\p,0)$ is an acceptable
distribution function, $f(\x,\p,t)$ will be an acceptable one at
all later time. In particular

\st\stq\be f(\x,\p,t) \geq 0; \,\,\,\,\,  \forall t\,, \ee

\stq\be \int f(\x,\p,t)d\Gamma = 1; \,\,\,\forall t\,. \ee See
Balescu \cite{Balescu75}.

In this section, we obtained classical Liouville's equation in
general form. To solve the equation for specific problem, we must first
define the Hamiltonian of the system.

\subsection{Properties of $\Lcl$}
In this section we review some important properties of $\Lcl$. For
more details see
\cite{Sob89a,Sob89b,SDe92,
DSo93,SSa89,
SSo95,Deh92}.

\noindent{\bf The Hilbert space:} An axiomatic study of the
eigensolutions of classical Liouville's equation requires
introduction of a Hilbert space. A Hilbert space, $\H$, is
defined to be the space of complex square integrable functions of
phase coordinates $(\x, \p)$ that vanish at the phase space
boundary of the system:

\st\be \H :{f(\x, \p); \int f^*f d\Gamma={\rm finite},~~ f({\rm
boundary})=0}\,. \ee  Integrations in $\H$ are over the volume of
the phase space available to the system.

\noindent{\bf Hermiticity:} $\Lcl$ is Hermitian in $\H$, i.e.

\st\be \int g^*(\Lcl f)~d\Gamma =\int(\Lcl g)^*f~d\Gamma;~~~~~~ g,
f \in \H\,. \label{2.8}\ee This is proved by integrating by (\ref{2.8})
by parts and letting the integrated
terms vanish at boundary.

\noindent{\bf Real eigenfrequency:} The eigenfunctions $f_n(\x,
\p)$ and eigenvalues $\omega_n$ are defined by

\st\be \Lcl f_n=\omega_n f_n\,.\ee Hermiticity of $\Lcl$ ensures
that $\omega_n$'s are real and eigenfunctions belonging to
distinct eigenvalues are orthogonal.

\noindent{\bf Completeness of eigensolutions:} We can further impose
the normalization condition on $f_n$'s,

\st\be \int f^*_m f_n d\Gamma=\delta_{mn}\,,\ee  and obtain
an orthonormal set.   We shall also assume that
they also form a complete set.

\noindent Since classical Liouville operator is purely imaginary,
${\Lcl}^*=-\Lcl$, its eigensolutions have following properties:
\begin{itemize}
\item[(a)] Eigensolutions belonging to non-zero eigenvalues are
complex, ie.

\st\be f(\x,\p)=u(\x, \p)+iv(\x,\p)\,.\ee

\item[(b)] If $(\omega, f)$ is an eigensolution, $(-\omega, f^*)$ is
another eigensolution.

\item[(c)]  $f^* f$ is an integral of motion, ie. $\Lcl (f^* f)=0$.

\item[(d)]  $[(n-m)\omega, {f^*}^m f^n ]$ is an eigensolution with
$n,m=$ positive integer.

\item[(e)] Eigenfunctions belonging to non-zero eigenvalues integrate
to zero:

\st\be \int fd\Gamma =0,~~~~~\omega\neq 0\,.\ee

\end{itemize}

\subsection{Linearized Liouville-Poisson's equation}
In applications to astrophysical problems, many investigators
\cite{Ant62}-\cite{KSy85}, have often used the linearized
Liouville-Poisson equation to study stability of the perturbed a
self-gravitating stellar system.

In this section, we follow Sobouti \cite{Sob84,Sob85,Sob86},
Sobouti and Samimi \cite{SSa89}, and Samimi and Sobouti
\cite{SSo95} to introduce the classical linearized Liouville
equation.

For a collisionless self-gravitating stellar system the classical
Liouville's equation, Eqs. (\ref{2.4}), for distribution function,
$F(\x, \p, t)$, becomes

\st\bea\stq && i{\partial F\over \partial t}=\Lcl F\,,\\
&&\Lcl=-i\lp p_i{\partial\over
\partial x_i}-{\partial U \over \partial x_i}{\partial\over
\partial p_i}\rp\nonumber\,,\label{L1} \eea
where the potential $U(\x, t)$ is the solution of Poisson's
equation

\st\be U(\x, t)=-G\int F(\xp, \p', t)\mid
\x-\xp\mid^{-1}d\Gamma'\,.\label{L2}\ee   The Hamiltonain used here
is the energy of the system, $E={1\over 2}p^2+U(\x ,t)$.
It is easy to see that the energy is an integral of $\Lcl$ in an
equilibrium state.   Furthermore, for spherically symmetric
potentials, the angular momentum, $\l_i=\varepsilon_{ijk}x_jp_k$,
is another integral of $\Lcl$.

To find the linearized equation, let $F\rightarrow F(E)+\d F(\x,
\p, t)$, where $F(E)$ is an equilibrium distribution function, and
$\mid \d F\mid~<~F(E)$ for all $(\x, \p, t)$ is a perturbation on $F(E)$.
Accordingly, the potential splits into a large and small terms,
$U(r)+\d U(\x, t)$ where $r=\mid\x\mid$.  Substituting in Eqs.
(\ref{L1}) and (\ref{L2}), in the first order we find

\st\bea  i{\partial\d F\over \partial t}&&=\Lcl \d F+i{\partial
F\over \partial p_i}{\partial\d U\over\partial x_i}\,,\nonumber\\
&&=\Lcl \d F+GF_E\Lcl \int \d F(\xp, \p', t)\mid
\x-\xp\mid^{-1}d\Gamma'\,, \label{L3}\eea

\st\be \d U(\x, t)=-G\int \d F(\xp, \p', t)\mid
\x-\xp\mid^{-1}d\Gamma'\,,\label{L4}\ee  where $\Lcl$ is now
constructed with the time-independent potential $U(r)$ and
$F_E=dF/dE$.

\noindent Let $\d F=\mid F_E\mid^{1/2}f(\x, \p, t)$, see Sobouti
\cite{Sob84}, then Eq. (\ref{L3}) can be written as

\st\bea  &&\hspace{-1cm} i{\partial f\over \partial t}={\cal A} f\,,\nonumber\\
&&\\ &&\hspace{-1cm} {\cal A}f=\Lcl f + ~G~{\rm sign}(F_E)\mid
F_E\mid^{1/2}\Lcl \int  \mid F_E\mid^{1/2} f(\xp, \p', t)\mid
\x-\xp\mid^{-1}d\Gamma'\,.\nonumber \label{L5}\eea   It is easy to
show that for the linearized equation the classical energy,
$E={1\over 2}p^2+U$, is not an integral, but the angular momentum
is.  Conservation of angular momentum means that
the operator ${\cal A}$ has O(3) symmetry.

\noindent Let $f=f_-(\x, \p, t) +i f_+(\x, \p, t)$, where $f_-$
and $f_+$ are odd and even in $\p$, respectively. Substituting this in
Eq. (\ref{L5}), and decomposing it into odd and even components, we
find

\st\stq\bea  &&~~ {\partial f_-\over \partial t}={\cal A}
f_+\,,\label{L6a}\\\stq && -{\partial f_+\over \partial t}={\cal
A} f_-\,,\label{L6b}\eea  where ${\cal A}$ is odd in $\p$.
Eliminating $f_+$ we obtain a wave equation for $f_-$

\st\be   -{\partial^2 f_-\over \partial t^2}={\cal A}^2
f_-\,,\label{L7}\ee where

\stq\be {\cal A}^2f_-={\Lcl}^2 f_- + ~G~{\rm sign}(F_E)\mid
F_E\mid^{1/2}\Lcl \int  \mid F_E\mid^{1/2} {\Lcl}' f_-(\xp, \p',
t)\mid \x-\xp\mid^{-1}d\Gamma'\,.\\ \label{L7a}\ee

\noindent Equations (\ref{L7}) are Antonov's equation.  $f_-$ and
$f_+$, calculated from Eqs. (\ref{L7}) and (\ref{L6b}), give a
solution of the linearized Liouvolle-Poisson equation.  Assuming sinusoidal
time dependence for $f_-(\x, \p, t)=f_-(\x, \p) e^{i\omega t}$, we find

\st\be   {\cal A}^2
f_-=\omega^2 f_-\,.\label{L71}\ee   Equation (\ref{L71}) is an eigenvalue
problem with eigenfrequancy $\omega$.
Sobouti
and Samimi \cite{SSa89} proved that the ${\cal A}$ is not
Hermitian in $\H$, i.e. ${\cal A}\neq{\cal A}^\dagger$.  However, by
decomposing the Hilbert space $\H$ into odd and even subspaces in
$\p$, they showed that ${\cal A}^2$ is Hermitian on odd subspace:

\st\bea && \int f_-^*{\cal A}^2 f_- d\Gamma = \int \mid \Lcl
f_-\mid^2 d\Gamma + ~G~{\rm sign}(F_E) \times\nonumber\\
&&\hspace{1cm}\times\int \mid F_E\mid^{1/2}\Lcl f_- \mid
F'_E\mid^{1/2} {\Lcl}' f'_- \mid \x-\xp\mid^{-1}d\Gamma
d\Gamma'={\rm real}\,.\nonumber\\\label{L8} \eea
Equation (\ref{L8}) ensures that the eigenfrequancies are real.

\noindent {\bf O(3) symmetry of ${\cal A}$:}  For spherically
symmetric potentials, the invariance of ${\cal A}$ under rotation
of both $\x$ and $\p$ coordinates was established by Sobouti and
Samimi \cite{SSa89}. The corresponding angular momentum operator in phase space
is

\st\be J_i=J_i^{\dagger}=-i\varepsilon_{ijk}\lp
x^j\frac{\partial}{\partial x^k}+ u^j\frac{\partial}{\partial
u^k}\rp\,, \label{L9}\ee with the angular momentum algebra

\st\bea \label{L10} \stq &&[J_i, J_j]=i\varepsilon_{ijk}
J_k\,,\label{L10a} \\\stq && [J^2, J_z]=0\,.\label{L10b}\eea We
note that $J_i$ rotates simultaneously both $x$ and $p$
coordinates. Commutation of $J_i$ with $\Lcl$ was first proved by
Sobouti \cite{Sob89a}:

\st\be [\Lcl, J_i]=0\,.\label{L11}\ee  Sobouti and Samimi
\cite{SSa89} extended the same to ${\cal A}$,

\st\be [{\cal A}, J_i]=0\,\label{L12}\ee An important consequence of
Eqs. (\ref{L10}) and (\ref{L12}) is the mutual commutation of the
following set of operators

\st\be\label{L13} [{\cal A}^2, J^2, J_z]=0\,.\ee The implication
of Eq. (\ref{L13}) is clear.  The eigenfunctions of ${\cal A}^2$
can simultaneously be the eigenfunctions of $J^2$ and $J_z$.  In
other words, the eigenfunctions of ${\cal A}^2$ can be classified
into classes specified by eigennumbers $j$, $m$ of $J^2$ and
$J_z$. See for more details \cite{SSa89}.

\section{Relativistic Liouville's equation}
In section 2.1, we introduced classical Liouville and the
linearized Liouville-Poisson equations. We reviewed some
properties of these equations, that help one to extract their
eigenfunctions and eigenfrequencies.  The goal of the present section is
to introduce the distribution function and its equation of
evolution in general relativity. For this purpose we need to
introduce the pahse space on which such a function is defined.

\subsection{Distribution function} \label{spar21} Consider a single
test particle with mass $m$ which moves in a gravitationally curve
spacetime.
Its motion is determined by the geodesic equation \st
\be
p^\mu=\frac{dx^\mu}{d\lambda};\quad \frac{Dp^\mu}{d\lambda}\equiv
\frac{dp^\mu}{d\lambda}+\Gamma^\mu_{\nu\rho}p^\nu p^\rho=0,
\label{MOTION} \ee where $\lambda$ is an affine parameter defined
by the requirement that $p^\mu$ be the 4-momentum. Hereafter,
$\Gamma^\mu_{\nu\rho}$ are Christoffel symbols associated with the
metric $g_{\mu\nu}$.
 Note that if
there are non gravitational forces (e.g. electromagnetic forces)
then we have to modify this equation.

The rest mass of the particle is defined as \st \be m^2=-p^\mu
p_\mu.\label{MASS}\ee Thus, according to Eq. (\ref{MOTION}), the
state of the particle is determined by the pair $(x^\mu,p^\mu)$.
The phase space is then the tangent bundle over the spacetime
manifold, i.e. \st
\be
\T=\left\lbrace(x^\mu,p^\mu), x^\mu\in{\cal M},
p^\mu\in\T_x\right\rbrace,\label{TGBUN} \ee where ${\cal M}$ is
the space-time and $\T_x$ is the tangent space to ${\cal M}$ at
$x^\mu$. From now on, Greek indices run from 0 to 3 and Latin
indices run from 1 to 3.

The volume element on $\T_x$ supported by the displacements
$dp_1,dp_2,dp_3,dp_4$ (with components $dp_1^\alpha$ etc.) is \st
\be
\pi(p^\mu)=\epsilon_{\alpha\beta\gamma\delta} dp_1^\alpha
dp_2^\beta dp_3^\gamma dp_4^\delta, \label{eVOL1} \ee
 where $\epsilon_{\lambda\alpha\beta\gamma}$ is
the totally antisymmetric tensor such that
$\epsilon_{0123}=\sqrt{-g}$. We also define $\pi_+(p^\mu)$, the
volume element corresponding to the subspace of $\T_x$ such that
$p^\mu$ is non-spacelike and future directed, \st
\be
\pi_+(p^\mu)=H(-p_\mu u^\mu)H(-p^2)\pi(p^\mu),\label{EVOL11} \ee
where $H$ is the heavyside step function
\[H(x)=\left\{\begin{array}{ll}
                   1& {\rm if}~~~~ x>0,\\
                   0&{\rm otherwise},
                 \end{array}\right.\]
and $u^\mu$ an arbitrary timelike vector field.\\ $\T_x$ is sliced
in hypersurfaces, $\P_m$, of constant $m$ called the mass-shell,
and defined by \st
\be
\P_m(x^\mu)=\left\lbrace p^\mu\in\T_x, p^\mu p_\mu=-m^2, p^\mu
u_\mu>0\right\rbrace.\label{MASSHELL} \ee The volume element of
Eq. (\ref{eVOL1}) on $\T$ can then be decomposed into a volume
element, $m\pi_m$, on $\Pm$ by \st
\be
\pi_+(p^\mu)=m\pi_m(p^\mu)dm.\label{eVOL2} \ee The factor $m$
allows one to include particles of zero rest mass (see Ehlers
\cite{Ehl71}). This defines the induced volume element
$m\pi_m(p^\mu)$ on $\P_m$. If we introduce an arbitrary future
directed unit timelike vector $u^\mu$ (i.e. satisfying $u_\mu
u^\mu=-1$), the 3-volume supported by the three displacements
$dx_1, dx_2, dx_3$ (with components $dx_1^\alpha$ etc.) in the
hypersurface perpendicular to $u^\mu$ is \st
\be
dV(u^\mu)=\epsilon_{\lambda\alpha\beta\gamma}u^\lambda dx_1^\alpha
dx_2^\beta dx_3^\gamma.\label{DVU} \ee We now consider a single
fluid composed of particles of all masses. The distribution
function, $f(x^\mu,p^\mu)$ will be defined as the mean number of
particles (on a statistical set) in a volume $dV$ around $x^\mu$
and $\pi(p^\mu)$ around $p^\mu$ measured by an observer with
4-velocity $u^\mu$, \st
\be
dN(x^\mu,p^\mu)=f(x^\mu,p^\mu)(-p^\mu u_\mu)dV(u^\mu)\pi(p^\mu).
\label{DEFF} \ee The assumptions involved in its existence have
been discussed in details by Ehlers \cite{Ehl77}. Synge \cite{Syn}
has demonstrated that $(-p^\mu u_\mu)dV(u^\mu)$ is independent of
$u^\mu$.  This implies that the distribution function is a scalar.
Moreover, $f(x^\mu,p^\mu)\geq0$ for all $x^\mu$ and all allowed
$p^\mu$.

For a gas, $dN$ is the number of particles in a volume
$dV\pi(p^\mu)$ thus the smoothness of $f$ depends on the existence
of a sufficient number of particles.

\subsection{Relativistic Liouville opertaor}
The equations of motion, Eq. (\ref{MOTION}), define on $\T$
the Liouville operator,
\st\be
{\cal L}=p^\mu\frac{\partial}{\partial
x^\mu}+\frac{dp^\mu}{d\lambda} \frac{\partial}{\partial
p^\mu}=\frac{d}{d\lambda}, \label{DEFLIOU} \ee which characterizes
the rate of change of $f$ along the particle's worldlines. Using
(\ref{MOTION}), this operator can be rewritten as \st
\be
\L f=\lp p^\mu\frac{\partial}{\partial x^\mu} -\Gamma^\mu_{\nu\rho}
p^\nu p^\rho\frac{\partial}{\partial p^\mu}\rp f.\label{DEFLIOU2}
\ee
The fact that the mass $m$ of Eq. (\ref{MASS})
is a scalar constant on each phase orbit leads to
\st
\be
\L (m^2)=0.\label{MASSLIOU}
\ee
The Boltzmann equation states that this rate of change is equal to the
rate of change due to collisions, i.e. that
\st
\be
\L f=C[f]. \label{DEFEQBOL}
\ee
$C[f]$ is the collision term and encodes the information about the
interactions between the particles of the fluid.\\

If we now consider a system of $N$ fluids (labelled by $i,j...$),
each of which is described by its distribution function $f_i(x^\mu,p^\mu)$,
the Boltzmann equation for a given fluid $i$ becomes
\st
\be
\L f_i=\sum_{j}C_j[f_i,f_j]\equiv C_i[f_i],
\ee
$C_j[f_i,f_j]$ is the collision term  describing the
interaction between the fluid $i$ and the fluid $j$. For elastic
collisions, it must satisfy the symmetry
\st
\be
C_j[f_i,f_j]=C_i[f_i,f_j],\label{SIMCOLL}
\ee
which means that in a collision between $i$ and $j$ the two distribution
functions undergo the same change.   If we assume the gas is dilute (i.
e. the mean free path of particles is much greater than the range of
interactions between them) such that we can neglect collisions between
its particles, the RHS of Eq. (\ref{DEFEQBOL}) will be zero.   Therefore we
find
\st
\be
\L f=0. \label{LIO}
\ee
This is the general relativistic Liouville's equation.   The equation describes
the evolution of distribution function, $f$, of a collisonless gas.  In the
mass-shell hypersurface, $\P_m$, which is defined in Eq. (\ref{MASSHELL}),
Liouville's equation for all particles with the constant mass $m$ reduces to
\st
\be
\L_m f=
\lp p^\mu\frac{\partial}{\partial x^\mu} -\Gamma^i_{\mu\nu}
p^\mu p^\nu\frac{\partial}{\partial p^i}\rp f =0,\label{DEFLIOU3}
\ee
where $\L_m$ is the Liouville operator, $\L$, in the mass-shell $\P_m$.

\section{Macroscopic quantities}\label{par3}
In section 2.2, we have introduced distribution functions and the basic
evolution equations of the system in general relativity.
The goal of this section is to
define a set of macroscopic quantities from the distribution function
and the collision term and then find the relations between these quantities.
  Given a distribution function $f$,
at any point $x^\mu$, one can introduce, following Ellis et al.
\cite{El83}, a set of macroscopic quantities associated with fluid
by \st \bea {X}_a^{\mu_1...\mu_n}(x^\mu)&&=\int_{\T_x}\lp -p^\mu
p_\mu\rp^{a/2} p^{\mu_1}...
p^{\mu_n}f_i(x^\mu,p^\mu)\pi_+(p^\mu)\nonumber \\ &&=\int_{\P_m}
m^ap^{\mu_1}... p^{\mu_n}f_i(x^\mu,p^\mu)\pi_m, \label{eDEF1} \eea
where $m$ is the mass of the particles (defined by Eq.
(\ref{MASS})) and $a$ an integer. If the particles of a given
fluid have different rest mass (this is the case e.g. when one is
dealing with a fluid of stars or of galaxies) the above equation
should be modified by an integration over $m$ (see Uzan
\cite{Uza}).   We assume that each distribution function vanishes
at infinity on the mass shell rapidly enough so that all these
integrals converge.

Among all these quantities, some are important in many applications,
\st
\bea
&&n^\mu\equiv X_0^\mu =\int p^\mu f(x^\mu, p^\mu)\pi_m,\\
\st
&&T^{\mu\nu}\equiv X_0^{\mu\nu} =\int p^\mu p^\nu f(x^\mu,
p^\mu)\pi_m.\label{Tmn} \eea
The vector $n^\mu$ is the number flux vector which is used
to define the average number flux velocity vector $v^\mu$ and the proper
density $n$ measured by an observer comoving with the fluid by
\be
n^\mu=nv^\mu,\quad v^\mu v_\mu=-1.
\ee
$T_{\mu\nu}$ is the energy-momentum tensor.
In terms of the timelike unit vector field $u^\mu$, chosen as
time direction, we can
split the energy-momentum tensor under the general form
\st
\be T_{\mu\nu}=(\rho + P) u_\mu u_\nu+P g_{\mu\nu}+ 2(q_\mu u_\nu + q_\nu
u_\mu )+ \pi_{\mu\nu},\label{eTMUNU}\ee
the quantities $\rho, P, q^\mu$ and $\pi_{\mu\nu}$ being defined as
\st
\bea
&&\rho\equiv T_{\mu\nu}u^\mu u^\nu,\label{2r1}\\
\st
&&P\equiv\frac{1}{3}T_{\mu\nu} h^{\mu\nu},\label{2r2}\\
\st
&&q_\mu\equiv- h^\nu_\mu T_{\nu\alpha} u^\alpha,\label{2r3}\\
\st
&&\pi_{\mu\nu}\equiv h^\alpha_\mu h^\beta_\nu T_{\alpha\beta}-
P h_{\mu\nu},\label{2r4}
\eea
where $h_{\mu\nu}=g_{\mu\nu}+u_\mu u_\nu$.
This decomposition is the most general splitting with respect to the
arbitrary vector field $u^\mu$ of a tensor of rank 2.
The four quantities $\rho$, $P$, $q_\mu$, and $\pi_{\mu\nu}$ are respectively
called the energy density, the pressure,
the energy flux vector, and the anisotropic stress tensor.  For the latters one
can  verify, from Eqs. (\ref{2r1}) -(\ref{2r4})

\st
\begin{equation} q_\mu u^\mu=\pi^\mu_\mu=\pi_{\mu\nu}u^\mu=0.
\end{equation}

The energy momentum tensor, (\ref{Tmn}), appears in the RHS of
Einstein's field equations \st\be
R_{\mu\nu}-\frac{1}{2}g_{\mu\nu}R=-\frac{8\pi
G}{c^4}T_{\mu\nu},\label{eins} \ee where $R_{\mu\nu}$ and $R$ are
the Ricci tensor and the scalar curvature respectively. Therefore,
in applications to self gravitating stars and stellar systems, one
should combine Einstein's field equations and Liouville's
equation, (\ref{DEFLIOU3}).   The resulting nonlinear equations
can be solved in certain approximations.

In the next chapter we adopt the post-Newtonian
approximation to study a self gravitating system.
We derive Liouville's equation in this
approximation by using the standard post-Newtonian expansion.

\setcounter{sub}{0}
\setcounter{subeqn}{0}
\renewcommand{\theequation}{3.\thesub\thesubeqn}

\chapter{
Liouville's equation in post Newtonian
approximation}

Solutions of general relativistic Liouville's equation ($grl$) in
a prescribed space-time have been considered by some
investigators. Most authors have sought its solutions as functions
of the constants of motion, generated by Killing vectors of the
space-time in question.   See for example Ehlers \cite{Ehl77}, Ray
and Zimmerman \cite{RZi}, Mansouri and Rakei \cite{MRa}, Ellis,
Matraverse and Treciokas \cite{El83}, Maartens and Maharaj
\cite{MMah}, Maharaj and Maartens \cite{MMaa}, Maharaj \cite{Mah},
and Dehghani and Rezania \cite{DRe96}.

In application to self gravitating stars and stellar systems, however, one
should combine Einstein's
field equations and $grl$.   The resulting nonlinear equations can
be solved in certain approximations.
Two such methods are available;
 the {\it post-Newtonian (pn) approximation} and the {\it weak-field
} one. In this chapter we adopt the first approach to study a self
gravitating system imbeded in an otherwise flat space-time.  In
section \ref{sec31}, we derive the $pn$ approximation of the Liouville
equation ($pnl$). In section \ref{sec32} we find two integrals of
$pnl$ that are the $pn$ generalizations of the energy and angular
momentum integrals of the classical Liouville's equation.
Post-Newtonian polytropes, as simultaneous solutions of $pnl$ and
Einstein's equations, are discussed and calculated in section
\ref{sec33}. %Section \ref{sec34} is devoted to concluding remarks.

The main objective of this chapter, however, is to set the stage for the
chapter 4.     There, we study a class
of
non static oscillatory solutions of $pnl$, which in their
hydrodynamical behavior are different from the conventional
$p$ and $g$ modes of the system.      They are a class of toroidal motions
driven by $pn$ force terms and are accompanied by oscillatory variations of
certain components of the space-time metric.

\section{Liouville's equation in post-Newtonian
approximation, General}\label{sec31}
The one particle distribution function
of a gas of collisionless particles with identical mass $m$, in
the restricted seven dimensional phase space \st
\be
P(m):\;\;g_{\mu \nu} U^\mu U^\nu = -c^2
\ee
satisfies $grl$:
\st
\be
\L_U F = (U^\mu \frac{\partial}{\partial x^\mu} - \Gamma_{\mu\nu}^i U^\mu
U^\nu \frac{\partial}{\partial U^i}) F(x^\mu,U^i) = 0,
\ee
where $(x^\mu,U^i)$ is the set of configuration and velocity coordinates in
 $P(m)$, $F(x^\mu,U^i)$ is a distribution function,
$\L_U$
is Liouville's operator in the $(x^\mu,U^i)$ coordinates, $\Gamma^i
_{\mu\nu}$ are Christoffel's symbols, and $c$ is the speed of light.
Greek indices run from 0 to 3 and Latin indices from
1 to 3.
  The four-velocity of the particle and
its classical velocity are related as
\st
\be
U^\mu = U^0 v^\mu;\;\;\;\;  v^\mu = (1, v^i = dx^i/dt),
\ee
where $U^0(x^\mu,v^i)$ is to be determined from Eq. (3.1).
In $pn$ approximation, we need an expansion of
$\L_U$ up to the order $(\bar{v}/c)^4$, where $\bar{v}$ is a
typical Newtonian speed.   To achieve this goal we
transform $(x^\mu,U^i)$ to $(x^\mu,v^i)$.
Liouville's operator transforms as
\st
\be
\L_U = U^0 v^\mu (\frac{\partial}{\partial x^\mu} + \frac{\partial v^j}
{\partial x^\mu} \frac{\partial}{\partial v^j}) - \Gamma^i _{\mu\nu}U^{0^2}
v^\mu v^\nu \frac{\partial v^j}{\partial U^i} \frac{\partial}{\partial v^j},
\ee
where  $\partial v^j/ \partial x^\mu$ and $\partial v^j/ \partial U^i$ are
determined
from the inverse of the transformation matrix (see appendix B).
Thus,
\st
\stq
\bea
&&\frac{\partial v^j}{\partial x^\mu} = - \frac{U^0}{2Q} v^j
\frac{\partial g_{\alpha \beta}}{\partial x^\mu} v^\alpha v^\beta,\\
\stq
&&\frac{\partial v^j}{\partial U^i} =
\;\; \frac{1}{Q} v^j (g_{0i} + g_{ik} v^k);\hspace{2cm}\;\;\mbox{for $i
\neq j$},\nonumber\\
&&\\
&&\hspace{.81cm}=-\frac{1}{Q} (U^{0^{-2}} + \sum_{k \neq i} v^k (g_{0 k}
+ g_{kl} v^l));\;\;  \mbox{for $i = j$},\nonumber
\eea
where
\stq
\be
Q = U^0 (g_{0 0} + g_{0 l} v^l).
\ee
Substituting Eqs. (3.5) in Eq. (3.4) gives
\st
\stq
\be
\L_U F =U^{0} \L_v F=0,
\ee
or
\stq
\be
\L_v F(x^\mu,v^i) = 0,
\ee
where
\stq
\bea
&&\hspace{-.8cm}\L_v= v^\mu (\frac{\partial}{\partial x^\mu} - \frac{U^0}{2Q} v^j
\frac{\partial g_{\alpha \beta}}{\partial x^\mu} v^\alpha
v^\beta\frac{\partial}{\partial v^j})
- \Gamma^i _{\mu\nu}U^0v^\mu v^\nu\{\sum_{j\neq i} \frac{1}{Q} v^j (g_{0i}
+ g_{ik} v^k)  \frac{\partial}{\partial v^j}  \nonumber \\
&&\hspace{4cm} - \frac{1}{Q} (U^{0^{-2}} + \sum_{k \neq i} v^k (g_{0 k}
+ g_{kl} v^l)) \frac{\partial}{\partial v^i}   \},
\eea
We caution that the post-Newtonian hydrodynamics is obtained from
integrations
of Eq. (3.6a) over the $\v$-space rather than Eq. (3.6b) (see appendix C).
Next we expand
$\L_v$ up to order $(\bar{v}/c)^4$.  For this purpose, we need
expansions of Einstein's field equations, the metric tensor, and the affine
connections up to various orders.
Einstein's field equation with harmonic coordinate
conditions, $g^{\mu\nu}\Gamma^{\lambda}_{\mu\nu}=0$, yields
(see appendix A):
\st
\stq
\bea
&&\nabla^2\; ^2g_{00} = -\frac{8\pi G}{c^4}\; ^0T^{00},\\
&&\nabla^2\; ^4g_{00} = \frac{\partial^2\; ^2g_{00}}{c^2\;\partial t^2} +
\;^2g_{ij} \frac{\partial^2\; ^2g_{00}}{\partial x^i \partial x^j} -
(\frac{\partial\; ^2g_{00}}{\partial x^i})(\frac{\partial\; ^2g_{00}}
{\partial x^i})\nonumber\\
\stq
&&\hspace{2cm}-\frac{8\pi G}{c^4} (\;^2T^{00} - 2\; ^2g_{00}\; ^0T^{00} +\;
^2T^{ii}),\\
\stq
&&\nabla^2\; ^3g_{i 0} =\frac{16\pi G}{c^4}\; ^1T^{i0},\\
\stq
&&\nabla^2\; ^2g_{ij} = -\frac{8\pi G}{c^4} \delta_{ij}\; ^0T^{00}.
\eea
The symbols $^ng_{\mu\nu}$ and $^nT^{\mu\nu}$ denote the $n$th order terms in
${\bar v/c}$ in
the metric and in the energy-momentum tensors, respectively.    Solutions of
Eqs. (3.7) are
\st
\stq
\bea
&&^2g_{00} = - 2 \phi/c^2,\\
\stq
&&^2g_{ij} = - 2 \delta_{ij} \phi/c^2, \\
\stq
&&^3g_{i 0} = \xi_i/c^3,   \\
\stq
&&^4g_{00} = -2(\phi^2+ \psi)/c^4,
\eea
where
\st
\stq
\bea
&&\phi(\x,t) = -\frac{G}{c^2} \int \frac{^0T^{00}(\xp,t)}{\vert{\bf
x} - \xp \vert} d^3 x', \\
\stq
&&\xi^i (\x,t) = -\frac{4G}{c} \int \frac{^1T^{i 0} (\xp,t)}{\vert
\x - \xp \vert} d^3 x',\\
\stq
&&\psi(\x,t) = - \int \frac{d^3 x'}{\vert \x - \xp \vert }
\left[ \frac{1}{4\pi} \frac{\partial^2 \phi(\xp,t)}{\partial t^2}
+G\; ^2T^{00}(\xp,t)\right. \nonumber\\
\stq
&&\hspace{6cm}\left. +G\;^2T^{ii}(\xp, t)\frac{}{}\right],
\eea
where a bold character denotes a three-vector.
Substituting Eqs. (3.8) and (3.9) in (3.6c) gives
\st
\bea
\L_v &= &\Lcl +\Lpn  \nonumber\\
                  & =&\frac{\partial}{\partial t} + v^i
\frac{\partial}{\partial x^i} - \frac{\partial \phi}{\partial x^i}
\frac{\partial}{\partial v^i} \nonumber\\
     &&- \frac{1}{c^2}[(4\phi + \v^2) \frac{\partial \phi}{\partial x^i} -
\frac{\partial \phi}{\partial x^j} v^i v^j -  v^i \frac{\partial
\phi}{\partial t} + \frac{\partial \psi}{\partial x^i} \nonumber\\
                  &&+ (\frac{\partial \xi_i}{\partial x^j} -  \frac{\partial
\xi_j}{\partial x^i}) v^j+\frac{\partial\xi_i}{\partial t}]
\frac{\partial}{\partial v^i}
\eea
where $\Lcl $ and $\Lpn $ are the
classical Liouville
operator and its post-Newtonian correction,
respectively.      Equation (3.6b) for the
distribution function $F(x^\mu,v^i)$ becomes
\st
\be
(\Lcl +\Lpn ) F(t, x^i ,v^i) = 0. \ee The classical Liouville's
equation and its symmetries have been studied extensively by
Sobouti \cite{Sob89a,Sob89b,Sob84,Sob85,Sob86}; Sobouti and Samimi
\cite{SSa89}; Samimi and Sobouti \cite{SSo95}; Sobouti and
Dehghani \cite{SDe92}; Dehghani and Sobouti \cite{DSo93,DSo95}.\\
The three scalar and vector potentials $\phi,\psi$ and $\xxi$ can
now be given in terms of the distribution function.
    The energy-momentum tensor in terms of $F(x^\mu, U^i)$ is
\st
\be
T^{\mu\nu}(x^\lambda)=\int \frac{U^\mu U^\nu}{\mid U_0 \mid } F(x^\lambda, U^i)
\sqrt{-g}d^3U,
\ee
where $g=det(g_{\mu\nu})$.     For various orders of $T^{\mu\nu}$
 one finds
\st
\stq
\bea
&&^0T^{00}(x^\lambda) =c^2 \int  F(x^\lambda,v^i) d^3 v, \\
\stq
&&^2T^{00}(x^\lambda) = \int (v^2 + 2\phi(x^\lambda))
F(x^\lambda,v^i) d^3 v,\\
\stq
&&^2T^{ij}(x^\lambda)= \int v^i v^j F(x^\lambda,v^i) d^3 v, \\
\stq
&&^1T^{0i}(x^\lambda) =c \int v^i F(x^\lambda,v^i) d^3 v.
\eea
Substituting Eqs. (3.13) in (3.9) gives
\st
\stq
\bea
&&\phi(\x,t) =-G\int \frac{F(\xp,t,\vp)}{\vert
\x - \xp \vert} d \Gamma',\\
\stq
&&\xxi (\x,t) = -4G \int
\frac{
\vp F(\xp,t,\vp)}{\vert \x - \xp \vert} d\Gamma'\\
&&\psi(\x,t) = \frac{G}{4 \pi} \int \frac{\partial^2
F(\xp,t,\vpp)/\partial t^2 }{
\vert \x - \xp \vert \vert
\x - \xp \vert} d^3x'd\Gamma'' \nonumber\\
&&\hspace{2cm}- 2G \int \frac{\vp^2 F(\xp,t, {\bf
v'})}{\vert \x - \xp \vert} d \Gamma'\nonumber\\
\stq
&&\hspace{2cm}+2G^2 \int \frac{F(\xp,t,\vp) F(\xpp
,t,\vpp)}{\vert \x - \xp
\vert \vert \xp - \xp \vert} d \Gamma' d
\Gamma'',
\eea
where $d\Gamma=d^3xd^3v$.      Equations (3.11) and (3.14) complete the
$pn$ order of Liouville's equation for self gravitating systems embeded in
a flat space-time.
\section{Post-Newtonian Liouville's equation: Static solutions}\label{sec32}
In the last section we obtained Liouville's equation in $pn$ approximation.  In
this section we seek static solutions of $pnl$, $F(\x, \v)$.   In this
time-independent regime
macroscopic velocities along with the vector potential
$\xxi$ vanish.  Equations (3.10)
and (3.11) reduce to
\st
\bea
&&(\Lcl +\Lpn )F(\x,\v)=[(v^i\frac{\partial}
{\partial x^i}-
\frac{\partial\phi}{\partial x^i}\frac{\partial}{\partial v^i})\nonumber\\
&&\hspace{1cm}-\frac{1}{c^2}\lp\frac{\partial\phi}{\partial x^i}(4\phi+v^2)-
\frac{\partial\phi}{\partial x^j}v^iv^j+\frac{\partial\psi}{\partial
x^i}\rp \frac{\partial}{\partial v^i}]F=0,
\eea
One easily verifies that the following, a generalization of the classical
energy integral, is a solution of Eq. (3.15)
\st
\bea
&&E=\frac{1}{2}v^2+\phi+(2\phi^2+\psi)/c^2.
\eea
The first two terms is exactly classical energy integral and the other terms
come out from $pn$ correction.
Furthermore, if $\phi(\x)$ and $\psi(\x)$ are spherically symmetric,
which actually is the
case for an isolated nonrotating system in an asymptotically flat space-time,
the following generalization of angular momenta are also integrals of
Eq. (3.15)
\st
\bea
&& l_i=\varepsilon_{ijk}x^jv^k\; exp(-\phi/c^2)\approx
\varepsilon_{ijk}x^jv^k(1-\phi/c^2),
\eea
where $\varepsilon_{ijk}$ is the Levi-Cevita symbol.      Static distribution
functions maybe constructed as functions of $E$ and even functions of
$l_i$.   The reason for restriction
to even functions of $l^{pn}_i$ is to ensure the vanishing of $\xi^i$,
the condition for validity of Eq. (3.15).

\section{Post-Newtonian polytropes}\label{sec33}
In addition to hydrodynamics equations, one need an equation of
state to determine comletely a theoretical model for a star.
The equation of state describe a relation between the mass density and
the pressure of the system.
In order of choosing equation of state, the theoretical models are different
for a star.  Polytropic model is a simple theoretical model to describe the
equilibrium of star.   It relates the pressure to the mass density to power of
$\Gamma$, the adiabatic index.   Classical polytropic model are studied by
Eddington \cite{Edd}.

As in classical polytropes we
consider the distribution
function for a polytrope of index $n$ as
\st
\bea
&&F_n(E)=\frac{\alpha_n}{4\pi\sqrt{2}}(-E)^{n-3/2};\;\;
\mbox{for}\;\; E< 0, \nonumber\\
&&\hspace{1.3cm}=0\hspace{3.00cm}\mbox{for}\;\; E> 0,
\eea
where $\alpha_n$ is a constant.
By Eqs. (3.13) the corresponding orders of the energy-momentum tensor are
\st
\stq
\bea
&&^0T^{00}_n=\alpha_n\beta_n c^2 (-U)^n,\\
\stq
&&^2T^{00}_n=2\alpha_n\beta_n\phi (-U)^n +2\alpha_n\gamma_n
(-U)^{n+1},\\ \stq
&&^2T^{ii}_n\;\;=\delta_{ij}\;^2T^{ij}=2\alpha_n\gamma_n (-U)^{n+1},\\
\stq
&&^1T^{0i}_n=0,
\eea
where
\st
\bea
&&\beta_n=\int^1_0(1-\zeta)^{n-3/2}\zeta^{1/2}d\zeta=\Gamma(3/2)\Gamma(n-1
/2)/\Gamma(n+1)\,,~~~~~~~\\
\st
&&\gamma_n=\int^1_0(1-\zeta)^{n-3/2}\zeta^{3/2}d\zeta=\Gamma(5/2)\Gamma(n-1/
2)/\Gamma(n+2)\,,~~~~~~~~~
\eea
and $U=\phi+2\phi^2/c^2+\psi/c^2$ is the gravitational potential in $pn$ order.
It will be chosen zero at the
surface of the stellar configuration.     With this choice, the escape velocity
$v_e=\sqrt{-2U}$ will mean escape to the boundary of the system rather
than to infinity.   Einstein's equations, Eqs. (3.7), (3.8) and (3.9), lead to
\st
\bea
&&\nabla^2\phi=\frac{4\pi G}{c^2}\; ^0T^{00}=
4\pi G\alpha_n\beta_n(-U)^n,\\
\st
&&\nabla^2\psi=4\pi G (^2T^{00}+^2T^{ii})
=8\pi G\alpha_n\beta_n\phi (-U)^n\nonumber\\
&&\hspace{1cm}+16\pi G\alpha_n\gamma_n(-U)^{n+1}.
\eea
Expanding $(-U)^n$ as
\st
\be
(-U)^n=(-\phi)^n[1+n(2\phi+\frac{\psi}{\phi})/c^2],
\ee
and substituting it in Eqs. (3.22) and (3.23) gives
\st
\bea
&&\nabla^2\phi=4\pi
G\alpha_n\beta_n[(-\phi)^n-2n(-\phi)^{n+1}/c^2-n(-\phi)^{n-1}\psi/c^2
]\,,~~~~~~~~~~ \\
\st
&&\nabla^2\psi=4\pi
G\alpha_n\beta_n
(4\frac{\gamma_n}{\beta_n}-2) (-\phi)^{n+1}.~~~~~~~~
\eea
For
further reduction we introduce the
dimesionless quantities
\st
\stq
\bea
&&~~~~~~x\equiv a\; \zeta,\\
\stq
&&-\phi (x)\equiv \lambda \theta (\zeta),\\
\stq
&&-\psi (x)\equiv \lambda^2 \Theta (\zeta),\\
\stq
&&-\xi^i (x)\equiv \lambda^{3/2}\eta^i (\zeta),
\eea
where, in terms of $\rho_c$, the central density,
$\lambda=(\rho_c/\alpha_n\beta_n )^{1/n}$ and
$a^{-2}=4\pi G\rho_c/\lambda $.      Equations
(3.25) and (3.26) reduce to
\st
\stq
\bea
&&\nabla_\zeta^2\theta+\theta^n = qn(2\theta^{n+1} - \theta^{n-1}
\Theta),\\
\stq
&&\nabla_\zeta^2\Theta+(4\frac{\gamma_n}{\beta_n}-2) \theta^{n+1}=0,
\eea
where
$\nabla_\zeta^2= \frac{1}{\zeta^2}
\frac{d}{d\zeta}(\zeta^2\frac{d}{d\zeta})$.
The
dimensionless  $pn$ expansion parameter $q$ emerges as
\st
\be
q=\frac{4\pi G \rho_c a^2}{c^2}=\frac{R_s}{R} \frac{1}{2\zeta_1\mid\theta'
(\zeta_1)\mid},
\ee
where $R_s$ is the Schwarzschild radius, $R=a \zeta_1$ is the radius of system,
and $\zeta_1$ is the first zero of $\theta (\zeta)$, $\theta (\zeta_1)=0$.
The order of magnitude of $q$ varies from $10^{-5}$  for white dwarfs to
$10^{-1}$ for neutron stars.       For future reference, let us also note that
\st
\be
-U=\lambda[\theta + q (\Theta - 2\theta^2)].
\ee
We use a forth-order Runge-Kutta method to find
numerical solutions of the two coupled nonlinear differential Eqs. (3.28).
At the center we adopt
\st
\be
\theta (0)=1;\;\;\;\theta' (0)=\frac{d\theta}{d\zeta}\mid_0=0. \ee
In tables 3.1 and 3.2, we summarize the numerical results for the
Newtonian and post-Newtonian polytropes for different polytropic
indices and $q$ values. The $pn$ corrections tend to reduce the
radius of the polytrope. The larger the polytropic index and/or
$q$ the larger this reduction.

%end{document}

%\section*{Table Captions}
%\typeout{Table Captions}
%\addcontentsline{toc}{section}{Table 3.1}

%\noindent Table 3.1.  A comparison of the Newtonian and post-Newtonian
%polytropes at certain selected radii for $n$=1, 2, 3, 4 and 5,  and
%different values of $q$.\vspace{.5cm}\\
%\noindent Table 3.2.  Same as Table 3.1.
% $n$= 4 and 5.
%\vspace{.5cm}\\
%\newpage

%\setcounter{table}{1}
\begin{table*}

\caption{
A comparison of the Newtonian and post-Newtonian
polytropes at certain selected radii for $n$=1, 2, 3, 4 and 5,  and
different values of $q$.}
\begin{center}
\begin{tabular}{|c|c|c|c|c|c|}\hline
n &Polytropic & Newtonian   &\multicolumn{3}{c|}{$pn$ polytrope,
$\theta+q(\Theta-2\theta^2)$}\\\cline{4-6}
  &radius, $\zeta$ &polytrope, $\theta$& $q=10^{-5}$ & $q=10^{-3}$ &
$q=10^{-1}$ \\\hline   & 0.0000000  & 1.00000 & 1.00000 & 1.00000 & 1.00000\\
  & 1.0000000  & 0.84147 & 0.84147 & 0.84156 & 0.85043\\
  & 2.0000000  & 0.45465 & 0.45465 & 0.45470 & 0.46069\\
1 & 3.0383400  & 0.03393 & 0.03392 & 0.03358 &  0.00000    \\
  & 3.1403800  & 0.00039 & 0.00038 & 0.00000 &         \\
  & 3.1415800  & 0.00001 & 0.00000 &         &      \\
  & 3.1415930  & 0.00000 &         &         &      \\\hline
  & 0.0000000  & 1.00000 & 1.00000 & 1.00000 & 1.00000\\
  & 2.0000000  & 0.52984 & 0.52984 & 0.53005 & 0.55904\\
  & 4.0000000  & 0.04885 & 0.04884 & 0.04858 & 0.02500\\
2 & 4.1451500  & 0.02776 & 0.02775 & 0.02746 &  0.00000     \\
  & 4.3501500  & 0.00035 & 0.00033 & 0.00000 &      \\
  & 4.3528000  & 0.00001 & 0.00000 &         &       \\
  & 4.3529000  & 0.00000 &         &         &       \\\hline
  & 0.0000000  & 1.00000 & 1.00000 & 1.00000 & 1.00000\\
  & 3.0000000  & 0.38286 & 0.38286 & 0.38315 & 0.41848\\
  & 6.0000000  & 0.04374 & 0.04373 & 0.04338 & 0.01817\\
3 & 6.2838000  & 0.02854 & 0.02853 & 0.02816 & 0.00000\\
  & 6.8862000  & 0.00044 & 0.00043 & 0.00000 &      \\
  & 6.8964000  & 0.00001 & 0.00000 &         &              \\
  & 6.8967000  & 0.00000 &         &         &              \\\hline
\end{tabular}
\end{center}
\end{table*}
\newpage

\begin{table*}

\caption{
Same as Table 3.1}
\begin{center}
\begin{tabular}{|c|c|c|c|c|c|}\hline
n &Polytropic & Newtonian   &\multicolumn{3}{c|}{$pn$ polytrope,
$\theta+q(\Theta-2\theta^2)$}\\\cline{4-6}
  &radius, $\zeta$ &polytrope, $\theta$& $q=10^{-5}$ & $q=10^{-3}$ &
$q=10^{-1}$ \\\hline   & 0.0000000  & 1.00000 & 1.00000 & 1.00000 & 1.00000\\
  & 3.0000000  & 0.44005 & 0.44005 & 0.44022 & 0.46949\\
  & 6.0000000  & 0.17838 & 0.17838 & 0.17818 & 0.17746\\
  & 9.0000000  & 0.07955 & 0.07954 & 0.07919 & 0.06496 \\
4 &12.5013000  & 0.02350 & 0.02349 & 0.02304 & 0.00000 \\
  &14.0000000  & 0.00802 & 0.00801 & 0.00753 &         \\
  &14.8625000  & 0.00051 & 0.00050 & 0.00000 &         \\
  &14.9705000  & 0.00001 & 0.00000 &         &      \\
  &14.9713400  & 0.00000 &         &         &      \\\hline
  & 0.0000000  & 1.00000 & 1.00000 & 1.00000 & 1.00000\\
  & 5.0000000  & 0.28480 & 0.28480 & 0.28482 & 0.29394\\
  &10.0000000  & 0.11894 & 0.11894 & 0.11862 & 0.10940\\
4.5 &12.2000000& 0.08779 & 0.08779 & 0.08743 & 0.00000\\
  &15.0000000  & 0.06125 & 0.06125 & 0.06085 &       \\
  &20.0000000  & 0.03231 & 0.03230 & 0.03185 & \\
  &25.0000000  & 0.01498 & 0.01492 & 0.01444 & \\
  &30.0000000  & 0.00334 & 0.00333 & 0.00284 & \\
  &31.2256000  & 0.00107 & 0.00106 & 0.00000 &\\
  &31.7847000  & 0.00001 & 0.00000 &         &    \\
  &31.7878400  & 0.00000 &         &         &  \\\hline
\end{tabular}
\end{center}
\end{table*}

\setcounter{sub}{0}
\setcounter{subeqn}{0}
\renewcommand{\theequation}{4.\thesub\thesubeqn}

\chapter{
The post Newtonian modes}

In the last chapter we obtained Liouville's equation in $pn$ approximation.
Furthermore, we found the integrals of $pnl$, generalization of the classical
energy and angular momentum,
and constructed an equilibrium distribution function for the system.
In this chapter, we study the non-equilibrium state of a stellar
system in $pn$ approximation.   We assume a small perturbation in the system,
i.e. in the distribution function, and obtain the linearized Liouville's
equation.   Finally, using the linearized equation, we study normal modes of
the system in $pn$ approximation.
In this chapter all quantities are dimensionless.

In section \ref{sec41} we give the $pn$ order
of the linearized
Liouville equation that governs the evolution of small perturbations from an
equilibrium state.       In sections \ref{sec42} and \ref{sec43} we extract the
equation for a sequence of new modes that are generated solely by $pn$ force
but are absent in classical regime.
In section \ref{sec44} we explore the O(3) symmetry of the modes and classify
them on basis of this symmetry.
In section \ref{sec45} we study hydrodynamics of these modes.       In
section \ref{sec46} we seek a variational approach to the calculation of $pn$
modes and
give numerical values for polytropes.  %  Section \ref{sec47} is devoted to
%concluding remarks.

\section{Post Newtonian  Liouville'e equation, Linearized}\label{sec41}
In chapter 3 we obtained Liouville's equation in the post-Newtonian
approximation ($pnl$) for the
one particle distribution of a gas of collisionless particles as
\stepcounter{sub}
\begin{equation}
(-i\frac{\partial}{\partial t}+ {\cal L})F({\bf x}, {\bf u}, t)=
(-i\frac{\partial}{\partial t}+ {\cal L}^{cl} +
q{\cal
L}^{^{ pn }})F({\bf x}, {\bf u}, t)=0,
\end{equation}
where $({\bf x}, {\bf u})$ are phase
space coordinates, $q$ is a small
post-Newtonian expansion parameter, the ratio of Schwarzchild radius
to a typical spatial dimension of the system, Eq. (3.29).
The classical and post-Newtonian operators, ${\cal L}^{cl}$ and ${\cal
L}^{pn}$, respectively, are
\stepcounter{sub}
\stepcounter{subeqn}
\begin{eqnarray}
&&\hspace{-.8cm}{\cal L}^{cl}=-i(
u^i\frac{\partial}{\partial
x^i} +\frac{\partial \theta}{\partial x^i}\frac{\partial}{\partial u^i}),\\
\stepcounter{subeqn}
&&\hspace{-.8cm}{\cal L}^{^{ pn }}=-i[({\bf u}^2-4\theta)\frac{\partial\theta}{\partial
x^i} -u^i u^j\frac{\partial\theta}{\partial x^j}
-u^i\frac{\partial\theta}{\partial t}
+\frac{\partial\Theta}{\partial x^i}
+u^j(\frac{\partial\eta_i}{\partial x^j}-
\frac{\partial\eta_j}{\partial x^i})
+\frac{\partial\eta_i}{\partial t}]
\frac{\partial}{\partial u^i}.\nonumber\\
\end{eqnarray}
The imaginary factor $i$ is included for later convenience.
The potentials
$\theta ({\bf x}, t)$, $\Theta ({\bf x}, t)$ and
$\eta\hspace{-.2cm}\eta ({\bf x}, t)$, solutions of
Einstein's equations in  $pn$  approximation, are
\stepcounter{sub}
\stepcounter{subeqn}
\stepcounter{subeqn}
\begin{eqnarray}
&&\theta({\bf x},t) = \int \frac{F({\bf x'},t,{\bf u')}}{\vert
{\bf x} - {\bf x'} \vert} d \Gamma',\;\;\;\;\;\;
\eta\hspace{-.2cm}\eta ({\bf x},t) = 4 \int \frac{{\bf u'} F({\bf
x'},t, {\bf u')}}{\vert {\bf x} - {\bf x'} \vert} d\Gamma',
\hspace{.2cm}\;{\rm (4.3a,b)}\nonumber\\ &&\Theta({\bf x},t) =
-\frac{1}{4 \pi} \int \frac{\partial^2 F({\bf x''},t,{\bf
u''})/\partial t^2 }{ \vert {\bf x} - {\bf x'} \vert \vert  {\bf
x'} - {\bf x''} \vert} d^3x'd\Gamma'' + 2 \int \frac{{\bf u'}^2
F({\bf x'},t, {\bf u'})}{\vert {\bf x} - {\bf x'} \vert} d
\Gamma'\nonumber\\ \stq &&\hspace{2cm}-2\int \frac{F({\bf
x'},t,{\bf u'}) F({\bf x''},t,{\bf u''})}{\vert {\bf x} - {\bf x'}
\vert \vert {\bf x'} - {\bf x''} \vert} d \Gamma' d \Gamma'',\hspace{-1cm}
\end{eqnarray}
where $d\Gamma=d^3xd^3u$.    See chapter 3
for details.    In an equilibrium state, $F({\bf x}, {\bf u})$ is
time-independent. If, further, it is isotropic in ${\bf u}$,
macroscopic velocities along with the vector potential
$\eta\hspace{-.2cm}\eta $ vanish.    It is also shown in
chapter 3 that the following generalizations of the classical
energy and classical angular momentum are integrals of $pnl$:
\stepcounter{sub}
\stepcounter{subeqn}
\begin{eqnarray}
&&e=e^{cl}+qe^{^{ pn }}=\frac{1}{2}u^2-\theta+q(2\theta^2-\Theta),\\
\stepcounter{subeqn}
&& l_i=\varepsilon_{ijk}x^ju^kexp(q\theta)\approx l_i^{cl}(1+q\theta),
\end{eqnarray}
for spherically symmetric $\theta(r)$ and
$\Theta(r)$.
Equilibrium distribution functions in  $pn$  approximation can be constructed
as appropriate functions
of these
integrals.   In chapter 3 the $pn$ models of polytrope were studied in
this spirit.

Here we are interested in the time evolution of small deviations from
a static solution.    Let $F\rightarrow \;F(e)+ \delta F({\bf x}, {\bf u}, t)$,
$\mid\delta F\mid \ll F, \;\;\;\;\forall~~ {\bf x}, {\bf u}, t$.
Accordingly, the potentials split
into large and small components, $\theta(r)+\delta \theta({\bf x},
t),\;\; \Theta(r)+\delta \Theta({\bf x}, t)$ and
$\delta\eta\hspace{-.2cm}\eta({\bf x}, t)$  where
$r=\vert {\bf x}\vert$.    Both the large and small components, can be read
out from Eqs. (4.3).    Substituting this splitting in Eq. (4.1) and keeping
terms linear in $\delta F$ gives
\stepcounter{sub}
\begin{equation}
i\frac{\partial }{\partial t}\delta F=
{\cal L}\delta F
+\delta {\cal L}F(e),
\end{equation}
where ${\cal L}$ is now calculated from Eqs. (4.2) with
$\theta(r),\;\Theta(r)$ and $\eta\hspace{-.2cm}\eta=0$.   Thus
\stepcounter{sub}
\stepcounter{subeqn}
\begin{eqnarray}
&&~{\cal L}={\cal L}^{cl}+q {\cal L}^{pn},\\
\stepcounter{subeqn}
&& {\cal L}^{cl}=-i\left(u^i\frac{\partial}{\partial
x^i}+\frac{\theta'}{r}x^i\frac{\partial}{\partial u^i}\right)~~~~~~~
\theta'=d\theta/dr, \\
\stepcounter{subeqn}
&&{\cal L}^{pn}=-\frac{i}{r}\left\{[(u^2-4\theta)
\theta'+\Theta']x^i-\theta'({\bf x}\cdot{\bf u})u^i\right\}
\frac{\partial}{\partial u^i}.
\end{eqnarray}
For $\d\L$ Eqs. (4.2), similarly, give
\stepcounter{sub}
\stepcounter{subeqn}
\begin{eqnarray}
&&\d \L=\d \Lcl+q\d \Lpn,\\
\stepcounter{subeqn}
&&\delta {\cal L}^{cl}F(e)=-iF_eu^i\frac{\partial\delta\theta}{\partial
x^i}~~~~~~~~~ F_e=dF/de,\\
\stepcounter{subeqn}
&&\d\Lpn F(e)=-iF_e\left[u^i\frac{\partial}{\partial x^i}(\delta\Theta-
4\theta\delta \theta)-u^2\frac{\partial\delta\theta}{\partial
t}+u^i\frac{\partial\delta\eta_i} {\partial t}\right].
\end{eqnarray}
Equations (4.5)-(4.7) are the generalizations of the linearized
classical Liouville-Poisson equations to $pn$ order. The classical
case was studied briefly by Antonov \cite{Ant62}.     He separated
$\delta F$ into even and odd components in ${\bf u}$ and extracted
an eigenvalue equation for $\delta F_{odd}$. Sobouti
\cite{Sob89a,Sob89b,Sob84,Sob85,Sob86} elaborated on this
eigenvalue problem, studied some of its symmetries and approaches
to its solution.    Sobouti and Samimi \cite{SSa89}, and Samimi
and Sobouti \cite{SSo95} showed that Antonov's equation has an
O(3) symmetry and its oscillation modes can be classified by a
pair of eigennumbers $(j, m)$ of a pair phase space angular
momentum operators $(J^2, J_z)$.  In analyzing Eqs. (4.5)-(4.7) we
have heavily relied on these studies.

\section{The Hilbert space}\label{sec42}
Let $\H$ be the space of complex square integrable functions of phase
coordinates $(\x, \u)$ that vanish at the phase space boundary of the system:
\st
\be
\H :{f(\x, \u); \int f^*f\sqrt{-g} d\Gamma={\rm finite},~~ f({\rm
boundary})=0}, \ee where $\sqrt{-g}=1+2q\theta$ in $pn$ order.
Integrations in $\H$ are over the volume of the phase space
available to the system.   In particular the boundedness of the
system sets the upper limit of $u$ at the escape velocity
$\sqrt{2\theta}$, where $=\theta(\x)$ is the gravitational
potential at $\x$.   Thus, $f(\x, \sqrt{2\theta(\x)})=0$.\\ {\it
Theorem }: $\L =\Lcl +q\Lpn$ of Eqs. (4.6) is Hermitian in $\H$,

\st\be \int g^*(\L f)~(1+2q\theta)d\Gamma =\int(\L
g)^*f~(1+2q\theta)d\Gamma;~~~~~~ g, f \in \H \ee Proof:
Substituting Eqs. (4.6) in (4.9), carrying out some integrations
by parts over the $\x$ and $\u$ coordinates and letting the
integrated parts vanish on the pahse space boundary: \stq\bea \int
g^*(\L f)~(1+2q\theta)d\Gamma&&\hspace{-.5cm}=\int g^*(\Lcl+q\Lpn)
f~(1+2q\theta)d\Gamma,\nonumber\\ &&\hspace{-.5cm}=\int g^* \Lcl f d\Gamma +
q\lp 2\int \theta~g^* \Lcl f d\Gamma+\int g^*\Lpn f
d\Gamma\rp.\nonumber\\ \eea At the classical order, the classical
Liouville operator, $\Lcl$, is Hermitian in $\H$, \cite{Sob89a}:
$$ \int g^*(\Lcl f)~d\Gamma =\int(\Lcl g)^*f~d\Gamma;~~~~~~ g, f
\in \H $$ Therefore the first two terms in RHS of Eq. (4.9a) will

be \stq\bea &&\int (\Lcl g)^* f d\Gamma + 2q \int [\Lcl~( g
\theta)]^* f d\Gamma,\nonumber\\ &&=\int (\Lcl g)^* f d\Gamma + 2q
\int (\Lcl g )^* f \theta d\Gamma+2iq\int \x\cdot\u~\theta' g^* f
d\Gamma.~~~~~ \eea The third term, the post-Newtonian Liouville operator, $\Lpn$, at
the $pn$ order is not Hermitian, then \stq\be \int g^*\Lpn f
d\Gamma=\int (\Lpn g)^* f d\Gamma -2iq \int \x\cdot\u~\theta' g^*
f d\Gamma \ee The proof will be completed by adding Eqs. (4.9b)
and (4.9c), QED.

The term $\d\L$ is not, in general, Hermitian.   Nonetheless, one may proceed
as
Antonov did with the classical case and obtain a second order differential
operator (almost square of $\L +\d\L$) in some subspace of $\H$.
We are, however, pursuing a much simpler problem here in which $\d\L$ term
vanishes identically leaving Eq. (4.5) as an eigenvalue problem governed with
the Hermitian operator $\L$ alone.

\section{The post-Newtonian modes}\label{sec43}
The effect of $pn$ corrections on the classical solutions of Eq. (4.5) can be
analyzed by the usual perturbation techniques.   Whatever the procedure, the
first order corrections on the known modes will be small and will not change
their nature.  We will not pursue
such issues here.    The main interest of this work is to study a new
class of solutions of Eq. (4.5) that originate solely from the  $pn$  terms
and have no precedence in classical theories.    It is not difficult to
anticipate the existence of such modes.     Perturbations on an equilibrium
state, that are functions of classical integrals (energy and angular momentum,
say) do not disturb the equilibrium of the system at classical level.    That
is
they do not induce restoring forces in the system.    They, however, do so in
the $pn$ regime, and make the system oscillate about the $pn$ equilibrium
state.    Such perturbations may be considered as a class of infinitely
degenerate zero frequency modes of the classical system.      The $pn$ forces
unfold this degeneracy and turn them into a sequence of non zero frequency
modes distinct and uncoupled from the other classical modes.     We have termed
them as $pn$ modes.

A hydrodynamic interpretation of $pn$ modes is the following. In
spherically symmetric fluids, toroidal motions are neutral.
Sliding one spherical shell of fluid over the other is not opposed
by a restoring force.     The $pn$ forces or for that matter a
small magnetic field or a slow rotation (mainly through Coriolis
forces) gives rigidity to the system.     The fluid resists
against such displacements and a sequence of well defined toroidal
modes of oscillation develop.    See Sobouti \cite{Sob80}, Hasan
and Sobouti \cite{HSo87}, Nasiri and Sobouti \cite{NSo89}, and
Nasiri \cite{Nas92} for examples and typical calculations in the
case weak magnetic fields and slow rotations.

In the Fourier time transform of Eq. (4.5),
\st\stq\be
\L\d F+\d\L F(e)=\omega \d F,
\ee
we split $\d F$ into even and odd terms in $\u$.  Thus,
\stq\be
\d F(\x, \u)=G_-(\x, \u)+G_+(\x, \u),~~~~G_{\pm}(\x,\u)=\pm G_\pm(\x,\pm \u).
\ee
Considering the fact that both $\L$ and $\d\L$ are odd in $\u$, Eq. (4.10a)
splits accordingly:
\st\stq\bea
&&\L G_- +q\omega F_e u^2\d\theta=\omega G_+,~~~~~~~~~\\
\stq
&&\L G_+ -i F_e u^i\frac{\partial}{\partial x^i}\ls\d\theta +q (\d\Theta-4
\theta\d\theta)\rs-q\omega F_e u^i\d\eta_i=\omega G_-,~~~~~~~~~~~
\eea
where
\stepcounter{sub}
\stepcounter{subeqn}
\stepcounter{subeqn}
\begin{eqnarray}
&&\d\theta = \int \frac{G_+({\bf x'},{\bf u')}}{\vert {\bf x} -
{\bf x'} \vert} d \Gamma',~~~~ \eta\hspace{-.2cm}\eta = 4 \int
\frac{{\bf u'} G_-(\xp,\up)}{\vert \x - \xp \vert} d\Gamma',
\hspace{2cm}(4.12{\rm a},{\rm b})\nonumber\\ &&\Theta(\x,t) =
\frac{\omega^2}{4 \pi} \int \frac{ G_+(\xpp,\upp) }{ \vert \x -
\xp \vert \vert {\bf x'} - {\bf x''} \vert} d^3x'd\Gamma'' + 2
\int \frac{ {u'}^2 G_+({\bf x'}, {\bf u'})}{\vert {\bf x} - {\bf
x'} \vert} d \Gamma'\nonumber\\ \stepcounter{subeqn}
&&\hspace{2cm}-2\int \frac{G_+({\bf x'},{\bf u'}) F(e'')+F(e')
G_+({\bf x''},{\bf u''}) }{\vert {\bf x} - {\bf x'} \vert \vert
{\bf x'} - {\bf x''} \vert} d \Gamma' d \Gamma'',\hspace{-1cm}
\end{eqnarray}
Operating on Eq. (4.11a) by $\L$ and substituting for $\L G_+$
from Eq. (4.11b) gives a second order differential equation for
$G_-$: \st\stq\be \L^2 G_-=\omega^2 G_- + i\omega F_e
u^i\frac{\partial}{\partial x^i}\ls\d\theta +q (\d\Theta-4
\theta\d\theta)\rs+q\omega^2 F_e u^i\d\eta_i -q\omega F_e
\L(u^2\d\theta). \ee We now seek a solution of Eq. (4.13a) in the
form of classical energy and angular momentum integrals, $G_-(\x,
\u)=G_-(e^{cl}, l_i^{cl})$.    In the next section, after we
discuss the O(3) of Eq. (4.13a), we show that such solutions can
be chosen from among the eigenfunctions of a pair of phase space
angular momentum operators, ($J^2, J_z$).  We also show that for
such solutions $\d\theta$ and $\d\Theta$ vanish identically
reducing Eq. (4.13a) to \stq\be \L^2 G_-=\omega^2\lp G_- + q F_e
u^i\d\eta_i\rp. \ee Multiplying Eq. (4.13b) by $G^*_-$,
integrating over the phase space volume of the system, and
considering the facts that $\L=\Lcl +q\Lpn$ is Hermitian and $\Lcl
G_-(e^{cl}, l^{cl}_i)=0$, gives \st\stq\bea \int (\L G_-)^*\L G_-
(1+2q\theta) d\Gamma && =q^2 \int (\Lpn G_-)^*\Lpn G_-
(1+2q\theta) d\Gamma\nonumber\\ &&\hspace{-3cm}=\omega^2\ls\int G_-^*G_-
(1+2q\theta)d\Gamma +q\int G^*_- F_e u^i\d\eta_i (1+2q\theta)
d\Gamma \rs.\nonumber\\ \eea Equation (4.14a) shows that $\omega$
is of the same order of smallness as $q$. Thus, eliminating the
terms of order $q^3$, $\omega^2 q$ and higher reduces Eq. (4.14a)
to \stq\be \int (\Lpn G_-)^*\Lpn G_-  d\Gamma
=\frac{\omega^2}{q^2}\int G_-^*G_- d\Gamma. \ee Equation (4.14b)
provides a variational expression for $\omega^2$ and will be used
as such to calculate the allowable $\omega^2$.   The frequencies,
$\omega$, are real meaning that the corresponding deviations from
the equilibrium state are stable oscillation modes.
Furthermore, these perturbations will be different from the
conventional classical modes, for they are excited by $pn$ terms
in the equations of motion that are absent at classical level.

\section{O(3) symmetry of $\L=\Lcl+ q\Lpn$}\label{sec44}
For spherically symmetric potentials, $\theta (r)$ and $\Theta
(r)$, both $\Lcl$ and $\Lpn$ depend on the angle between $\x$ and
$\u$ and their magnitudes.    Simultaneous rotations of the $x$
and $u$ coordinates about the same axis by the same angle leaves
these operators form invariant.    The generator of such
simultaneous infinitesimal rotations on the function space $\H$ is
\st\be J_i=J_i^{\dagger}=-i\varepsilon_{ijk}\lp
x^j\frac{\partial}{\partial x^k}+ u^j\frac{\partial}{\partial
u^k}\rp, \ee which has the angular momentum algebra \st\be [J_i,
J_j]=i\varepsilon_{ijk} J_k. \ee Commutation of $J_i$ with $\Lcl$
was first established by Sobouti \cite{Sob89a}. Here we confine
the discussion to the symmetry of $\Lpn$. Straightforward
calculations reveal that \st\be [\Lpn, J_i]=0, \ee since \beas
&&\Lpn J_i =-\frac{1}{r}\varepsilon_{ijk}\{[( u^2-4\theta) \theta'
+ \Theta' ] x^j -\theta' (\x\cdot\u) u^j\}
\frac{\partial}{\partial u^k},\\ &&J_i \Lpn
=-\frac{1}{r}\varepsilon_{ijk}\ls \{[(u^2-4\theta) \theta' \right.
+ \Theta' ] x^j -\theta' (\x\cdot\u) u^j\}
\frac{\partial}{\partial u^k}\\ &&~~~~~~~~~~~~~~~~~~~~~~\left. +(2
u^j u^k -x^ju^k-x^ku^j)\theta' u^m\frac{\partial}{\partial u^m}
\rs. \eeas Thus, it is possible to choose the eigensolutions,
$G_-$ of Eq. (4.14b) simultaneously with those of $J^2$ and $J_z$.
The eigensolutions of the latter pair of operators are worked out
in the appendix D. They are of the form
$f(e^{cl},l_i^{cl})\Lambda_{jm}$; $j, m$ integers, where $f$ is an
arbitrary function of the classical integrals and $\Lambda_{jm}$
is a complex polynomial of order $j$ of the components of the
classical angular momentum, $l_i^{cl}$. The $x$ and $u$ parity of
$\Lambda_{jm}$ is that of $j$.  See appendix D for proofs this
statement.

We are now in a position to point out an interesting feature of the eigenmodes.
Both $\omega^2$ and $\L^2$ in Eq. (4.13b) and the integrals in Eq. (4.14b) are
real.    Thus, $G_-$ can be chosen real or purely imaginary.   By Eq. (4.11a),
then $G_+$ will be purely imaginary or real.    That is, an
eigensolution $\d F=G_- +G_+$ belonging to a nonzero $\omega$ is a complex
function of phase coordinates in which both the $x$ and $u$ parities of the
real and imaginary parts are opposite to each other.   This feature is shared
by the classical modes of the classical Liouville's and Antonov's equation.

In section \ref{sec46} we will take a variational approach to solutions of Eq.
(4.14b). As variational trial functions we will consider the following
\stepcounter{sub}
\begin{equation}
G_-=f_{jm}=f(e)Re\;\Lambda_{jm}=[\sum_{n=j+1}^N c_n (-e)^n] Re
\Lambda_{jm},\;\;\;\;\;j={\rm odd},~~ c_n={\rm
consts}.\label{pertdist}
\end{equation}
Combining
this with its corresponding even counterpart from Eq. (4.10a) we obtain
\stepcounter{sub}
\begin{equation}
\delta F_{jm}({\bf x}, {\bf u}, t)=(1+\frac{q}{\omega}{\cal L}^{pn})f_{jm}e^{-i
\omega t}.
\end{equation}
At this
stage let us note an important property of Liouville's equation.    If a pair
$(\omega, \delta F)$ is an eigensolution of Liouville's equation,
$(-\omega, \delta F^*)$ is another eigensolution.   This can be verified by
taking the complex conjugate of Eq. (4.10a).    These solutions, being complex
quantities, cannot serve as physically meaningful distribution functions.
Their real or imaginary parts, however, can.  With
no loss of generality we will adopt the real part.    Thus,
\stepcounter{sub}
\begin{equation}
Re\;\delta F_{jm}({\bf x}, {\bf u},
t)=f(e)Re\;\Lambda_{jm}\cos\omega t +i\frac{q}{\omega}{\cal L}^{pn}(f(e)
Re\;\Lambda_{jm})\sin\omega t.
\end{equation}

The eigenmodes of Eq. (4.10a) are $m$-independent.   By
$m$-independence we mean a) the eigenvalues $\omega$ do not depend
on $m$ and are $2j+1$ fold degenerate, and b) the expansion
coefficients, $c_n$, of Eq. (4.12) do not depend on $m$. {\it
Proof:}  From the appendix D, Eq. (D.4), $J_{\pm}=J_x\pm i J_y$
are ladder operators for $\Lambda_{jm}$.     Operating on $f_{jm}$
of Eq. (4.18) by $J_{\pm}$ will give the mode $f_{j,m\pm 1}$
without changing the expansion coefficients.     Secondly,
substituting $J_{\pm}f_{jm}=\sqrt{(j\mp m)(j\pm m+ 1)}f_{j,m\pm
1}$ in Eq. (4.14a) instead of $f_{jm}$, and noting that $f_{jm}$'s
can be normalized for all $m$'s, $\omega^2$ will remain unchanged.

\section{Hydrodynamics of $pn$ modes}\label{sec45}
In this section we calculate the density fluctuations, macroscopic velocities,
and the perturbations in the space-time metric generated by a $pn$ mode.
It was pointed out earlier that for $j$ an odd integer, $f_{jm}({\bf x}, {\bf
u})$ of Eq. (4.18) is odd while ${\cal L}^{pn}f_{jm}$ is even in both ${\bf x}$
and ${\bf u}$.
The macroscopic velocities are obtained by multiplying Eq. (4.20) by ${\bf u}$
and integrating over the u-space.     Only the odd component of $\delta F_{jm}$
contributes to this bulk motion,
\stepcounter{sub}
\begin{equation}
\rho {\bf v}=\int f(e) Re\;\Lambda_{jm}{\bf u} d^3u\; \cos\omega t.
\end{equation}
In appendix D, Eqs. (D.11), we show that $\rho {\bf v}$ is a
toroidal spherical harmonic vector field. In spherical polar
coordinates it has the following form \stepcounter{sub}
\stepcounter{subeqn}
\begin{equation}
\rho (v_r,\; v_{\vartheta},\; v_{\varphi})=r^jG(v_{es})(0,\;
Re\;\frac{-1}{\sin\vartheta}\frac{\partial}{\partial\varphi}Y_{jm}(\vartheta,
\varphi),\; Re\;
\frac{\partial Y_{jm}}{\partial\vartheta}(\vartheta, \varphi))\;\cos\omega t,
\end{equation}
where
\stepcounter{subeqn}
\begin{equation}
G(v_{es})=\int_0^{v_{es}}f(e)u^{j+3}du,
\end{equation}
and $v_{es}=\sqrt{2\theta}$ is the escape velocity from the potential $\theta
(r)$.
The macroscopic density,
generated by the even component of Eq. (4.20), is
\stepcounter{sub}
\begin{eqnarray}
&&\delta \rho ({\bf x}, t)=i\frac{q}{\omega}\int
{\cal L}^{pn}(f(e)Re\;\Lambda_{jm})d^3u \sin\omega t\nonumber\\
&&\hspace{1.4cm}=2\frac{q}{\omega}\frac{\theta'}{r}{\bf x}\cdot\int
f(e)Re\;\Lambda_{jm}{\bf u}d^3u\; \sin\omega t =0.
\end{eqnarray}
The second integral is obtained by an integration by parts.    The vanishing of
it comes about because of the fact that the radial vector
${\bf x}$ is orthogonal to the toroidal vector $\rho {\bf v}$.
One also notes that $\nabla\cdot (\rho {\bf v})=0$.
It can further be verified that, the continuity
equation is satisfied at both classical and $pn$ level.

To complete the reduction of Eqs. (4.13) we should also show that
$\d\theta$ and $\d\Theta$ vanish.  The former is zero because $\d\rho=0$.  For
the latter, from Eq. (4.3c) and Eq. (4.20) for
$\delta F$, one has
\stepcounter{sub}
\begin{eqnarray}
&&\delta\Theta = \frac{\omega^2}{4\pi} \int \frac{\delta\theta
({\bf x'})}{
\vert {\bf x} - {\bf x'} \vert} d^3x'
-2\int\frac{\rho (r')\delta\rho ({\bf x''})+\delta\rho({\bf
x'})\rho (r'')}{\vert {\bf x} - {\bf x'}
\vert \vert {\bf x'} - {\bf x''} \vert} d^3x' d^3x''\nonumber\\
&&\hspace{2cm}+ 2\int\frac{d^3x'}{\vert {\bf x} - {\bf x'} \vert}\int
{\bf u'}^2 \delta F({\bf x'}, {\bf u'}) d^3u'=0.
\end{eqnarray}
The vanishing of the first two terms is obvious.
The third term vanishes because the integral over ${\bf u'}$ has the same form
as in $\delta\rho$ except for the additional scalar factor ${\bf u'}^2$.
Like $\delta\rho$ it can be reduced to the inner product of
the radial
vector ${\bf x}$ and a toroidal vector. QED.

The toroidal motion
described here
slides one spherical shell of the fluid over
the other without perturbing the density, the Newtonian gravitational field
and,
therefore, the hydrostatic equilibrium of the classical fluid.    In doing so,
it does not
affect and is not affected by the conventional classical modes of the fluid at
this first $pn$ order.

Nonetheless, the $pn$ modes are associated with space time perturbations.
From Eq. (3.8c) and Eq. (4.3b), $g_{0i}$ component of
the metric tensor is
\stepcounter{sub}
\begin{equation}
g_{0i}=\eta_i=4\int\frac{\rho v_i({\bf x'})}{\vert {\bf x}-{\bf x'}\vert}d^3x'.
\end{equation}
In spherical polar coordinates, one obtains
\stepcounter{sub}
\stepcounter{subeqn}
\begin{eqnarray}
&&\eta_r=0,\\
\stepcounter{subeqn}
&&\eta_{\vartheta}=-a_jRe\;\frac{1}{\sin\vartheta}
\frac{\partial}{\partial\varphi}Y_{jm}(\vartheta,\varphi)\cos\omega t,\\
\stepcounter{subeqn}
&&\eta_{\varphi}=a_j
Re\;\frac{\partial Y_{jm}}{\partial\vartheta}(\vartheta,\varphi)\cos\omega t,
\end{eqnarray}
where
\stepcounter{subeqn}
\begin{eqnarray}
&&\hspace{-1.5cm}a_j=\frac{16\pi}{2j+1}\left\{ \begin{array}{ll}
   (r/R)^jy_j(R)+(2j+1)r^j\int_r^R{r'}^{-j-1}y_j(r')dr'&{\rm for}\;r<R ~~\\
                               (R/r)^{j+1}y_j(R)&{\rm for}\;r>R
                                 \end{array}\right.         ~~~ \\
\stepcounter{subeqn}
&&\hspace{-1cm}y_j(r)=r^{-j-1}\int_0^r{r'}^{2j+2}G(\theta(r'))dr',\,\,\,\,\,\,\\
\stepcounter{subeqn} &&\hspace{-1cm}G(\theta
(r))=\int_0^{v_{es}}f(e)u^{j+3}du\nonumber  \\ &&\hspace{.5cm}
=2^{j/2+1}\Gamma(j/2+2) \Gamma(n+1)
\theta(r)^{n+j/2+2}/\Gamma(n+j/2+3),
\end{eqnarray}
where $R$ is the radius of the system and $\Gamma(n)$ is the gamma function.
The remaining components of the metric tensor remain unperturbed.

\section{Variational solutions of $pn$ modes}\label{sec46}
We substitute the trial function of Eq. (4.18) in Eq. (4.14b) and turn it into
a matrix equation.    Thus
\st\be
C^\dagger W C=\frac{\omega^2}{q^2} C^\dagger S C,
\ee
where $C=[c_n]$ is the column matrix of the variational coefficients of Eq.
(4.18), and the elements of $S$ and $W$ matrices are
\st
\stq
\bea
&&S_{pq}=\int(-e)^{p+q}\vert Re\;\Lambda_{jm}\vert^2 d\Gamma,\\
\stq
&&W_{pq}=\int({\cal L}^{pn}(-e)^p Re\;\Lambda_{jm})^*
({\cal L}^{pn}(-e)^q Re\;\Lambda_{jm})d\Gamma.
\eea
Minimizing $\omega^2$ with respect to variations of $C$ gives the following
matrix equation
\st
\be
WC=\frac{\omega^2}{q^2}SC.
\ee
Eigen $\omega$'s are the roots of the characteristic equation
\st
\be
\vert W-\frac{\omega^2}{q^2}S\vert=0.
\ee
For each $\omega$, Eq. (4.29) can then be solved for the eigenvector C.
This completes the Rayleigh-Ritz variational formalism of solving Eq. (4.14a).
In what follows we present some numerical values for polytropes.

\subsection{ pn Modes of polytropes belonging to $(j, m)=(1, m)$}
We analyze the case $m=0$, only.    From the $m$-independence of
eigenmodes (see theorem of section \ref{sec44}) the eigenvalue and
the expansion coefficients, $c_n$, for $m=\pm 1$ will be the same.
From Eqs. (D.9),
$\Lambda_{1\;0}=l_z=ru\sin\vartheta\sin\alpha\sin(\beta-\varphi)$,
where ($\vartheta,\varphi$) and ($\alpha,\beta$) are the polar
angles of ${\bf x}$, of ${\bf u}$, respectively. Substituting this
in Eqs. (4.28) and integrating over directions of ${\bf x}$ and
${\bf u}$ vectors and over $0<u<\sqrt{2\theta}$ gives \st \stq
\bea &&S_{pq}=\int_0^1\theta^{p+q+2.5}x^4dx,\\ \stq &&W_{pq}=\pi
G\rho_c\{(16a_{pq}-b_{pq})\int_0^1{\theta'}^2\theta^{p+q+3.5}x^4dx
\nonumber\\
&&\hspace{1.7cm}+(1-8a_{pq})\int_0^1\Theta'\theta'\theta^{p+q+2.5}x^4dx
\nonumber\\
&&\hspace{2.8cm}+a_{pq}\int_0^1{\Theta'}^2\theta^{p+q+1.5}x^4dx\},\\
\stq &&a_{pq}=\frac{pq(p+q+2.5)}{(p+q)(p+q-1)},\nonumber\\ &&\\ &&b_{pq}=
\frac{4(p+q)^2+9(p+q) -13}{(p+q-1)(p+q+ 3.5)},\;p,\;q\;=2, 3,
\cdots.\nonumber \eea Polytropic potentials $\theta$ and
$\Theta$ were obtained from integrations of Lane Emden equation
and Eqs. (3.28), respectively.    Eventually, the matrix elements
of Eqs. (4.31), the characteristic Eq. (4.30) and the eigenvalue
Eq. (4.29) were numerically solved in succession.   Tables 4.1-4.4
show some sample calculations for polytropes 2, 3, 4, and 4.9.
Eigenvalues are displayed in lines marked by an asterisks.     The
column following an eigenvalue is the corresponding eigenvector,
i. e. the values of $c_1,\;c_2,\; \cdots$, of Eq. (4.18).    To
demonstrate the accuracy of the procedure, calculations with six
and seven variational parameter are given for comparison. The
first three eigenvalues can be trusted up to two to four figures.
Convergence improves as the polytropic index, i.e. the central
condensation, increases.    Eigenvalues are in units of $\pi
G\rho_c q^2$ and increase as the mode order increases.

%\newpage
%end{document}

%\section*{Table Cpations}
%\typeout{Table Captions}
%\addcontentsline{toc}{chapter}{Table Captions}

%\noindent Table 4.1: $pn$ modes of polytrope n=2, belonging to $(j, m)=(1,0)$.
%Eigenvalues are in units $4\pi G\rho_c q^2$, $c_n$'s are the linear variational
%parameters of Eq. (22).
%A number $a\times 10^{\pm b}$ is
%written a $a\pm b$.    To appraise the accuracy of the computations two sets of
%data with six and seven variational parameters are given.     The first three
%eigenvalues are reliable up to three figures.     Characteristically, the
%accuracy deteriorates as one goes to higher order modes.
%\vspace{.5cm}\\
%\noindent Table 4.2: Same as Table 4.1. $n=3$ and $(j,m)=(1,0)$.
%\vspace{.5cm}\\
%\noindent Table 4.3: Same as Table 4.1. $n=4$ and $(j,m)=(1,0)$.
%\vspace{.5cm}\\
%\noindent Table 4.4: Same as Table 4.1. $n=4.9$ and $(j,m)=(1,0)$.
%\vspace{.5cm}\\
%\newpage

\begin{table*}

\caption{
$pn$ modes of polytrope n=2, belonging to $(j, m)=(1,0)$.
Eigenvalues are in units $4\pi G\rho_c q^2$, $c_n$'s are the linear variational
parameters of Eq. (4.22).
A number $a\times 10^{\pm b}$ is
written a $a\pm b$.    To appraise the accuracy of the computations two sets of
data with six and seven variational parameters are given.     The first three
eigenvalues are reliable up to three figures.     Characteristically, the
accuracy deteriorates as one goes to higher order modes.}

\begin{center}
\begin{tabular}{crrrrrrr}\hline
$\omega^2$& .1825+01& .4973+01& .6448+01& .1216+02& .3425+02& .1686+03&\\
&&&&&&&\\
$c_1$& .3113+02&-.8912+02& .1663+03& .1344+03& .7545+01&-.1399+04&\\
$c_2$& .3908+02& .1045+04&-.3234+04&-.9746+03&-.2392+04& .8484+04&\\
$c_3$&-.1420+03&-.6649+04& .1801+05& .4514+04& .7952+04&-.9647+04&\\
$c_4$& .5803+03& .1804+05&-.4351+05&-.7014+04&-.2607+03&-.2251+05&\\
$c_5$&-.9110+03&-.2210+05& .4724+05& .8324+03&-.1811+05& .5188+05&\\
$c_6$& .5252+03& .1020+05&-.1874+05& .2882+04& .1317+05&-.2717+05&\\
&&&&&&&\\
$\omega^2$&.1823+01&.4865+01&.5895+01&.9113+01&.1465+02&.4228+02&.3226+03\\
&&&&&&&\\
$c_1$& .3028+02&-.7086+02& .1529+03&-.3129+02& .1561+03&-.4624+02& .2042+04\\
$c_2$& .4812+02& .6908+03&-.2810+04& .1313+04&-.1513+04&-.2762+04&-.1461+05\\
$c_3$&-.1305+03&-.3993+04& .1702+05&-.5686+04& .6685+04& .1077+05& .2271+05\\
$c_4$& .2576+03& .8181+04&-.4788+05& .3425+04&-.3673+04& .1875+04& .4154+05\\
$c_5$& .1303+03&-.3086+04& .6823+05& .2433+05&-.2910+05&-.4718+05&-.1496+06\\
$c_6$&-.7534+03&-.7924+04&-.4771+05&-.4855+05& .5132+05& .5873+05& .1425+06\\
$c_7$& .5475+03& .6707+04& .1302+05& .2568+05&-.2386+05&-.2120+05&-.4423+05\\
\hline
&$pn_1$&$pn_2$&$pn_3$&$pn_4$&$pn_5$&$pn_6$&$pn_7$
\end{tabular}
\end{center}
\end{table*}
\newpage
\begin{table*}

\caption{
Same as Table 4.1. $n=3$ and $(j,m)=(1,0)$.}
\begin{center}
\begin{tabular}{crrrrrrr}\hline
$\omega^2$&.1534+01& .4836+01& .9473+01& .1938+02& .4083+02& .1128+03&\\
&&&&&&&\\
$c_1$& .9752+02&-.6975+02& .2464+03&-.2246+03&-.9102+03& .3169+04&\\
$c_2$& .3284+02&-.8725+03&-.1121+04&-.2590+04& .1713+05&-.2631+05& \\
$c_3$& .2096+03& .3859+04& .5591+04& .1444+05&-.1023+06& .6390+05& \\
$c_4$&-.5354+03&-.5728+04&-.1216+05&-.9903+04& .2599+06&-.3406+05& \\
$c_5$& .3941+03& .2528+04& .5215+04&-.2221+05&-.2933+06&-.4814+05& \\
$c_6$& .1803+01& .1125+04& .3307+04& .2153+05& .1208+06& .4268+05& \\
&&&&&&&\\
$\omega^2$&.1533+01& .4688+01& .7993+01& .9068+01& .1124+02& .1909+02& .1093+03\\
&&&&&&&\\
$c_1$& .9318+02&-.1440+03&-.1202+03&-.1069+04&-.5706+03&-.5482+02& .3703+04\\
$c_2$& .1121+03& .6997+03& .5482+04& .1856+05& .7685+04&-.5626+04&-.3381+05\\
$c_3$&-.2118+03&-.4506+04&-.2955+05&-.1063+06&-.4112+05& .3078+05& .1007+06\\
$c_4$& .2709+03& .9777+04& .5298+05& .2726+06& .7791+05&-.4371+05&-.1109+06\\
$c_5$& .1206+03&-.9309+03&-.6283+04&-.3375+06&-.1278+05&-.7049+03& .1239+05\\
$c_6$&-.7005+03&-.1574+05&-.7154+05& .1894+06&-.9027+05& .3228+05& .4581+05\\
$c_7$& .5309+03& .1200+05& .5087+05&-.3511+05& .5945+05&-.1218+05&-.1722+05\\
\hline
&$pn_1$&$pn_2$&$pn_3$&$pn_4$&$pn_5$&$pn_6$&$pn_7$
\end{tabular}
\end{center}
\end{table*}
\newpage
\begin{table*}

\caption{
Same as Table 4.1. $n=4$ and $(j,m)=(1,0)$.}
\begin{center}
\begin{tabular}{crrrrrrr}\hline
$\omega^2$&.7569+00& .2822+01& .5661+01& .8814+01& .1519+02& .6952+02&\\
&&&&&&&\\
$c_1$& .6291+03&-.1067+04& .2143+04&-.1949+04&-.6870+04& .1400+05&\\
$c_2$&-.9217+02& .1770+04&-.1693+05& .1131+05& .8373+05&-.2337+06&\\
$c_3$& .4162+03& .2808+04& .5682+05&-.3654+04&-.3195+06& .1293+07&\\
$c_4$&-.3883+04& .5860+04&-.1184+06&-.2807+05& .4791+06&-.3112+07&\\
$c_5$& .6427+04&-.2303+05& .1257+06&-.4668+04&-.2545+06& .3371+07&\\
$c_6$&-.3089+04& .1612+05&-.4514+05& .3416+05& .1251+05&-.1344+07&\\
&&&&&&&\\
$\omega^2$&.7569+00& .2813+01& .5021+01& .8747+01& .1272+02& .3322+02& .7683+02\\
&&&&&&&\\
$c_1$& .5590+03&-.8716+03& .2653+03&-.2421+04& .1881+04& .1412+05& .3376+05\\
$c_2$& .1189+04&-.2018+04& .1406+05& .1926+05&-.7436+04&-.2356+06&-.5191+06\\
$c_3$&-.6377+04& .2349+05&-.1057+06&-.4732+05&-.5363+05& .1298+07& .2528+07\\
$c_4$& .9376+04&-.3509+05& .2059+06& .6165+05& .2228+06&-.3112+07&-.4750+07\\
$c_5$& .5449+03&-.4645+04&-.2977+05&-.4272+05&-.7106+05& .3356+07& .2298+07\\
$c_6$&-.1192+05& .4364+05&-.2533+06&-.3854+05&-.4046+06&-.1333+07& .2455+07\\
$c_7$& .7228+04&-.2275+05& .1775+06& .5845+05& .3227+06&-.1382+03&-.2085+07\\
\hline
&$pn_1$&$pn_2$&$pn_3$&$pn_4$&$pn_5$&$pn_6$&$pn_7$
\end{tabular}
\end{center}
\end{table*}
\newpage
\begin{table*}

\caption{
Same as Table 4.1. $n=4.9$ and $(j,m)=(1,0)$.}
\begin{center}
\begin{tabular}{crrrrrrr}\hline
$\omega^2$&.4481+00& .1827+01& .4078+01& .6515+01& .1170+02& .1391+03&\\
&&&&&&&\\
$c_1$&-.2888+02& .1663+03&-.2794+03& .1593+03& .1405+03& .1081+05&\\
$c_2$&-.2440+03&-.7593+04& .2050+05&-.2099+05& .2665+05&-.2129+06&\\
$c_3$& .4933+05&-.2772+04&-.1400+06& .1883+06&-.3467+06& .1344+07&\\
$c_4$&-.1722+06& .1443+06& .2902+06&-.5138+06& .1372+07&-.3583+07&\\
$c_5$& .2124+06&-.2675+06&-.2194+06& .4871+06&-.2092+07& .4207+07&\\
$c_6$&-.8916+05& .1394+06& .5712+05&-.1179+06& .1073+07&-.1790+07&\\
&&&&&&&\\
$\omega^2$&.4380+00& .1805+01& .4006+01& .6190+01& .7980+01& .1439+02& .8964+02\\
&&&&&&&\\
$c_1$&-.1701+02& .1379+03&-.3341+03& .3427+03&-.3020+03& .7695+03& .8642+04\\
$c_2$&-.6649+03&-.6322+04& .2326+05&-.3097+05& .2196+05&-.1111+05&-.1534+06\\
$c_3$& .5135+05&-.1143+05&-.1601+06& .2940+06&-.2552+06& .1349+06& .8174+06\\
$c_4$&-.1667+06& .1599+06& .3264+06&-.9227+06& .1022+07&-.9712+06&-.1565+07\\
$c_5$& .1694+06&-.2551+06&-.1784+06& .1132+07&-.1574+07& .3018+07& .4968+06\\
$c_6$&-.1770+05& .8656+05&-.9582+05&-.4879+06& .7432+06&-.3959+07& .1421+07\\
$c_7$&-.3646+05& .3341+05& .9586+05& .2938+05& .8318+05& .1819+07&-.1047+07\\
\hline
&$pn_1$&$pn_2$&$pn_3$&$pn_4$&$pn_5$&$pn_6$&$pn_7$
\end{tabular}
\end{center}
\end{table*}

%***************************   Part II   *********************************

\part{The stability curve of r-modes}

\setcounter{sub}{0}
\setcounter{subeqn}{0}
\renewcommand{\theequation}{5.\thesub\thesubeqn}

\chapter{ Introduction}

Rotating neutron stars and black holes have been the objects of
many astrophysical studies in recent years. Their strong
gravitational fields make them ideal laboratories for testing
predictions of the theory of general relativity. Both types of
compact objects are interesting for different reasons. Black holes
are objects which have completely collapsed under their own
gravitational field. Since they are curved empty space, black
holes are relatively simple objects to describe. Most observed
phenomena from quasars (young, active galaxies) can be explained
consistently by assuming a central supermassive black hole exists
at the galaxy's core. In contrast neutron stars, possibly
densest configurations of matter which are stable to gravitational
collapse, have more complicated structures.   Their study against requires a
diverse range of physics. The interior structure of a neutron star
includes such features as a normal fluid coexisting with
superfluidity and superconductivity, various nuclear processes,
rapidly rotating configuration, strong magnetic fields, and many
other features.  They are one of the most fascinating
objects of theoretical investigation. Observational features, such
as strong X-ray emission, periodically  pulsating radio waves in a
relatively narrow beam, sudden spinning up of rotational frequency
(glitch), open a window for deep understanding of their internal
structure.

Neutron stars are probably among the most promising sources
of detectable gravitational waves by the future generation of gravitational wave
detectors.
Studies suggest that the
emitted gravitational radiation by  neutron stars in the
Virgo cluster could be detected by the laser interferometer
gravitational wave detectors such as GEO600, the advanced
Laser Interferometer Gravitational wave Observatory (LIGO),
and VIRGO \cite{Ste98}.  By 2001, both LIGO and GEO600
are expected to be sensitive to bursts of amplitude around
$10^{-21}$. VIRGO, with an even better sensitivity, could come
online in 2002 or later \cite{Sch99}.

Recently Andersson \cite{And98} numerically showed that in rotating systems
perturbations driven by Coriolis forces can be unstable at
arbitrarily small angular velocities. He considered a relativistic but slowly
rotating configuration and found
that toroidal perturbations, in the absence
of fluid viscosity, are unstable because of the emission of
gravitational radiation for $all$ rates of rotation.  These modes
which are the relativistic analouge of the Newtonaian $r$-modes
\cite{PPr78, Saio82}, are unstable even for very slowly rotating
perfect fluid stars.  Thus, the $r$-mode instability is different
to other mode instability like $f$-mode of the star that set at a
certain rate of rotation \cite{Com79}-\cite{SFr97}. Friedman and
Morsink \cite{FMo98} used the relativistic axisymmetric background
(with slowly rotating approximation) and showed that
analytically the instability is generic: `` every $r$-mode is in
principle unstable in every rotating star, in the absence of
viscosity ''. The mechanism of the instability can be understood
by the generic argument for the gravitational radiation driven
instability, so-called CFS-instability, after
Chandrasekhar
\cite{Cha70}, Friedman and Schutz \cite{FSc78}, and Friedman
\cite{Fri78}, who studied it first.

Further work which included the effect of viscosity showed the
gravitational radiation driven instability of the Coriolis modes
is important in the class of neutron stars which are born with rapid
rotations (such as the pulsar found in the supernova remnant
N157B).

The excitement over the r-mode instability has generated a large
literature in the past two years. Andersson et al. \cite{AKSc99}
and , Lindblom et al. \cite{LOM98} independently computed that
slowing down of a rapidly rotating, newly born star to typical periods of
Crab like pulsar ($\approx$ 19 ms) can be explained by the $r$-mode
instability.  This is due to the emission of current-quadrupole
gravitational radiations, which reduces the angular momentum of the
star.
Kokkotas and Stergioulas \cite{KSt98} investigated
analytically $r$-mode instability for a uniform density Newtonain
star and calculated the corresponding timescales and stability curve
associated with $r$-mode.
Lindblom, Owen, and Morsink \cite{LOM98}
also evaluated the $r$-mode
growing/damping timescale by considering fluid viscosity and
calculated critical angular velocities for a polytropic neutron
star model.  They
showed that the coupling of gravitational radiation to the $r$-modes is
sufficiently strong to overcome internal fluid dissipation effects and
so drive these modes unstable in hot young neutron stars.  This result which
has been verified by Andersson, Kokkotas, and Schutz \cite{AKSc99},
seemed somewhat surprising at first because the dominant
coupling of gravitational radiation to the $r$-modes is through the
{\it current multipoles} rather than the more familiar and usually dominant
mass multipoles.  But it is now generally accepted that gravitational
radiation does drive unstable any hot young neutron star with angular
velocity greater than about 5\% of the maximum (the angular velocity
where mass shedding occurs).  This instability therefore provides a
natural explanation for the lack of observed very fast pulsars
associated with young supernovae remnants.
Kojima
\cite{Koj98} suggested that, in contrast to Newtonian theory,
$r$-mode frequencies in general relativity become continuous.
This fact is verified mathematically by Beyer and Kokkotas
\cite{BKo99}.

The $r$-mode instability is also interesting as a possible source of
gravitational radiation.  In the first few minutes after the formation
of a hot young rapidly rotating neutron star in a supernova,
gravitational radiation will increase the amplitude of the $r$-mode
(with spherical harmonic index $m=2$) to levels where non-linear
hydrodynamic effects become important in determining its subsequent
evolution.  While the non-linear evolution of these modes is not well
understood as yet, Owen et al. \cite{Owe98} have developed a
simple non-linear evolution model to describe it approximately.  This
model predicts that within about one year the neutron star spins down
(and cools down) to an angular velocity (and temperature) low enough
that the instability is again suppressed by internal fluid
dissipation.  All of the excess angular momentum of the neutron star
is radiated away via gravitational radiation.  Owen et al.
 \cite{Owe98} estimated the detectability of the gravitational waves
emitted during this spindown, and found that neutron stars
spinning down in this manner may be detectable by the
second-generation (``enhanced'') LIGO interferometers out to the
Virgo cluster. Bildsten \cite{Bil98} and Andersson, Kokkotas, and
Stergioulas~\cite{AKSt98} have raised the possibility that the
$r$-mode instability may also operate in older colder neutron
stars spun up by accretion in low-mass x-ray binaries.  The
gravitational waves emitted by some of these systems (e.g.\ Sco
X-1) may also be detectable by enhanced LIGO \cite{BCr98}.  Thus,
the $r$-modes of rapidly rotating neutron stars have become a
topic of considerable interest in relativistic astrophysics.

Furthermore, the $r$-mode observations open a rich prospect for
gravitational astronomy.   Identifying hidden or unnoticed
supernovas would be the most exciting use of $r$-mode observation.
Another implication of $r$-mode observation is detection of
background radiation.  If we assume that most neutron stars are
born with rapid rotation, the background spectrum will reveal not
only the star formation rate in the early universe, but also tell
us about the distribution of initial rotation speeds of neutron
stars. Investigating $r$-mode events associated with known
supernovas is the other prospect of the $r$-mode observation. This
would give us some information about cooling rates, viscosity,
crust formation, the equation of state of neutron matter, and the
onset of superfluidity in neutron stars \cite{Owe98}.

%Andersson, Kokkotas
%and Stergioulas \cite{akst98}, Madsen \cite{mad98}, Hiscock \cite{h98},
%Lindblom and Ipser \cite{li98},
%Bildsten \cite{b98},
%Levin \cite{l98},
%Ferrari et al. \cite{fms98},
%Spruit \cite{s99},
%Brady and Creighton \cite{bc98},
%Lockitch and Friedman \cite{lf98},
%Lindblom et al. \cite{lmo99},
%,
%Kojima and Hosonuma \cite{kh99},
%Lindblom \cite{l99},
%Schneider et al. \cite{sfm99},
%Rezzolla et al. \cite{rea99},
%Rezzolla, Lamb and Shapiro
%Yoshida and Lee \cite{yl99})
%[Earlier studies of the axial-parity oscillations of
%models of the neutron star crust  were reported by van Horn \cite{vh80}
%and by Schumaker and  Thorne \cite{st83}); and Chandrasekhar and Ferrari
%discussed the resonant scattering of axial wave modes \cite{cf91b} and the
%coupling between axial and polar modes induced by stellar rotation
%\cite{cf91a}.]

In spite of the recent improvements in our understanding of this
instability, it seems that the fundamental properties of these
modes have not yet been sufficiently understood.  Previous
investigations of the $r$-modes are restricted to the case of
uniformly and slowly rotating, isentropic, Newtonian stars
\cite{AKSc99}. A few recent studies were done for relativistic
stars with slowly rotating and Cowling approximations
\cite{And98}. In this sense, it is interesting to study the
properties of the $r$-mode instability in the more general cases,
for example, differentially and rapidly rotating, non-isentropic
relativistic stars.

Furthermore,
they used a thermodynamic model for the neutron star fluid that is
not compatible with special relativity, they largely ignored superfluid and
magnetic field effects.

In the first work, we address one of $r$-mode's weaknesses, by
utilizing a
thermodynamic model for the neutron star fluid that takes the coupling between
vorticity and shear viscosity into account.
Navier-Stokes theory has been
used to calculate the viscous damping timescales and produce a
stability curve for $r$-modes in the $(\Omega,T)$ plane. In
Navier-Stokes theory, viscosity is independent of vorticity, but
kinetic theory predicts a coupling of vorticity to the shear
viscosity. We calculate this coupling and show that it can in
principle significantly modify the stability diagram at lower
temperatures. As a result, colder stars can remain stable at higher
spin rates \cite{RMa00}.

In the second one, we propose a possible solution of the unsolved
post-glitch relaxation of Crab by $r$-modes.
More than 30 years after the
discovery of the pulsar phenomenon and its identification with
neutron stars, there exists still a number of uncertainties and
open questions about the theoretical model for pulsars, mainly due
to the extremely dense state of matter in neutron stars.  After
two decades, the glitch phenomenon, a sudden increase of
angular velocity of the order of $\Delta\Omega/\Omega <
10^{-6}$, and the very long relaxation times, from months to
years, after the glitch, remain as one of the great mysteries of
pulsars.  The observed post-glitch relaxation of the Crab pulsar
has been unique in that the rotation frequency of the pulsar is
seen to decrease to values $less$ than its pre-glitch extrapolated
values.

The excitation of $r$-modes at a glitch and the resulting emission
of gravitational waves could, however, account for the required
``sink'' of angular momentum in order to explain the peculiar
post-glitch relaxation behaviour of the Crab pulsar. We show
that excitation of the $r$-modes at a glitch may provide a
solution to an unsolved observed effect in post-glitch relaxation
of the Crab pulsar \cite{RJa00}. Assuming that $r$-modes are excited
at a glitch, we show that this can conveniently describe
post-glitch relaxations of both Crab and Vela pulsars for a
reasonable initial amplitude of the excitation.     We use a
simple model for the total angular momentum of the star, as in
\cite{Owe98}, in which $r$-mode amplitude is independent of the
rotational frequency of the star.

In chapter 6, we review $r$-mode instability and CFS mechanism.
Further, we calculate the coupled shear viscosity-vorticity correction to
the $r$-mode timescale.   Finally we discuss the possible role of $r$-mode in
post-glitch relaxation of the Crab.

\setcounter{sub}{0}
\setcounter{subeqn}{0}
\renewcommand{\theequation}{6.\thesub\thesubeqn}

\chapter{ R-mode instabilities in neutron stars}

%\section{Introduction}

In this chapter we introduce the recently $r$-mode instability in
neutron stars.
Thses modes have been found to play an interesting and
important role in the evolution of hot young rapidly rotating neutron
stars.
Gravitational radiations tend to drive the $r$-modes unstable in
all rotating stars and spin down them.

%The $r$-mode instability has been the subject of about thirty papers
%over the past two years \cite{fl}.

In section 6.1 we briefly review the $r$-mode instability.  CFS intability,
the mechanism that governs the $r$-mode instability, is discussed in section 6.1.1.
In section 6.1.2
the equilibrium model for
a slowly and uniformly rotating background is elaborated.
For small perturbations,
the pulsation equations of a rotating star in the slow rotation
limit are extracted in section 6.1.3.
In Section 6.1.4 the mode eqautions are solved for
interesting $\ell = m$ $r$-modes.  The stability curve of $r$-mode is
discussed in section 6.1.5.

In section 6.2 using kinetic theory we calculate the effect of
shear viscosity-vorticity coupling  to
the stability curve of $r$-mode.
As an application, the possible role of $r$-mode in
post-glitch relaxation of the Crab is discussed in section 6.3.

\section{$r$-mode instability}
\subsection{CFS instability}
The $r$-mode instability is a member of the class of gravitational radiation
driven instabilities called
CFS instabilities---named for Chandrasekhar, who discovered it in a special
case \cite{Cha70}, and for Friedman and Schutz, who investigated it in
detail and found that it is generic to rotating perfect fluids
\cite{FSc78}.  The CFS instability allows some oscillation modes of a
fluid body to be driven rather than damped by radiation reaction,
essentially due to a disagreement between two frames of reference.

The mechanism can be explained as follows.  In a
non-rotating star, gravitational waves radiate positive angular
momentum from a forward-moving mode and negative angular momentum from
a backward-moving mode, damping both as expected.  However, when the
star rotates the radiation still lives in a non-rotating frame.  If a
mode moves backward in the rotating frame but forward in the
non-rotating frame, gravitational radiation still removes positive
angular momentum---but since the fluid sees the mode as having
negative angular momentum, radiation drives the mode rather than damps
it.

Mathematically, the criterion for the CFS instability is
\st\be
\label{crite}
\omega (\omega+m\Omega) < 0,
\ee
with the mode angular frequency $\omega$ (in an inertial frame) in
general a function of the azimuthal quantum number $m$ and rotation
angular frequency of star, $\Omega$.  For any set of modes of a perfect fluid,
there will be some modes unstable above some minimum $m$ and $\Omega$.
However, fluid viscosity generally grows with $m$ and also there is a
maximum value of $\Omega$ (known as the Kepler frequency $\Omega_K$)
above which a rotating star flies apart.  Therefore the instability is
astrophysically relevant only if there is some range of frequencies
and temperatures (viscosity generally depends strongly on temperature)
in which it survives.

The $r$-modes are a set of fluid oscillations with dynamics
dominated by rotation.  They are in some respects similar to the
Rossby waves found in the Earth's oceans and have been studied by
astrophysicists since the 1970s \cite{PPr78}.  The restoring force
is the Coriolis inertial ``force'' which is perpendicular to the
velocity.  As a consequence, the fluid motion resembles
(oscillating) circulation patterns. The (Eulerian) velocity
perturbation is \st\be \label{dv} \d\v = \alpha\Omega R(r/R)^\ell
\Y^B_{\ell\ell} e^{i\omega t} +O(\Omega^3), \label{deltav}\ee
where $\alpha$ is a dimensionless amplitude (roughly $\delta v/v$)
and $R$ is the radius of the star.   $\Y^B_{\ell m}$ is the
magnetic type vector spherical harmonic defined by \cite{Tho80}
\st\be \Y^B_{\ell m}=[\ell(\ell
+1)]^{-1/2}r\nab\times(r\nab Y_{\ell m}(\theta,
\phi). \ee Since $\d\v$ is an axial vector, mass-current
perturbations are large compared to the density perturbations. The
Coriolis restoring force guarantees that the $r$-mode frequencies
are comparable to the rotation frequency \cite{PPr78}, \st \be
\omega+m\Omega = {2\over m+1} \Omega
+O(\Omega^3),\,\,\,\,\ell=m.\label{rfreq} \ee In mid-1997 that
Andersson \cite{And98} noticed that the $r$-mode frequencies
satisfy the mode instability criterion~(\ref{crite}) for all $m$
and $\Omega$, and that Friedman and Morsink \cite{FMo98} showed
the instability is not an artifact of the assumption of discrete
modes but exists for generic initial data. In other words, {\it
all} rotating perfect fluids are subject to the instability.

%%%%%%%%%%%%%%%%%%%%%%%%%%%%%%%%%%%%%%%%%%%%%%%%%%%%%%%%%%%%%%%%%%%%%%%%%%%%%

\subsection{Slow rotation approximation}
To analyze the $r$-modes of rotating stars, we use the standard
expansion of the hydrodynamics equations as power series in the
angular velocity $\Omega$ of the star.
In
this section we follow the method presented in
\cite{FMo98,LMO99}, and describe how to solve the equilibrium
structure equations for uniformly rotating Newtonian and
barotropic stars for slow rotations.  The solutions
will be obtained here up to the terms of order $\Omega^2$. Here,
we use the standard spherical coordinates.

The general equations which describe the dynamical evolution of an
arbitrary state of a Newtonian self-gravitating perfect fluid are
the continuity equation \st\be \frac{\partial \rho}{\partial
t}+\nabla_a(\rho v^a)=0,\label{Cont} \ee the Euler's equation
\st\be \frac{\partial v^a}{\partial t}+v^b\nabla_b(\rho
v^a)=-\nabla^a[h(p)-\Phi]= -\nabla^a U, \label{Euler} \ee and the
gravitational potential equation \st\be \nabla^a\nabla_a\Phi =
-4\pi G\rho.\label{Grav} \ee The quantities $\rho$, $p$ are the
mass density and pressure of the fluid, respectively.   They are assumed to
satisfy a barotropic equation of state, $p=p(\rho)$; $v^a$,
$\Phi$, and $G$ are the fluid velocity, the gravitational
potential and the Newtonian gravitational constant, respectively.
Here $h(p)$ denotes the thermodynamic enthalpy of the barotropic
fluid in a comoving frame, \st\be h(p) = \int_0^p {dp'\over
\rho(p')}. \label{1.1} \ee This definition can always be inverted
to determine $p(h)$.

In equilibrium, we consider a rotating self-gravitating perfect fluid with
uniform
angular velocity, $\Omega$. The velocity of the fluid becomes
\st\be
v^a=\Omega\phi^a,
\ee
where $\phi^a$ is the rotational Killing vector field.
The equilibrium equations will be
\st\be
\nabla_a\bigl[h-\frac{1}{2}r^2\Omega^2-\Phi\bigr]=0,\label{1.2}
\ee
\st\be
\nabla^a\nabla_a\Phi =  -4\pi G\rho.\label{1.3}
\ee

We seek solutions to Eqs.~(\ref{1.2}) and (\ref{1.3}) as power series
in the angular velocity $\Omega$.  For a slowly rotating star,
\st\stq\be
h = h_0(r) + {\cal O}(\Omega^2),
\label{1.4}
\ee
\stq\be
\rho = \rho_0(r) + {\cal O}(\Omega^2),
\label{1.5}
\ee
\stq\be
p = p_0(r) + {\cal O}(\Omega^2),
\label{1.5a}
\ee
\stq\be
\Phi = \Phi_0(r) + {\cal O}(\Omega^2),
\label{1.6}
\ee
where $h_0$, $\rho_0$, $p_0$ and $\Phi_0$ values for the corresponding
non-rotating (spherical) equilibrium model.   Using these expressions,
the zero order solution to Eq.~(\ref{1.2}) is
\st\be
C_0 = h_0(r) -\Phi_0(r),\label{1.7}
\ee
where $C_0$ is constant.  The non-rotating model
can be determined in the usual way by solving the gravitational
potential equation,
\st\be
{1\over r^2} {d\over dr}\left(r^2{d\Phi_0\over dr}\right) = -
4\pi G \rho_0,\label{1.9}
\ee
together with Eq.~(\ref{1.7}).  The integration constant
can be shown to be $C_0=-GM_0/R_0$ by evaluating Eq.~(\ref{1.7})
at the surface of the star,  where the constants $M_0$ and $R_0$ are the
mass and radius of the non-rotating star.

%We have developed a computer code that solves these equations
%numerically for stars with an arbitrary equation of state.  We have
%tested this code against analytical expressions which can be obtained
%for a polytropic neutron star equation of state, $p=K\rho^2$, with $K$
%chosen so that a $1.4M_\odot$ model has a radius of $R_0=12.533$km.
%We find that the constants that determine the slowly rotating model
%for this polytropic case have the values $C_2=.09802 C_0$,
%$R_{20}=.15198 R_0$, and $R_{22} = -.37995 R_0$.  Our numerical
%results agree with the analytical to floating-point precision.

%%%%%%%%%%%%%%%%%%%%%%%%%%%%%%%%%%%%%%%%%%%%%%%%%%%%%%%%%%%%%%%%%%%%%%%%%%%%

\subsection{Pulsation equations} \label{sectionII}
In the last section, the equilibrium model of uniformly rotating
star in the slowly rotating approximation was discussed.  In this
section we assume a small perturbation in the fluid, to extract
the mode equations of the system. Using Ipser and Lindblom's
method \cite{ILi90}, one finds the pulsation equations in
general. In the next section we restrict our calculations to
$r$-mode only.

The modes of any rotating barotropic stellar model can be
described completely in terms of two scalar potentials
$\delta\Phi$ and
\st\be
\delta U\equiv {\delta p\over \rho}-\delta\Phi,\label{5.1} \ee
where $\delta p$ and $\d\Phi$ are the Eulerian pressure and
gravitational potential perturbations, respectively, and $\rho$ is
the unperturbed density of the equilibrium stellar model \cite{ILi90}. For a
barotropic equation of state, $p=p(\rho)$, Eq. (\ref{5.1}) reduces to
\st\be \delta U = {1\over \rho}{dp\over
d\rho}\d\rho-\d\Phi,\label{5.1a} \ee where $\delta\rho $ is the
Eulerian mass density perturbation. We assume here that the time
dependence of the mode is $e^{i\omega t}$ and that its azimuthal
angular dependence is $e^{im\varphi}$, where $\omega$ is the
frequency of the mode and $m$ is an integer.

Linearizing Euler's equation, Eq. (\ref{Euler}), about a uniformly
rotating background, we find \st\be iQ^{-1}_{ab}\d v^b\equiv\ls
i(\omega +m\Omega )g_{ab}+2\nabla_b v_a\rs \d v^b=-\nabla_a\d
U\,,\label{LinEuler} \ee
where $g_{ab}$ is
the Euclidean metric tensor (the identity matrix in Cartesian
coordinates).
The quantity $Q^{-1      }_{ab}$ can be
inverted to obtain an expression for the velocity perturbation in
terms of the potential $\d U$: \st\be \delta v^a =
iQ^{ab}\nabla_b\delta U.\label{5.2} \ee $Q^{ab}$ depends on the
frequency of the mode, and the angular velocity of the equilibrium
star: \st\be Q^{ab}={1\over (\omega+m\Omega)^2-4\Omega^2}
\Biggl[(\omega+m\Omega)g^{ab}-
       {4\Omega^2\over \omega+m\Omega}z^az^b - 2i\nabla^av^b \Biggr].
\label{5.3} \ee In Eq.~(\ref{5.3}) the unit vector $z^a$ points
along the rotation axis of the equilibrium star.
For real frequencies $\omega$, $Q^{ab}$ is
Hermitian, $Q^{ab}=Q^{*~ba}$, and covariantly constant, $\nabla_c
Q^{ab}=0$.

Replacing the perturbed mass density and fluid velocity in terms
of the potentials $\delta U$ and $\delta \Phi$, $\d\rho=\rho
(d\rho/dp)(\d U+\d\Phi)$ and using Eq. (\ref{5.2}),
Eqs.
(\ref{Cont}) and (\ref{Grav}) reduce to
\st\be \nabla_a(\rho
Q^{ab}\nabla_b\delta U)=-(\omega+m\Omega) \rho {d\rho\over
dp}(\delta U+\delta\Phi),\label{5.4} \ee \st\be
\nabla^a\nabla_a\delta\Phi = -4\pi G \rho{d\rho\over dp} (\delta U
+\delta\Phi).\label{5.5} \ee We note that in obtaining Eqs.
(\ref{5.4}) and (\ref{5.5}), the slow rotation approximation is not
assumed. Equations (\ref{5.4}) and (\ref{5.5}) are the master
equations that determine the properties of the oscillations of
{\it rapidly} rotating Newtonian stellar model with uniform spin
rate. They are a forth-order system of partial differential
equations for two potentials $\d U$ and $\d\Phi$.  These equations
form an eigenvalue problem with eigenfrequencies, $\omega$, with
appropriate boundary conditions at the surface of the star for $\d
U$ and at infinity for $\d\Phi$.

In the slow rotation limit we expand the potentials $\d U$ and
$\d\Phi$ as \st\be \delta U = R_0^2\Omega^2\delta U_0 + {\cal
O}(\Omega^4)\,,\label{5.11} \ee

\st\be \delta \Phi = R_0^2\Omega^2\delta \Phi_0 + {\cal
O}(\Omega^4).\label{5.12} \ee
 The normalizations of $\delta U$ and
$\delta\Phi$ have been chosen to make $\delta U_0$ and $\delta
\Phi_0$ dimensionless under the assumption that the lowest order
terms scale as $\Omega^2$. Using Eqs.(\ref{5.3}), (\ref{5.11}) and
(\ref{5.12}), Eqs. (\ref{5.4}) and (\ref{5.5}) in the lowest order
in $\Omega$ reduce to \st\be \nabla^a\Bigl[\rho_0 (\kappa_0^2\delta
^{ab} - 4z^az^b)\nabla_b\delta U_0\Bigr] + {2m\kappa_0\over
\xi}\xi^a\nabla_a\rho_0\,\, \delta U_0 =0, \label{5.14} \ee
\st\be \nabla^a\nabla_a\delta\Phi_0 = - 4\pi G \rho
\left({d\rho\over dp}\right)_0 (\delta U_0 + \delta \Phi_0).
\label{5.15} \ee The quantity $\kappa_0$ is the first order of
$\kappa$, $\kappa\Omega=\omega+m\Omega$.  The notation $\xi$ is
the cylindrical radial coordinate, $\xi = r\sin(\vartheta)$,
and $\xi^a$ denotes the unit vector in the $\xi$ direction.

The eigensolutions and eigenfunctions will be determined by
solving system of equations (\ref{5.14}) and (\ref{5.15}) with the
appropriate boundary conditions \cite{LMO99}.

%%%%%%%%%%%%%%%%%%%%%%%%%%%%%%%%%%%%%%%%%%%%%%%%%%%%%%%%%%%%%%%%%%%%%%%%
\subsection{$\ell=m$ $r$-modes} \label{sectionIII} In this
section we restrict our consideration on the $r$-modes which
contribute primarily to the gravitational radiation driven
instability. The reason for restriction to $\ell=m$ goes back to Provost et al.
\cite{PBR81}, who showed that for the barotropic equation
of state, only the $\ell=m$ $r$-mode exists in the rotating star.
The ``classical'' $r$-modes (which were studied first by
Papaloizou and Pringle~\cite{PPr78}) are generated by hydrodynamic
potentials of the form (see e.g. Lindblom and Ipser~\cite{LIp99})

\st \be \delta U = \alpha\left[{2 \ell\over
2\ell+1}\sqrt{\ell\over \ell+1} \right] \left({r\over
R}\right)^{\ell+1} Y_{\ell+1 \ell}(\cos(\vartheta))e^{im\varphi}.
\label{3.1} \ee Here, and after, we drop index $0$ for
brevity. It is straightforward to verify that this $\delta U$ is a
solution to Eq.~(\ref{5.14}) for the eigenvalue $\kappa$ has the
value

\st \be \kappa = {2\over \ell+1}.\label{3.2} \ee Then, the
frequency of the mode, $\omega$, can be obtained

\st \be \omega = -{(\ell-1)(\ell+2)\over
\ell+1}\Omega.\label{3.2a} \ee

\noindent The perturbed gravitational potential $\delta \Phi$ must
have the same angular dependence as $\delta U$, so

\st \be \delta\Phi = \alpha \delta \Psi(r) Y_{\ell+1 \ell
}(\cos(\vartheta)) e^{im\varphi}. \label{3.3} \ee Therefore, the
gravitational potential Eq.~(\ref{5.15}) reduces to an ordinary
differential equation for $\delta\Psi(r)$: \st \be
{d^2\delta\Psi\over dr^2}+{2\over r}{d\delta\Psi\over dr} +
\biggl[4\pi G \rho {d\rho\over dp}  - {(\ell+1)(\ell+2)\over r^2}
\biggr]\delta\Psi = -{8\pi G \rho \ell\over
2\ell+1}\sqrt{\ell\over \ell+1} {d\rho\over d} \left({r\over
R}\right)^{\ell+1}. \label{3.4} \ee Substituting Eqs. (\ref{3.1})
and (\ref{3.4}) into Eq. (\ref{5.1a}), the perturbed mass density
to order $\Omega^2$ becomes

\st\be {\delta \rho\over\rho} = \alpha R^2\Omega^2 {d\rho\over dp}
\left[{2 \ell\over 2\ell+1}\sqrt{\ell\over \ell+1} \left({r\over
R}\right)^{\ell+1}+\delta\Psi(r)\right] Y_{\ell+1 \ell}
\,e^{i\omega t}.\label{drho} \ee Furthermore, using
Eq.~(\ref{5.2}), the velocity perturbation expression, Eq.
(\ref{deltav}), will be recovered.

%%%%%%%%%%%%%%%%%%%%%%%%%%%%%%%%%%%%%%%%%%%%%%%%%%%%%%%%%%%%%%%%%%%%%%%%%%%%%

\subsection{$r$-mode timescale}
Our main interest here is to study the evolution of the $r$-modes due to the
dissipative influences of viscosity and gravitational radiation.  To achieve
it, we study the energy evolution of the mode which is affected by radiation
and viscosity.
The energy of the
mode measured in the rotating frame of the equilibrium star, $\tilde{E}$, is
\st\begin{equation}
\tilde{E}={1\over 2} \int\left[ \rho\, \d \v\cdot \d\v\,^*
+ \left({\delta p\over\rho} - \delta\Phi\right)  \delta \rho^*\right]d^3x.
\label{6}
\end{equation}
This energy evolves on the secular timescale of the dissipative
processes. The dissipation of energy due to the gravitational
radiation and viscosity can be estimated from general expression
\cite{CLi87}

\st\bea {d \tilde{E}\over dt}
&&=-\int\left(2\eta\delta\sigma^{ab}\delta\sigma_{ab}^*
+\zeta\delta\sigma \delta\sigma^*\right)d^3x\nonumber\\
&&\hspace{1cm} -\omega(\omega+m\Omega)\sum_{\ell\geq 2} N_\ell \omega^{2\ell}
\left(|\delta D_{\ell m}|^2+|\delta J_{\ell m}|^2\right)\,,
\label{7}
\eea
where thermodynamic functions $\eta$ and $\zeta$
are the shear and bulk viscosities of the fluid, respectively.  The
viscous forces are driven by the shear $\delta \sigma_{ab}$ and the
expansion $\delta\sigma$ of the perturbation,
defined by the usual expressions
\st\begin{equation}
\delta \sigma_{ab}={\scriptstyle {1\over 2}}
(\nabla_{\!a}\delta v_b+\nabla_{\!b}\delta v_a
-{\scriptstyle {2\over 3}}\delta_{ab}\nabla_{\!c}\delta v^c),
\label{8}
\end{equation}
\st\begin{equation}
\delta\sigma=\nabla_{\!a}\delta v^a.\label{9}
\end{equation}

\noindent
Gravitational radiation couples to the
evolution of the mode through the mass $\delta D_{\ell m}$ and current $\delta
J_{\ell m}$ multipole moments of the perturbed fluid,
\st
\begin{equation}
\delta D_{\ell m} = \int \delta\rho\, r^\ell Y^{*}_{\ell\,m} d^3x,\label{10}
\end{equation}
\st
\begin{equation}
\delta J_{\ell m}= {2\over c}\sqrt{\ell\over \ell+1}
\int r^\ell (\rho \,\delta\v +\delta \rho\, \v)\cdot
\Y^{B*}_{\ell\,m} d^3x,\label{11}
\end{equation}
with coupling constant
\st
\begin{equation}
N_\ell = {4\pi G\over c^{2\ell+1}} {(\ell+1)(\ell+2)\over
\ell(\ell-1)[(2\ell+1)!!]^2}. \label{12}
\end{equation}

\noindent The terms in the expression for $d\tilde{E}/dt$ due to
viscosity and the gravitational radiation generated by the mass
multipoles are well known \cite{ILi91}.  The terms involving
the current multipole moments have been deduced from the general
expressions given by Thorne \cite{Tho80}.

\noindent We can now use Eqs.~(\ref{6}) and (\ref{7}) to evaluate
the stability of the $\ell=m$ $r$-modes.  Viscosity, however,
tends to decrease the energy $\tilde{E}$, while gravitational
radiation may either increase or decrease it.  The sum that
appears in Eq.~(\ref{7}) is positive definite; thus the effect of
gravitational radiation is determined by the sign of
$\omega(\omega+\ell\Omega)$.  Using Eq. (\ref{rfreq}), this
quantity for $r$-modes is negative definite:

\begin{equation}
\omega(\omega+\ell\Omega)= -{2(\ell-1)(\ell+2)\over
(\ell+1)^2}\Omega^2<0.\label{13}
\end{equation}
Therefore gravitational radiation tends to increase the energy of
these modes.  In addition,
for small angular velocities the energy $\tilde{E}$ is
positive definite: the positive term $|\delta\v |^2$ in
Eq.~(\ref{6}) (proportional to $\Omega^2$) dominates the
indefinite term $(\delta p/\rho - \delta \Phi)\delta\rho^*$
(proportional to $\Omega^4$).  Thus, gravitational radiation tends to
make {\it every} $r$-mode unstable in slowly rotating stars.

\noindent To determine whether these modes are actually stable or unstable in
rotating neutron stars, therefore, we must evaluate the magnitudes of
all the dissipative terms in Eq.~(\ref{7}) and determine
the dominant one.

Here we estimate the relative importance of these dissipative effects in
the small angular velocity limit using the lowest order expressions
for the $r$-mode $\delta \v$ and $\delta \rho$ given in
Eqs.~(\ref{deltav}) and (\ref{drho}).  The lowest order expression for the
energy of the mode $\tilde E$ is

\st\begin{equation} \tilde{E} = {\scriptstyle {1\over 2}} \alpha^2
\Omega^2 R^{-2\ell+2}\int_0^R\rho\,r^{2\ell+2}dr .\label{14}
\end{equation}

The lowest order contribution to the gravitational radiation terms
in the energy dissipation comes entirely from the current
multipole moment $\delta J_{\ell\,\ell}$.  This term can be
evaluated to lowest order in $\Omega$ using Eqs.~(\ref{deltav})
and (\ref{11}):

\st\begin{equation} \delta J_{\ell\,\ell}={2\alpha \Omega\over
cR^{\ell-1}}\sqrt{l\over \ell+1} \int_0^R \rho\,r^{2\ell+2}
dr.\label{15}
\end{equation}

\noindent The other contributions from gravitational radiation to
the dissipation rate are all higher order in $\Omega$.  The mass
multipole moment contributions are higher order because the
density perturbation $\delta\rho$ from Eq.~(\ref{drho}) is
proportional to $\Omega^2$ while the velocity perturbation $\delta
\v$ is proportional to $\Omega$.  Furthermore, the density
perturbation $\delta\rho$ generates gravitational radiation at
order $2\ell+4$ in $\omega$ while $\delta\v$ generates
radiation at order $2\ell+2$.

The contribution of gravitational radiation to the imaginary part of the
frequency of the mode $1/\tau_{\scriptscriptstyle GR}$ can be computed
as follows,

\st\begin{equation}
{1\over \tau_{\scriptscriptstyle GR}}
=- {1\over 2 \tilde{E}}
\left({d\tilde{E}\over dt}\right)_{\scriptscriptstyle GR}.
\label{16}
\end{equation}

\noindent Using Eqs.~(\ref{14})--(\ref{16}) an explicit expression
for the gravitational radiation timescale associated with the
$r$-modes can be obtained:

\st\be {1\over \tau_{\scriptscriptstyle GR}} =-{32\pi
G\Omega^{2\ell+2}\over c^{2\ell+3}}{ (\ell-1)^{2\ell}\over
[(2\ell+1)!!]^2} \left({\ell+2\over \ell+1}\right)^{2\ell+2}
\int_0^R\rho\,r^{2\ell+2} dr.\label{17} \ee

The time derivative of the energy due to viscous dissipation is
given by the shear $\delta \sigma_{ab}$ and the expansion $\delta
\sigma$ of the velocity perturbation.  The shear can be evaluated
using Eqs.~(\ref{deltav}) and (\ref{8}) and its integral over the
spherical coordinates $\vartheta$ and $\varphi$. Using the
formulae for the viscous dissipation rate Eq.~(\ref{7}) and the
energy Eq.~(\ref{14}), the contribution of shear viscosity to the
imaginary part of the frequency of the mode is,

\begin{equation}
{1\over \tau_{\scriptscriptstyle SV}} = (\ell-1)(2\ell+1) \int_0^R
\eta r^{2\ell} dr \left( \int_0^R \rho\, r^{2\ell+2}
dr\right)^{-1}. \label{18}
\end{equation}

The bulk viscosity dissipation expression, $\d\sigma$, can be
re-expressed in terms of the density perturbation. The perturbed
continuity equation gives the relationship

\st \be\delta\sigma = -i(\omega+m\Omega)\Delta\rho/\rho, \ee where
$\Delta\rho$ is the Lagrangian perturbation in the density. The
perturbation analysis used here is not of sufficiently high order
(in $\Omega$) to evaluate the lowest order contribution to $\Delta
\rho$. However, we are able to evaluate the Eulerian perturbation
$\delta\rho$ as given in Eq.~(\ref{drho}).  We expect that the
integral of $|\delta \rho/\rho|^2$ over the interior of the star
will be similar to (i.e., within about a factor of two of) the
integral of $|\Delta \rho/\rho|^2$.  Thus, the magnitude of the
bulk viscosity contribution to the energy dissipation can be
estimated by

\st\begin{equation} {1\over \tau_{\scriptscriptstyle BV}} \approx
{(\omega+m\Omega)^2\over 2\tilde{E}}\int\zeta
{\delta\rho\,\delta\rho^*\over\rho^2}d^3x.\label{19}
\end{equation}

\noindent Using Eqs.~(\ref{drho}) and (\ref{14}) for
$\delta\rho/\rho$ and $\tilde{E}$, Eq.~(\ref{19}) becomes an
explicit formula for the contribution to the imaginary part of the
frequency due to bulk viscosity.

To evaluate the dissipative timescales associated with the
$r$-modes using the formulae in Eqs.~(\ref{17})--(\ref{19}), we
need models for the structures of neutron stars as well as
expressions for the viscosities of neutron star matter. The
$r$-modes timescales for $1.4M_\odot$ neutron star models based on
several realistic equations of state \cite{BFG96} are evaluated by
Lindblom et al. \cite{LOM98}.   The standard formulae for the
shear and bulk viscosities of hot neutron star matter
are\cite{CLi87}

\st\begin{equation} \eta=347\rho^{9/4}T^{-2} {\rm g cm}^{-1} {\rm
s}^{-1},\label{20}
\end{equation}

\st\begin{equation} \zeta = 6.0\times 10^{-59} \rho^2
(\omega+m\Omega)^{-2} T^6 {\rm g cm}^{-1} {\rm s}^{-1}, \label{21}
\end{equation}

The timescales for the more realistic equations of state are
comparable to those based on a simple polytropic model $p=k\rho^2$
with $k$ chosen so that the radius of a $1.4M_\odot$ star is 12.53
km.  The dissipation timescales for this polytropic model (which
can be evaluated analytically) are
$\tilde{\tau}_{\scriptscriptstyle GR}= -3.26$s,
$\tilde{\tau}_{\scriptscriptstyle SV}=2.52\times 10^8$s and
$\tilde{\tau}_{\scriptscriptstyle BV}=6.99\times 10^8$s for the
fiducial values of the angular velocity $\Omega=\sqrt{\pi G
\bar{\rho}}$ and temperature $T=10^9$K in the $\ell=2$ $r$-mode.
Here ${\bar \rho}=3M/4\pi R^3$ is the mean density of the star.
The gravitational radiation timescales increase by about one order
of magnitude for each incremental increase in $\ell$, while the
viscous timescales decrease by about 20\%.

The evolution of an $r$-mode due to the dissipative effects of
viscosity and gravitational radiation reaction is determined by
the imaginary part of the frequency of the mode,

\st\be {1\over \tau(\Omega)} = {1\over
\tilde{\tau}_{\scriptscriptstyle GR}} \left({\Omega^2\over \pi G
\bar{\rho}}\right)^{\ell+1} + {1\over
\tilde{\tau}_{\scriptscriptstyle SV}} \left({10^9 {\rm K}\over
T}\right)^2+ {1\over \tilde{\tau}_{\scriptscriptstyle BV}}
\left({T\over 10^9 {\rm K}}\right)^6 \left({\Omega^2\over \pi G
\bar{\rho}}\right). \label{22} \ee

\noindent Eq.~(\ref{22}) is displayed in a form that makes
explicit the angular velocity and temperature dependences of the
various terms. Dissipative effects cause the mode to decay
exponentially as $e^{-t/\tau}$ (i.e., the mode is stable) as long
as $\tau > 0$. From Eqs.~(\ref{17})--(\ref{19}) we see that
$\tilde{\tau}_{\scriptscriptstyle SV}>0$ and
$\tilde{\tau}_{\scriptscriptstyle BV}>0$ while
$\tilde{\tau}_{\scriptscriptstyle GR}<0$.  Thus gravitational
radiation drives these modes towards instability while viscosity
tries to stabilize them.  For small $\Omega$ the gravitational
radiation contribution to the imaginary part of the frequency is
very small since it is proportional to $\Omega^{2l+2}$.  Thus for
sufficiently small angular velocities, viscosity dominates and the
mode is stable.  For sufficiently large $\Omega$, however,
gravitational radiation will dominate and drive the mode unstable.
It is convenient to define a critical angular velocity $\Omega_c$
where the sign of the imaginary part of the frequency changes from
positive to negative: $1/\tau(\Omega_c) = 0$. If the angular
velocity of the star exceeds $\Omega_c$ then gravitational
radiation reaction dominates viscosity and the mode is unstable.

\begin{figure}
\begin{center}
\leavevmode
\hbox{%
\epsfxsize=4.5in
%\epsfysize=8.5in
%\epsffile{lin1.ps}}
\epsffile{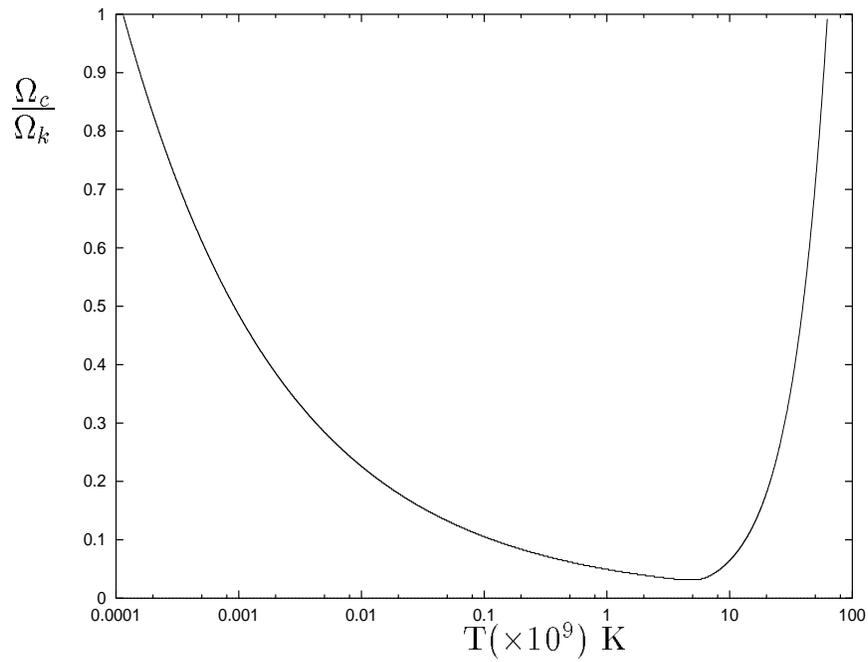}}
\end{center}
\caption{Critical angular velocities for a $1.4M_\odot$ polytropic
neutron star.} \label{fig1}
\end{figure}

\begin{figure}
\begin{center}
\leavevmode
\hbox{%
\epsfxsize=4.5in
%\epsffile{lin2.ps}}
\epsffile{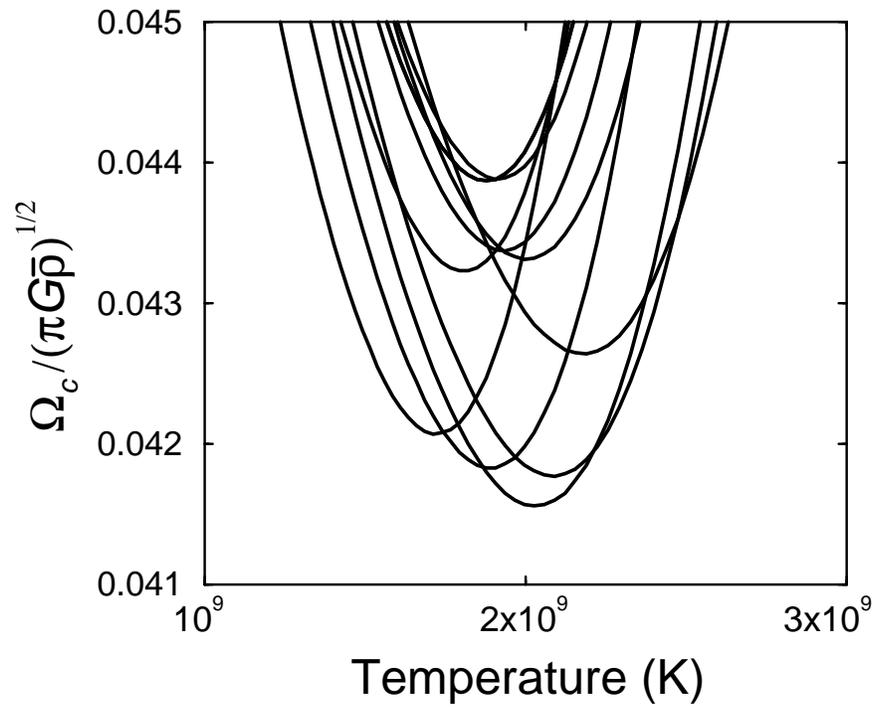}}

\end{center}
\caption{Critical angular velocities of realistic $1.4M_\odot$
neutron star models, Lindblom et al. \cite{LMO99}} .\label{fig2}
\end{figure}

For a given temperature and mode $l$ the equation for the critical
angular velocity, $0=1/\tau(\Omega_c)$, is a polynomial of order
$l+1$ in $\Omega_c^2$, and thus each mode has its own critical
angular velocity. However, only the smallest of these (always the
$l=2$ {\it r-}mode here) represents the critical angular velocity
of the star. Fig.~\ref{fig1} shows that the critical angular
velocity for a range of temperatures relevant for neutron stars
for the polytropic model discussed above. Fig.~\ref{fig2} depicts
the critical angular velocities for $1.4M_\odot$ neutron star
models computed from a variety of realistic equations of state
\cite{BFG96}. Fig.~\ref{fig2} illustrates that the minimum
critical angular velocity (in units of $\sqrt{\pi G \bar{\rho}}$)
is extremely insensitive to the equation of state. The minima of
these curves occur at $T \approx 2\times 10^9$K, with $\Omega_c
\approx 0.043\sqrt{\pi G \bar{\rho}}$.  The maximum angular
velocity for any star occurs when the material at the surface
effectively orbits the star. This `Keplerian' angular velocity
$\Omega_K$ is very nearly ${2\over 3}\sqrt{\pi G \bar{\rho}}$ for
any equation of state. Thus the minimum critical angular velocity
due to instability of the {\it r-}modes is about $0.065\Omega_K$
for any equation of state.

\newpage

\newpage
\newpage
\newpage
\newpage

\newpage
\newpage
\newpage
\newpage

%%%%%%%%%%%%%%%%%%%%%%%%%%%%%%%%%%%6%%%%%%%%%%%%%%%%%%%%%%%%%%%%%%%%%%%%%%%%%%%%%%%%%%%%%%%%

\break

\clearpage

\section{Vorticity-shear viscosity coupling}
In spite of the recent improvements in our understanding of the
$r$-mode instability, it seems that the fundamental properties of
these modes have not yet been sufficiently understood.  Previous
investigations of the $r$-modes are restricted to the case of
uniformly and slowly rotating, isentropic, Newtonian stars
\cite{AKSt98}. A few recent studies were done for relativistic
stars with slowly rotating and Cowling approximations
\cite{And98}. In this sense, it is interesting to study the
properties of the $r$-mode instability in the more general cases,
for example, differentially and rapidly rotating, non-isentropic
relativistic stars.

In addition, people have used the standard Navier-Stokes theory to
study $r$-mode stability enforced by viscosity, and have calculated the
corresponding timescales. This theory and its relativistic
generalization are non-causal and unstable.   An improved causal
dissipative fluid theory is based on kinetic theory \cite{ISt79}.
This latter theory has implications different from Navier-Stokes
theory. For example, kinetic theory predicts that the angular
velocity of the star couples to the viscosity and heat flux in
rotating stars.

In this section we investigate the possible effect of
vorticity-shear viscosity coupling on the stability of $r$-mode
and show that the coupling may have a significant effect on
viscous damping timescales of $r$-mode \cite{RMa00}. We find that
colder stars can remain stable at higher spin rates.

\subsection{Predicted effect}
In standard Navier-Stokes theory the viscous quantities are
defined by

\st\begin{equation}\label{ns}
\Pi=-\zeta\Theta\,,~q_a=-\kappa\nabla_a T\,,~
\pi_{ab}=-2\eta\sigma_{ab}\,,
\end{equation}
where $\Pi$ is the bulk viscous stress and $\zeta$ is the bulk
viscosity; $q_a$ is the heat flux and $\kappa$ is the thermal
conductivity; $\pi_{ab}$ is the shear viscous stress and $\eta$ is
the shear viscosity; $\Theta=\nabla_av^a$ is the volume expansion
rate of the fluid, $T$ is the temperature, and
$\sigma_{ab}=\nabla_{\langle a}v_{b\rangle}$ is the rate of shear
(where the angled brackets denote the symmetric tracefree part).
Its is clear that the fluid vorticity $\varp={1\over2}\nab\,\times\,\v$
does not enter Eq. (\ref{ns}), even when the
equilibrium state is rotating.

On physical grounds, one might expect that rotational
accelerations can couple with gradients of momentum and
temperature, so that there could in principle be couplings of
$\varpi_a$ to $q_a$ and $\pi_{ab}$. In the case of heat flux,
qualitative particle dynamics indicates \cite{MRt93} (p. 34) that
this coupling does exist as a result of a Coriolis effect, which
is in some sense analogous to the Hall effect in a conductor
subject to a magnetic field. The Coriolis effect on heat flux is
confirmed by molecular dynamics simulations \cite{HBML81}.
M\"uller \cite{Mul72} and Israel \& Stewart \cite{ISt79} showed
that the Boltzmann equation predicts in general a coupling of
vorticity to heat flux and shear viscous stress. The microscopic
and self-consistent kinetic approach is in contrast to the
continuum view, where a phenomenological principle of ``frame
indifference" is invoked to argue against any vorticity coupling.
(See \cite{MRt93,HBML81,JCL93} for further discussion.)

Using the Grad moment method to approximate the hydrodynamic
regime via kinetic theory, the relations in Eq. (\ref{ns}) are
modified to \cite{ISt79} (Eq. (7.1))

\st\stq\begin{eqnarray}
\Pi&=&-\zeta\left[\Theta+\beta_0\dot\Pi\right]\,,\label{r1}\\
\stq
q_a&=&-\kappa\left[\nabla_a
T+T\beta_1\left\{\dot{q}_a-\varpi_{ab}q^b
\right\}\right]\,,\label{r2}\\
\stq
\pi_{ab}&=&-2\eta\left[\sigma_{ab}+\beta_2\left\{\dot{\pi}_{\langle
ab\rangle} -2\varpi^c{}_{\langle a}\pi_{b\rangle
c}\right\}\right]\,, \label{r3}
\end{eqnarray}
where $\beta_A$, $A=0,1,2$, can
be evaluated in terms of collision integrals for specific gases,
an overdot denotes the comoving (Lagrangian) derivative, and the
vorticity tensor is given by
\begin{eqnarray*}
&&\varpi_{ab}=\nabla_{[a}v_{b]}=\varepsilon_{abc}\varpi^c\,,\\
&&\varp={1\over 2}\nab\,\times\,\v\,,
\end{eqnarray*}
where
square brackets on indices indicate the skew part.
Navier-Stokes theory is recovered from the M\"uller-Israel-Stewart
theory when $\beta_A=0$. However, kinetic theory gives $\beta_A$
values for simple gases which are definitely {\em not} zero.
Furthermore, if $\beta_A=0$, the equilibrium states are unstable
and dissipative signals can propagate at unbounded speed
\cite{ISt79,JCL93}.

The $\beta_A$-corrections will be very small except if there are
either high frequency oscillations (pumping up the time-derivative
terms) or rapid rotation (pumping up the vorticity-coupling
terms). In the context of rapidly rotating neutron stars, we
expect the vorticity-dissipative couplings to dominate the
time-derivative terms; this expectation is borne out by
calculations (see below). The vorticity-dissipative couplings will
be negligible if the unperturbed equilibrium state is
irrotational, i.e., if $\varpi_a=0$ in the background, so that the
coupling terms become second-order. However, for fast rotation,
$\varpi_a\neq0$ in the background and the coupling terms make a
first-order contribution to dissipation. In the words of Israel \&
Stewart \cite{ISt79}: ``these results will ultimately be of
practical interest in astrophysical and cosmological situations
involving fast rotation, strong gravitational fields or rapid
fluctuations (neutron stars, black hole accretion, early
universe), although it will probably be some time before the state
of the art in these fields makes such refinements necessary."\\ We
believe that recent and ongoing developments in rotating neutron
star physics have reached the stage where the
M\"uller-Israel-Stewart theoretical corrections to the
Navier-Stokes equations need to be examined, and our results
indicate that the corrections could be important.

We follow the standard assumption \cite{CLi87} that the heat flux
may be neglected relative to viscous stresses in calculating
damping timescales. Then the vorticity correction to Navier-Stokes
theory reduces to the coupling term $\varpi^c{}_{\langle
a}\pi_{b\rangle c}$. This term means that the angular momentum of
the star changes the shear viscosity timescale, and we find (for
axial $r$-modes) a correction proportional to $T^{-r}\Omega^2$,
where $r=9$ for a nonrelativistic fluid and $r=12$ for an
ultrarelativistic fluid.

Replacing $\d\sigma$ and $\d\sigma_{ab}$ by $\d\Pi$ and
$\d\pi_{ab}$ in Eq. (\ref{7}), respectively, the evolution of
dissipation energy contained in small fluctuations is given by

\st\be \frac{d\tilde E}{dt}=-\int\lp{\delta\Pi\d\Pi^*\over\zeta}+
{\d\pi^{ab}\d\pi_{ab}^*\over 2\eta} \rp d^3x -\lp\frac{d\tilde
E}{dt}\rp_{\!\rm\sc gr}\,,\label{ENERGY} \ee where $(d\tilde
E/dt)_{\rm\sc gr}$ is the energy flux in gravitational radiation
(see Eq. (\ref{7})), $\delta\Pi=\Pi-\bar\Pi$ and
$\delta\pi_{ab}=\pi_{ab}-\bar{\pi}_{ab}$, with an overbar denoting
background quantities. In this case, $\bar\Pi=0=\bar{\pi}_{ab}$.
The normal modes of the star are damped by dissipation, and the
damping rate can be determined by Eq. (\ref{ENERGY}). For a normal
mode with time dependence $e^{i\omega t}$, the energy has time
dependence $\exp[-2{\rm Im}(\omega) t]$. Then by Eq.
(\ref{ENERGY}), the characteristic damping time $\tau=1/{\rm
Im}(\omega)$ of the fluid perturbation is given by

\st\be \frac{1}{\tau}=-{1\over2\tilde{E}}{d\tilde{E}\over dt}=
\frac{1}{\tau_{\rm\sc bv}}+\frac{1}{\tau_{\rm\sc sv}} +\frac{1}
{\tau_{\rm\sc gr}}\,, \label{TOTALTIME} \ee where $\tau_{\rm\sc
bv}$, $\tau_{\rm\sc sv}$, and $\tau_{\rm\sc gr}$ are the bulk
viscous, shear viscous, and gravitational radiation timescales
respectively.

To evaluate the vorticity-corrected shear viscous timescale, we
use Eq. (\ref{r3}) in Eqs. (\ref{ENERGY}) and (\ref{TOTALTIME}).
To lowest order

\[
\delta\pi_{ab}=-2\eta\left[\delta\sigma_{ab}-2i\omega\eta\beta_2
\delta \sigma_{ab} +4\eta\beta_2\delta\sigma^c_{\left<a\right.}
\varpi_{\left. b\right>c}\right]\,, \] where $\varpi_a$ is the
background vorticity (the background shear vanishes). Then

\begin{eqnarray*}
\delta\pi^{ab}\delta\pi^*_{ab}&=&4\eta^2\left\{\delta\sigma^{ab}
\delta\sigma^*_{ab} +4\gamma^2\left[\omega^2\delta\sigma^{ab}
\delta\sigma^*_{ab} \right.\right.\\
&&\left.\left.{}+4\lp\delta\sigma^{ab}\delta\sigma^*_{ab}
\varpi^c\varpi_c
-\delta\sigma^{ca}\delta\sigma^*_{da}\varpi_c\varpi^d \rp
\right]\right\}\,,
\end{eqnarray*}
where $\gamma=\eta\beta_2$. The first term is the usual term in
Navier-Stokes theory, while the following terms are the
M\"uller-Israel-Stewart corrections. The $\omega^2$ term arises
from $\dot{\pi}_{ab}$ in Eq. (\ref{r3}), and is negligible
relative to the $\varpi^2$ terms which arise from the
$\varpi^c{}_{\langle a}\pi_{b\rangle c}$ term in Eq. (\ref{r3}).
The energy dissipation rate through shear viscosity will be

\st\bea \lp\!\frac{d\tilde E}{dt}\!\rp_{\!\!\rm\sc sv}&=&
-2\!\int\! \eta\left\{ \delta\sigma^{ab}\delta\sigma^*_{ab}
-4\gamma^2\left[\omega^2\delta\sigma^{ab}
\delta\sigma^*_{ab}\right.\right.\nonumber\\
&&\left.\left.{}+4\lp\delta\sigma^{ab}\delta\sigma^*_{ab}
\varpi^c\varpi_c
-\delta\sigma^{ca}\delta\sigma^*_{da}\varpi_c\varpi^d\rp
\right]\right\}d^3x. \label{ENERGYSHEAR} \eea

In order to proceed further, we need expressionx for the shear
viscosity $\eta$ and the coupling coefficient $\beta_2$. For the
various interactions, $\eta(\rho,T)$ is calculated in
\cite{FLt76,FLt79}, where it is shown that electron-electron
scattering is more important for shear viscosity than other
interactions. The expression for $\eta$ is given in \cite{CLi87},
in good agreement with \cite{FLt76,FLt79}, as

\st\be \eta=1.10\times 10^{16}\left(\!{ \rho\over 10^{14} {\rm
g/cm}^3}\! \right)^{\! 9/4}\left(\!{10^9{\rm K}\over
T}\!\right)^{\! 2} \rm{g/cm\,s}\,.\label{ETA} \ee For a
Maxwell-Boltzmann gas, the coefficient $\beta_2$ is found in
\cite{ISt79}, but we require the expression for a degenerate Fermi
gas. This has been found by Olson \& Hiscock \cite{OHi89} in the
case of strong degeneracy:

\st\be \beta_2={15\pi^2\hbar^3\over
m^4gc^5}\frac{(1+\nu)}{(\nu^2+2\nu)^{5/2}} +{\cal
O}\left[\left({kT\over mc^2\nu}\right)^{\!2}\right]\,, \ee
where
$m$ is the particle mass, $g$ is the spin weight, and
$mc^2\nu/kT\gg1$. The dimensionless thermodynamic potential
$\nu=(\rho+p)/nm-mc^2s/kT-1$, where $s$ is the specific entropy,
is equal to the nonrelativistic chemical potential per particle
divided by the particle rest energy. For a strongly degenerate
gas, the nonrelativistic chemical potential is proportional to
$T$, so that
\[
\nu\approx{\bar\alpha} \,{kT\over mc^2}\,,
\]
where ${\bar\alpha}\gg1$ is a dimensionless constant measuring the
degree of degeneracy. The nonrelativistic regime is obtained for
$\nu\ll1$, while the ultrarelativistic case corresponds to $\nu\gg
1$.

For temperatures below $10^{10}$ K, neutrons in the neutron star
are nonrelativistic, while electrons are ultrarelativistic
\cite{FLt76}. The nonrelativistic limit of $\beta_2$ is

\st\be (\beta_2)_{{\rm\sc nr}}\approx 3.16 \times 10^{-5}({\bar\alpha}
T)^{-5/2}~ \mbox{cm s}^{2}/\rm{g}\,,\label{BETA2NON} \ee
and its
ultrarelativistic limit is

\st\be (\beta_2)_{{\rm\sc ur}}\approx 6.45 \times 10^{15}({\bar\alpha}
T)^{-4}~ \mbox{cm s}^{2}/\rm{g}\,. \label{BETA2RE} \ee
Using Eqs.
(\ref{ETA}), (\ref{BETA2NON}) and (\ref{BETA2RE}), we have

\st\bea &&\gamma_{{\rm\sc nr}}\approx {1.10\times
10^{-11}\over{\bar\alpha}^{5/2}} \left(\!{ \rho\over 10^{14} {\rm
g/cm}^3}\!\right)^{\!9/4} \left(\!{10^9{\rm K}\over
T}\!\right)^{\!9/2} \rm{s}\,,\label{r4}\\ \st &&\gamma_{{\rm\sc
ur}} \approx {7.08\times10^{-5}\over{\bar\alpha}^4} \left(\!{ \rho\over
10^{14} {\rm g/cm}^3}\!\right)^{\!9/4} \left(\!{10^9{\rm K}\over
T}\!\right)^{\!6} \rm{s}\,. \label{r5} \eea In these calculations, we
have used the same relation for $\eta$ in both cases, because in the
high-density regime ($\rho > 10^{14}$g/ cm$^{3}$) for both
electron-electron scattering and electron-neutron scattering,
$\eta$ is proportional to $T^{-2}$, with nearly equal
proportionality factor \cite{FLt76}. For typical values of the
temxerature, $T=10^9$ K, and density, $\rho= 3\times 10^{14}$
g/cm$^{3}$, we find that $\gamma_{{\rm\sc ur}}\sim
{\bar\alpha}^{-4}\times10^{-4}$ s, while $\gamma_{{\rm\sc
nr}}\sim{\bar\alpha}^{-5/2}\times 10^{-10}$ s.

\subsection{$r$-mode instability curve}
In this section we calculate the predicted effect, Eq.
(\ref{ENERGYSHEAR}), for the $r$-mode
instability. We assume that the background is a uniformly rotating
star, so that the equilibrium fluid velocity is
$v^a=\Omega\phi^a$, where $\phi^a$ is the rotational Killing
vector field \cite{LF}. The vorticity vector of the equilibrium
state is

\st\be \varp=\frac{\Omega}{2r}\ls\cot\vartheta\,,
-1\,,0\rs\,. \label{VORTICE1} \ee The $r$-modes of rotating
barotropic Newtonian stars have Eulerian velocity perturbations
given by Eq. (\ref{5.2})

\st\be \delta\v= \alpha R \Omega \lp\frac{r}{R}\rp^\ell
\Y^B_{\ell\ell}\exp(i \omega t)\,,\label{VELOCITY} \ee where $C$
is an arbitrary constant, $R$ is the unperturbed stellar radius,
and $\omega=2\Omega/(\ell+1)$. The magnetic-type vector
spherical harmonics $\Y^B_{\ell m}$ are defined by

\st\be
\Y^B_{lm}=\frac{r}{\sqrt{\ell(\ell+1)}}\nab\times\left[
 r\nab Y_{\ell m}(\vartheta,\varphi)\right]\,.
\ee The shear of the perturbed star is given by

\st\begin{equation} \delta\sigma_{ab}=\nabla_{\langle a}\delta
v_{b\rangle}\,.\label{r6}
\end{equation}
Substituting Eqs. (\ref{VORTICE1})--(\ref{r6}) into
Eq. (\ref{ENERGYSHEAR}), we find the shear viscosity timescale for
$\ell=m$:

\st\be \frac{1}{\tau_{\rm s}} \approx
Q_\ell\left[(\ell-1)(2\ell+1)\int^R_0\eta r^{2\ell}\,dr +\Omega^2
{\cal S}_\ell\right]\,, \label{NEWSHEAR} \ee where $Q_\ell^{-1}=
\int_0^R\rho r^{2\ell+2}\,dr$. The first term in brackets is in
agreement with the expression calculated in \cite{LOM98}, and
${\cal S}_\ell$ is the correction term:

\st\begin{eqnarray} {\cal S}_\ell
&\approx&16{(\ell-1)(2\ell+1)\over(\ell+1)^{2}}U_0
+\frac{\ell(\ell-2)![(2\ell-1)!!]^2}{(\ell+1)(2\ell-1)(2\ell)!}\,
\frac{\Gamma({{1\over2}})} {\Gamma(\ell-{{1\over2}})}
\times\nonumber\\ &&{}\times\left[ (2\ell^3-8\ell^2-3\ell-6)U_2
+12(\ell^3-\ell^2-\ell+1)U_3 \right.\nonumber\\
&&\left.{}+2(4\ell^4-\ell^3-9\ell^2+5\ell+1)U_4\right]\,,
\label{s}\end{eqnarray} where $U_k(T)\equiv R^{k}\int^R_0
\gamma^2~\eta r^{2\ell-k}\,dr$.

For the $\ell=2$ modes, Eqs. (\ref{NEWSHEAR}) and (\ref{s}) give

\st\be \frac{1}{\tau_{\rm s}}= 5Q_2 \int^R_0\eta r^4 \,dr\,
+{\textstyle{1\over9}}Q_2\Omega^2 \left[80U_0+ 93U_2
+54U_3-42U_4\right]\,. \label{NEWSHEAR1} \ee For comparison with
previous calculations based on Navier-Stokes viscosity (see, e.g.,
\cite{LMO99}), we use an $n=1$ polytrope with mass $M=1.4
M_{\odot}$ and radius $R=12.53$ km to evaluate the integrals in
Eq. (\ref{NEWSHEAR1}). The bulk viscous and gravitational
radiation timescales are unaffected by the vorticity correction,
and we obtain
 \st\bea &&\frac{1}{\tau (\Omega,T)}
=\frac{1}{\tilde\tau_{\rm\sc gr}}\lp\frac{\Omega}{\Omega_{\rm\sc
k}}\rp^6 +\frac{1}{\tilde\tau_{\rm\sc bv}}\lp\frac{T}{10^9\rm
K}\rp^6\lp\frac{\Omega}{\Omega_{\rm\sc k}}\rp^2\nonumber\\
&&~~~~~~~~~~~~~~~~~{}+ \frac{1}{\tilde\tau_{\rm\sc
sv}}\lp\frac{10^9\rm K}{T}\rp^2\ls 1+
q{\bar\alpha}^{4-r}\lp\!\frac{10^9\rm K}{T}\!\rp^{\!r}\!
\lp\!\frac{\Omega}{\Omega_{\rm\sc k}}\!\rp^{\!2}\rs\,,
\label{RMODE} \eea where $\Omega_{\rm\sc k}=\sqrt{\pi G
\bar\rho}$, which is ${3\over2}$ times the Keplerian
(mass-shedding) frequency, and the vorticity correction factors
are

\st\begin{equation} q= \left\{
\begin{array}{l}
1.36 \times 10^{-23}\,,\\ 5.67\times 10^{-10}\,,
\end{array}
\right. ~r= \left\{  \begin{array}{ll}
         9 & {\rm nonrel}\,,\\
         12 & {\rm ultrarel}\,.
         \end{array}\right.
\label{r7}
\end{equation}
The standard result (see, e.g., \cite{LMO99}) is regained for
$q=0$, with
\[
\tilde\tau_{\rm\sc gr}=-3.26\,{\rm s}\,,~\tilde\tau_{\rm\sc
bv}=2.01\times 10^{11} \,{\rm s}\,,~ \tilde\tau_{\rm\sc
sv}=2.52\times 10^8\,{\rm s}\,.
\]
We note that the contribution from the $\dot{\pi}_{ab}$ term in
Eq. (\ref{r3}) to the $q$-correction is less than 1\% of the
contribution from the $\varpi^c{}_{\langle a}\pi_{b\rangle c}$
term.

\begin{figure}
\begin{center}
\leavevmode
\hbox{%
\epsfxsize=4.5in
\epsffile{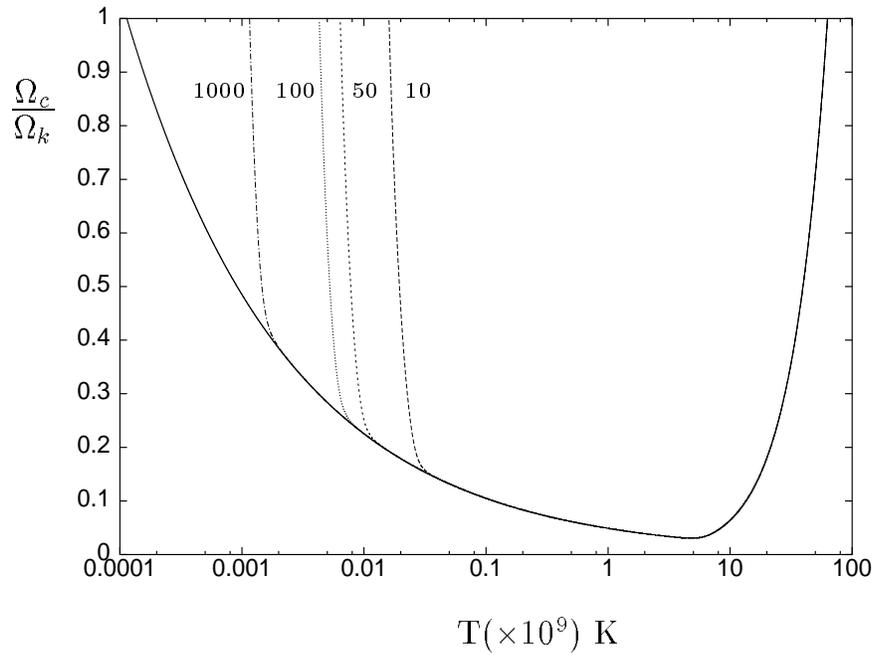}}
\end{center}

\caption{Critical angular velocity versus temperature ($n=1$
polytrope with mass 1.4$M_\odot$ and radius 12.57 km).  The
stability region is below the curves. The solid curve shows the
standard result, with no coupling of viscosity to vorticity.
Broken curves (labelled by the degeneracy parameter ${\bar\alpha}$) show
how the instability region is reduced by the kinetic-theory
coupling of shear viscosity to vorticity, for an
ultra-relativistic degenerate Fermi fluid (electron-electron
viscosity).}
\label{fig3}

\end{figure}

Now we are able to determine from Eq. (\ref{RMODE}) the critical
angular velocity $\Omega_{\rm\sc c}$, defined by $1/\tau
(\Omega_{\rm\sc c},T)=0$, which governs stability of the star: if
$\Omega>\Omega_{\rm\sc c}$, then dissipative damping cannot
overcome the gravitational radiation-driven instability. In
Fig.~\ref{fig3} we plot $\Omega_{\rm\sc c}/\Omega_{\rm\sc k}$
against temperature $T$, showing how the vorticity-viscosity
coupling affects the standard result (see, e.g., \cite{LMO99}).
Electrons are assumed to dominate the shear viscosity, and they
are ultrarelativistic over the range of temperatures.

It is clear from Fig.~\ref{fig3} that the vorticity correction is
only appreciable at temperatures $T\leq 10^8$ K, but that for
these lower temperatures, the correction can be large, especially
for smaller ${\bar\alpha}$. As the degree of degeneracy increases (i.e.,
with increasing ${\bar\alpha}$), the correction is confined to lower and
lower temperatures. The effect of the vorticity-viscosity coupling
is to increase the stable region, so that cooler stars can spin at
higher rates and remain stable. This may modify recent results
\cite{cold} which suggest that $r$-mode instability could stall
the spin-up of accreting neutron stars with $T\geq 2\times10^5$
K; if the vorticity correction operates, then the stability region
is increased, so that spin-up could be more effective, especially
for lower degeneracy parameter ${\bar\alpha}$.

We note that, here, we have used for our $r$-mode calculations
solutions that assume slow rotation. Thus the
$\Omega/\Omega_{\rm\sc k}\geq 0.3$ part of Fig. \ref{fig3} is
an extrapolation to high spin rates, in common with previous
stability diagrams. Recent calculations of $r$-modes for rapid
rotation \cite{LIp99} should be used in future calculations of the
vorticity correction. Since $f$-modes are unstable at high spin
rate, the effect of the vorticity correction on these modes would
also be interesting to calculate.

\newpage

\clearpage
\newpage
\newpage
\newpage
\newpage

%%%%%%%%%%%%%%%%%%%%%%%%%%%%%%%%%%%%%%%%%%%%%%%%%%%%%%%%%%%%%%%%%%%%%%%%%%%%%%%%%%%%%%%%%%%%%

\section{Post-glitch relaxation of Crab}
As we discussed before, the $r$-mode instability opens a wide
window for gravitational wave astronomy.   It would give us some
information about the interior matter of neutron stars.
Determining cooling rates, viscosity, crust formation, the
equation of state of neutron matter, the onset of superfluidity in
neutron stars, and several other features of neutron stars are
more interesting implications of this instability. In this
section, we discuss one of possible $r$-mode implications.

More than 30 years after the discovery of the pulsar phenomenon
and its identification with neutron stars, there exists still a
number of uncertainties and open questions about the theoretical
model for pulsars, mainly due to the extremely dense state of
matter in neutron stars.  During the past two decades, the glitch
phenomenon, a sudden increase of angular velocity of the order of
$\Delta\Omega/\Omega\leq 10^{-6}$, and the very long
relaxation times, from months to years, after the glitch, remain
as one of the great mysteries of pulsars.  The observed
post-glitch relaxation of the Crab pulsar has been unique in that
the rotation frequency of the pulsar is seen to decrease to values
$less$ than its pre-glitch extrapolated values. So far, two
mechanisms have been suggested to account for the observed excess
loss of angular momentum during post-glitch relaxations of the
Crab. The first mechanism, in the context of the vortex creep
theory of Alpar et~al. \cite{Alp84, Alp96}, invokes generation, at
a glitch, of a so called ``capacitor'' region within the pinned
superfluid in the crust of a neutron star, resulting in a
permanent decoupling of that part of the superfluid. Neverthless,
this suggestion has been disqualified (Lyne et al. \cite{Lyn93})
since the moment of inertia required to have been decoupled
permanently in such regions during the past history of the pulsar
is found to be much more than that permitted for {\it all} of the
superfluid component in the crust of a neutron star. In another
attemp, Link et~al. \cite{LEB92} have attributed the excess loss
of angular momentum to an increase in the electromagnetic braking
torque of the star, as a consequence of a sudden increase, at the
glitch, in the angle between its magnetic and rotation axes. As
they point out, such an explanation is left to future
observational verification since it should also accompany other
observable changes in the pulsar emission, which have not been
detected, so far, in any of the resolved glitches in various
pulsars. Moreover, the suggestion may be questioned also on the
account of its long-term consequences for pulsars, in general.
Namely, the inclination angle would be expected to show a
correlation with the pulsar age, being larger in the older pulsars
which have undergone more glitches. No such correlation has been
deduced from the existing observational data. Also, and even more
seriously, the assumption that the braking torque depends on the
inclination angle is in sharp contradiction with the common
understanding of pulsars spin-down process. The currently inferred
magnetic field strengths of all radio pulsars are in fact based on
the opposite assumption, namely that the torque is independent of
the inclination angle. The well-known theoretical justification
for this, following Goldreich \& Julian \cite{GJu69}, is that the
torque is caused by the combined effects of the magnetic dipole
radiation and the emission of relativistic particles, which
compensate each other for the various angles of inclination (see,
eg., Manchester \& Taylor \cite{MTa77}; Srinivaran \cite{Sri89}).

The excitation of r-modes at a glitch and the resulting emission
of gravitational waves could, however, account for the required
``sink'' of angular momentum in order to explain the peculiar
post-glitch relaxation behavior of the Crab pulsar. As is shown in
Figs.~\ref{fig4} and \ref{fig5}, for values of $ \alpha_0 \geq 0.04 $ the predicted time
evolution of $\Delta \Omega \over \Omega$ and $\Delta\dot \Omega
\over\dot\Omega $ during the 3--5 years of the inter-glitch
intervals in Crab, might explain the observations. That is, the
predicted total change in the rotation frequency of the star,
$|{\Delta\Omega\over\Omega}|$,  is much larger than the
corresponding jump $\frac{\Delta\Omega}{\Omega}\sim 10^{-8}$ at
the glitch, which explains why the post-glitch values of $\Omega$
should fall below that expected from an extrapolation of its
pre-glitch behavior. Also, the predicted values of ${\Delta \dot
\Omega \over\dot\Omega} \sim 10^{-4}$, after a year or so
(Fig.~\ref{fig5}), are in good agreement with the observed persistent
shift in the spin-down rate of the Crab (Lyne et~al.
\cite{Lyn95}).

The predicted increase in the spin-down rate would be however
diminished as the excited modes at a glitch are damped out,
leaving a permanent negative offset in the spin frequency. Hence
the above so-called persistent shift in the spin-down rate of the
Crab may be explained in terms of the effect of r-modes, as long
as it persists during the inter-glitch intervals of 2-3 years. It
may be noted that a really persistent shift in the spin-down rate
at a glitch may be caused by a sudden decrease in the moment of
inertia of the star. However this effect, by itself, could not
result in the observed {\it negative} offset in the spin
frequency.

The same mechanism would be expected to be operative during the
post-glitch relaxation in the other {\it colder} and {\it slower}
pulsars, as well. However, for the similar values of $\alpha_0$,
ie. the same initial amplitude of the excited modes, the effect is
not expected to become ``visible'' in the older pulsars.
Particularly, for the Vela its initial jump in frequency at a
glitch, $\frac{\Delta\Omega}{\Omega}\sim 10^{-6}$, is seen from
Fig.~\ref{fig4} to be much larger (ie. by some four orders of
magnitudes) than that of the above effect due to the r-modes. In
other words, while the predicted loss in the stellar angular
momentum due to the excitation of r-modes result in a negative
$\Delta \Omega \over \Omega$ which, in the case of Crab,
overshoots the initial positive jump at a glitch, however for the
Vela and older pulsars it comprises only a negligible fraction of
the positive glitch-induced jump. A more detailed study should,
however, take into account the added complications due to internal
relaxation of various components of the star, which is highly
model dependent. The observed initial rise in $\Omega$ need not be
totally compensated for by the losses due to r-modes which we have
discussed, since part of it could be relaxed internally (by a
transfer of angular momentum between the ``crust'' and other
components, and/or temporary changes in the effective moment of
inertia of the star) even in the absence of any real sink for the
angular momentum of the star. Such considerations would not only
leave the above conclusions valid but also allow for even smaller
values of the initial amplitude of the excited modes, compared to
our presently adopted value of $\alpha_0 \sim 0.04$.

The suggested effect of the r-modes in the post-glitch relaxation
of pulsars should be understood as one operating in addition to
that of the internal relaxation which is commonly invoked. While
the latter could account {\it only} for a relaxation back to the
extrapolated pre-glitch values of the spin frequency, the
additional new effect due to the r-modes may explain the {\it
excess} spin-down observed in the Crab pulsar as well.

It is further noted that the above estimates are for an adopted
value of \( Q = 9.4\times 10^{-2}\), which corresponds to the
particular choice of the polytropic model star. Differences in the
structure among pulsars, in particular between Crab and Vela,
which have also been invoked in the past (see, eg., Takatsuka \&
Tamagaki \cite{TTa89}), could be further invoked to find a better
agreement with the data for the above effect due to r-modes as
well. Also, the initial amplitude of the excited modes need not be
the same in all pulsars. It is reasonable to assume that in a
hotter and faster rotating neutron star, as for the Crab, larger
initial amplitudes, ie. larger values of $\alpha_0$, are realized
than in the colder--slower ones.

\subsection{Predicted Effect}
In order to estimate the effect of the $r$-mode instability, in a
``stable'' neutron star, on its post-glitch relaxation, we have
used the model described by Owen et~al. \cite{Owe98}. The total
angular momentum of a star is parameterized in terms of the two
degrees of freedom of the system. One, is the uniformly rotating
equilibrium state which is represented by its angular velocity
$\Omega_{\rm eq}$. The other, is the excited $r$-mode that is
parameterized by its magnitude $\alpha$ which is bound to an upper
limiting value of $\alpha=1$, in the linear approximation regime
treated in the model. Thus, the total angular momentum $J$ of the
star is written as a function of the two parameters $\Omega_{\rm
eq}$ and $\alpha$: \stepcounter{sub}
\begin{equation}
J=I_{\rm eq} \Omega_{\rm eq} + J_c,\label{rj1}
\end{equation}
where $I_{\rm eq}=\tilde{I} M R^2$ is the moment of inertia of the
equilibrium state, and $J_c=-{3 \over 2} \tilde{J} \alpha^2
\Omega_{\rm eq} M R^2$ is the canonical angular momentum of the
$l=2$ $r$-mode, which is negative in the rotating frame of the
equilibrium star. The dimensionless constants

\st\bea\label{rj2a}
&& \tilde{I}={8\pi\over 3MR^2}\int_0^R\rho r^4\,dr \\
 \st
&&\tilde{J}={1\over MR^4}\int_0^R\rho r^6\,dr, \label{rj2b}\eea
 depend on
the detailed structure of the star, and for the adopted $n=1$
polytropic model considered have values $\tilde{I}=0.261$ and
$\tilde{J}=0.01635$. Also $R=12.54$~km and $M=1.4 \ M_\odot$ are
the assumed radius and mass of the star, for the same polytropic
model.

Eq. (\ref{rj1}) above implies that an assumed instantaneous
excitation of $r$-modes at a glitch would cause a sudden increase
in $\Omega_{\rm eq}$. For definiteness, we define the ``real''
observable rotation frequency $\Omega$ of the star as $\Omega = {J
\over I}$, where $I$ is the moment of inertia of the real star.
The two are equal, $\Omega =\Omega_{\rm eq}$, in the absence of
the $r$-modes, ie. before the excitation of the modes at a glitch
and after the modes are damped out. If there were no loss of
angular momentum (by gravitational radiation) accompanying the
post-glitch damping of the modes (by viscosity)  $\Omega_{\rm eq}$
would recover its extrapolated pre-glitch value; ie. its initial
rise would be compensated exactly. However due to the net loss of
angular momentum by the star, the post-glitch decrease of
$\Omega_{\rm eq}$ {\it overshoots} its initial rise. The negative
offset between values of $\Omega_{\rm eq}$ before the excitation
of the modes and after they are damped out is the quantity of
interest for our discussion. The question of whether the
instantaneous rise in the value of $\Omega_{\rm eq}$ at a glitch,
due to the excitation of the $r$-modes, is  observable or not is a
separate problem, and its resolution would have no consequence for
the net loss of angular momentum from the star which is the
relevant quantity here. It is noted that the distinction between
$\Omega$ and $\Omega_{\rm eq}$, in the presence of modes, is
quantitatively negligible, in all cases of interest, and is
usually disregarded. Also, one might dismiss an {\em increase} in
$\Omega_{\rm eq}$ as implied by Eq. (\ref{rj1}) to be observable
as a spin-up of the star since for an inertial outside observer
the $r$-modes rotate in the prograde direction and their
excitation should result, if at all, in a {\it spin-down} of the
star. Moreover, an excitation of the $r$-modes should not result,
by itself, in any real change of the rotation frequency of the
star at all. Because one could not distinguish two physically
separate parts of the stellar material such that the two
components of angular momentum in Eq. (\ref{rj1}) may be assigned
to the two parts separately.

The total
angular momentum $J$ of the star in terms of $\Omega_{eq}$ and $\alpha$ is

\st\be J(\Omega_{\rm eq},\alpha)=\lp{\tilde I}-{3\over 2}{\tilde
J} \alpha^2\rp\Omega_{\rm eq} M R^2.\label{rj3}\ee

The perturbed star loses angular momentum primarily through the
emission of gravitational radiation.  Thus, the evolution of
$J(\Omega_{\rm eq}, \alpha)$ can be computed by using the standard
multipole expression for angular momentum loss. Then, for the
$\ell=m=2$ case

\st\be {dJ\over dt}=-{c^3\over 16\pi G}\lp{4\Omega_{\rm eq}\over
3}\rp^5(S_{22})^2, \label{rj4}\ee where $c$ is the speed of light
and $\ell=m=2$ current multipole, $S_{22}$, is given by
\cite{Tho80}

\st\be S_{22}=\sqrt{2}{32\pi\over 15}{G M\over c^5}\alpha
\Omega_{\rm eq} R^3 {\tilde J}. \label{rj5}\ee Combining Eq.
(\ref{rj4}) for the angular momentum evolution of the star with
Eqs. (\ref{17}) and (\ref{rj5}), we obtain one equation for the
evolution of the parameters $\Omega_{\rm eq}$ and $\alpha$ that
determine the state of the star:

\st\be \lp{\tilde I}-{3\over 2}{\tilde J} \alpha^2\rp{d\Omega_{\rm
eq}\over dt}-3\alpha\Omega_{\rm eq}{d\alpha\over
dt}={3\alpha^2\Omega_{\rm eq}{\tilde J}\over \tau_{\rm\sc gr}}.
\label{rj6}\ee

In addition to radiating angular momentum from the star via
gravitational radiation, the mode will also lose energy through
gravitational radiation and neutrino emission (from bulk
viscosity).  Furthermore the mode energy is deposited into the
thermal state of the star by shear viscosity.  Therefore the
energy balance equation should be considered together with Eq.
(\ref{rj6}) to determine the parameters $\Omega_{\rm eq}$ and
$\alpha$.  For the $\ell=2$ $r$-mode $\tilde E$, Eq. (\ref{14}),
is given by

\st\be {\tilde E}={\scriptstyle \frac{1}{2}}\alpha^2\Omega^2_{\rm eq} M R^2
{\tilde J}.\label{rj7}\ee

\noindent The time derivative of $\tilde E$, Eq.
(\ref{TOTALTIME}), is

\st\be {d\tilde{E}\over dt}= -2{\tilde E}\lp\frac{1}{\tau_{\rm\sc
v}}+\frac{1}{\tau_{\rm\sc gr}}\rp\,, \label{rj8} \ee where
$\tau_{\rm\sc v}$ and $\tau_{\rm\sc gr}$ are the  viscous and
gravitational radiation timescales respectively. Combining Eqs.
(\ref{rj7}) and (\ref{rj8}), the second evolution equation for
$\Omega_{\rm eq}$ and $\alpha$ is given by

\st\be \Omega_{\rm eq}{d\alpha\over dt}+\alpha{d\Omega_{\rm
eq}\over dt}=-\alpha\Omega_{\rm eq}\lp\frac{1}{\tau_{\rm\sc
v}}+\frac{1}{\tau_{\rm\sc gr}}\rp. \label{rj9} \ee

\noindent Therefore the time evolution of the quantities $\alpha$
and $\Omega_{\rm eq}$ can be determineded from the coupled
equations (Owen et~al. \cite{Owe98}):

\st\label{rj10}
\stq
\begin{eqnarray}\label{rj10a}
&&\frac{{\rm d}\Omega_{\rm eq}}{{\rm d}t}= -\frac{2\Omega_{\rm
eq}}{\tau_{\rm\sc v}}\frac{\alpha^2 Q}{1+\alpha^2 Q},\\
\stepcounter{subeqn} &&\frac{{\rm d}\alpha}{{\rm d}t}=
-\frac{\alpha}{\tau_{\rm\sc gr}}-\frac{\alpha}{\tau_{\rm\sc v}}
\frac{1-\alpha^2 Q}{1+\alpha^2 Q}\label{rj10b} ,
\end{eqnarray}
where $Q= {3 \over 2}{\tilde{J} \over \tilde{I}} = 0.094$, for the
adopted equilibrium model of the star.  The viscous time has two
contributions from the shear and bulk viscousities with
corresponding timesacels $\tau_{\rm sv}$ and
 $\tau_{\rm bv}$, respectively. The overall ``damping''
timescale $\tau$ for the mode, which is a measure of the period
over which the excited mode will persist, is defined as
\stepcounter{sub}
\begin{equation}
\frac{1}{\tau}=\frac{1}{\tau_{\rm\sc v}}+\frac{1}{\tau_{\rm\sc gr}}=
\frac{1}{\tau_{\rm\sc sv}}+\frac{1}{\tau_{\rm\sc bv}}+\frac{1}{\tau_{\rm\sc
gr}}\label{rj11}
\end{equation}
Following Owen et~al. \cite{Owe98} we use \(\tau_{\rm\sc
sv}=2.52\times 10^8 ({\rm s}) T_9^2 \), \(\tau_{\rm\sc bv}=4.92\times
10^{10} ({\rm s}) T_9^{-6} \Omega_3^{-2} \), and \(\tau_{\rm\sc
gr}=-1.15\times 10^6 ({\rm s}) \Omega_3^{-6}\), where $T_9$ is the
temperature, $T$, in units of $10^9$~K, and $\Omega_3$ is in units
of $10^3 {\rm rad \ s}^{-1}$. These estimates do not however
include the role of superfluid mutual friction in damping out the
oscillations. We have further taken into account the damping due
to the mutual friction using the associated damping time as given
by Lindblom and Mendell \cite{LMe99}.  The effect of the mutual
friction is nevertheless seen to be negligible and the computed
curves shown below remain almost the same in the presence of
mutual friction.

\begin{figure}
\begin{center}
\leavevmode
\hbox{%
\epsfxsize=4.5in
\epsffile{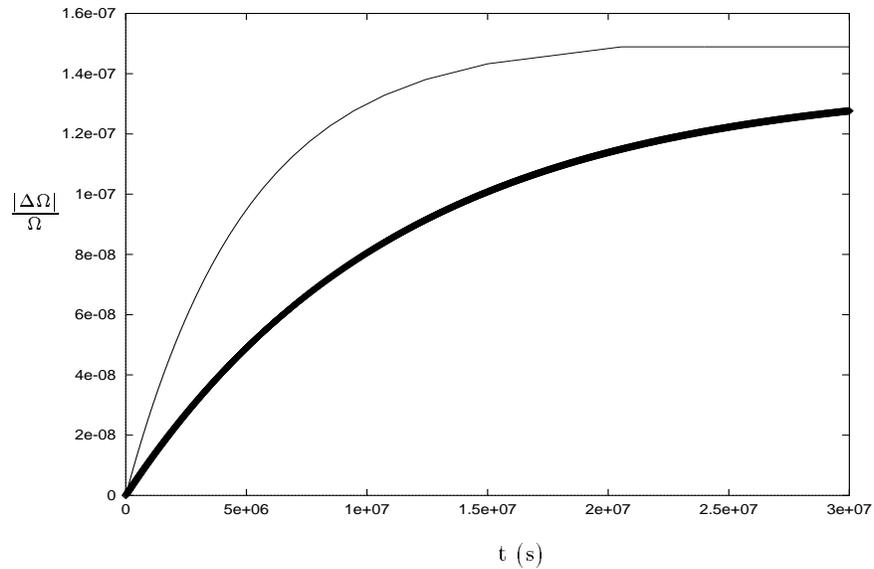}}
\end{center}

\caption{  The post-glitch time evolution of the
spin
frequency of a pulsar, caused by its loss of angular
momentum due to gravitational waves driven by the
r-modes that are assumed to be excited at the glitch
epoch, $t=0$, with an initial amplitude of
$\alpha_0=0.04$.
The two curves correspond to assumed values of
$T_9=0.3$ and $\Omega_3 =0.19$, for
the Crab ({\it thick} line), and
$T_9=0.2$ and $\Omega_3 =0.07$, for
the Vela ({\it thin} line). Note that the curve for
Vela represents the results {\bf after being multiplied}
by a factor of $10^3$.}

\label{fig4}

\end{figure}

\begin{figure}
\begin{center}
\leavevmode
\hbox{%
\epsfxsize=4.5in
\epsffile{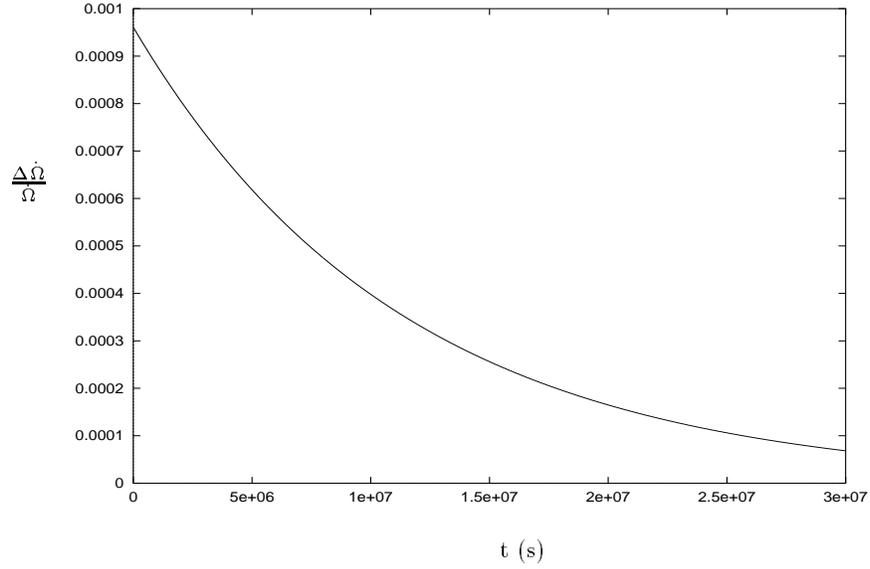}}
\end{center}

\caption{Time evolution of the
spin-down rate of a pulsar, caused by its loss of
angular momentum due to the excitation of r-modes at $t=0$.
A value of $\dot\Omega= 2.4 \times 10^{-9}$rad s$^{-2}$, and
other parameter values same as in Fig. \ref{fig4} for the Crab have
been assumed.}

\label{fig5}
\end{figure}

By integrating Eqs. (\ref{rj10a}) and (\ref{rj10b}), numerically,
for a given initial value of $\alpha$, one may therefore follow
the time evolution of $\alpha$ and $\Omega_{\rm eq}$ which
together with Eq. (\ref{rj1}) determine the time evolution of the
total angular momentum, $J$, and hence the time evolution of
$\Omega$. Figs.~\ref{fig4} and \ref{fig5} shows the computed time evolution for the
absolute value of the resulting (negative) fractional change
$\Delta \Omega \over \Omega$ in the spin frequency (Fig.~\ref{fig4}) and
also the change $\Delta \dot \Omega \over\dot \Omega$ in the
spin-down rate of the star (Fig.~\ref{fig5}), starting at the glitch
epoch which corresponds to time $t=0$. The results in Figs.~\ref{fig4} and \ref{fig5} are
for a choice of an initial value of \(\alpha_0=0.04 \), and for
the assumed values of $T$ and $\Omega$ corresponding to the Crab
and Vela pulsars, as indicated. Fig.~\ref{fig4} shows that for the same
amplitude of the r-modes assumed to be excited at a glitch the
resulting loss of angular momentum through gravitational radiation
would be much larger in Crab than in Vela, ie. by more than 3
orders of magnitudes. (Note that the curve for Vela in Fig.~\ref{fig4}
represents the results after being multiplied by a factor of
$10^3$.) Furthermore, for the adopted choice of parameter values,
the magnitude of the corresponding decrease in $\Omega$ for the
Crab, is \(|{\frac{\Delta\Omega}{\Omega}}| \sim 10^{-7} \)
(Fig.~\ref{fig4}). The observational consequence of such an effect would
nevertheless be closely similar to what has been already observed
during the post-glitch relaxations of, {\it only}, the Crab
pulsar.

\begin{figure}
\begin{center}
\leavevmode
\hbox{%
\epsfxsize=4.5in
\epsffile{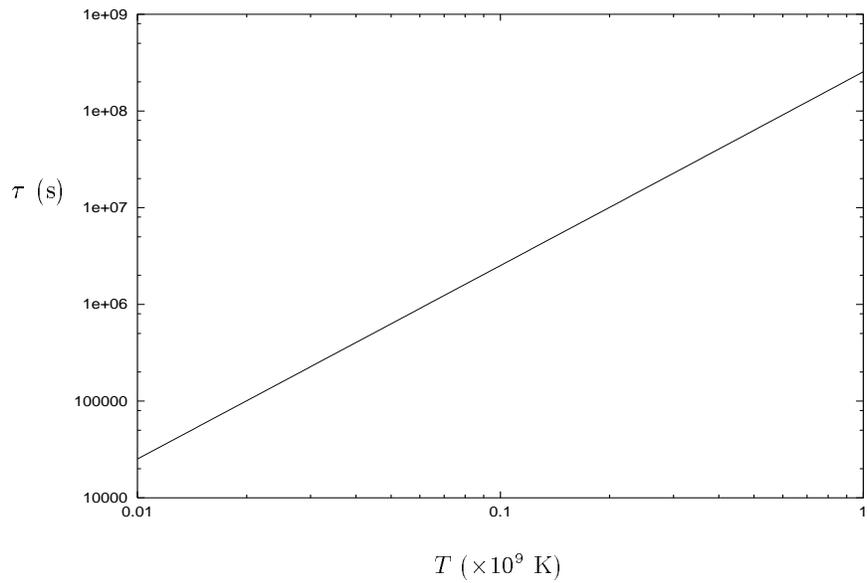}}
\end{center}

\caption{ Dependence of the total damping
timescale of r-modes on the internal
temperature of a neutron star.
Parameter values same as for the Crab in Fig. \ref{fig4} have
been assumed.}

\label{fig6}
\end{figure}

\begin{figure}
\begin{center}
\leavevmode
\hbox{%
\epsfxsize=4.5in
\epsffile{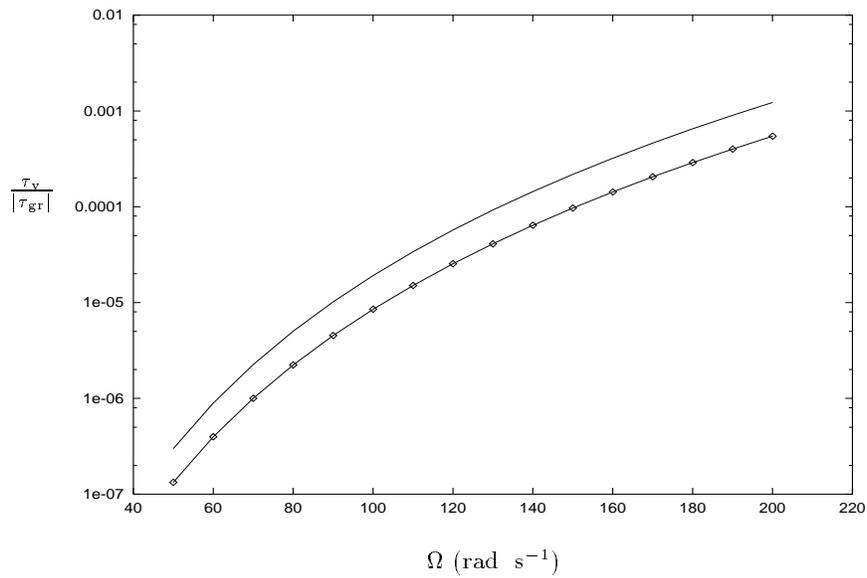}}
\end{center}

\caption{ Dependence of the
net post-glitch decrease in the rotation
frequency, on the rotation frequency of a neutron star.
The two curves are to show the dependence on the temperature,
where $T_9=0.3$ ({\it bare} line) and
$T_9=0.2$ ({\it dotted} line) have been used.}

\label{fig7}
\end{figure}

Before proceeding further with Crab, we note that the post-glitch
effects of excitation of r-modes would however have not much
observational consequences for the Vela, and even more so for the
older pulsars, which are colder and rotate more slowly. This has
two, not unrelated, reasons: in the older pulsars r-modes a) are
damped out faster (ie. have smaller values of $\tau$), and b)
result in less gravitational radiation. The dependence of $\tau$
on the stellar interior temperature is shown in Fig.~\ref{fig6}. For the
colder, i.e. older, neutron stars the r-modes are expected to die
out very fast. The damping timescale for a pulsar with a period \(
P \sim 1 {\rm s} \), being colder than $10^8$~K, could be as short
as a few hours (Fig.~\ref{fig6}), and r-modes would have been died out
at times longer than that after a glitch. For the hot Crab pulsar,
on the other hand, r-modes are expected to persist for 2-3 years
after they are excited, say, at a glitch. The value of $\tau$
decreases for older pulsars due to both their longer periods as
well as lower temperatures, but the effect due to the latter
dominates by many orders of magnitudes, for the standard cooling
curves of neutron stars (Urpin et. al. \cite{UPr93}). The second
reason, ie. the loss of angular momentum being negligible in older
pulsars, was already demonstrated in Fig.~\ref{fig4}, by a comparison
between Crab and Vela pulsars. We have verified it also for the
case of pulsars older than Vela. It may be also demonstrated
analytically from Eqs. (\ref{rj10a}) and (\ref{rj10b}), in the
limit of $\alpha^2 Q << 1$. The initial increase in $\Omega_{\rm
eq}$ due to excitation of r-modes with a given initial amplitude
$\alpha_0$ is seen from Eq.~(\ref{rj1}) to be $|{\Delta\Omega_{\rm
eq}\over\Omega_{\rm eq}}|_0= \alpha_0^2 Q$. The subsequent damping
of the modes result in secular decrease in $\Omega_{\rm eq}$, and
the total decrease at large $t\rightarrow\infty$ would be
$|{\Delta\Omega_{\rm eq}\over\Omega_{\rm eq}}|_\infty \sim {\tau
\over \tau_{\rm\sc v}}\alpha_0^2 Q$, which is true for
$|{\Delta\Omega_{\rm eq}| << \Omega_{\rm eq}}$. Note that in the
absence of gravitational radiation losses (ie. ${1 \over \tau_{\rm\sc
gr}}=0; \tau = \tau_{\rm\sc v}$) the total decrease would be the same
as the initial increase, which is expected for the role of viscous
damping alone. The difference between these two changes (total
decrease minus initial increase) in $\Omega_{\rm eq}$ would
correspond to the total loss of angular momentum from the star,
hence to the net decrease in its observable rotation frequency,
ie. \stepcounter{sub}
\begin{equation}
|{\Delta\Omega \over\Omega}|_\infty= {\tau_{\rm\sc v} \over
|\tau_{\rm\sc gr}|} \alpha_0^2 Q\,,\label{rj12}
\end{equation}
which is valid in the limit of ${\tau_{\rm\sc v} \over |\tau_{\rm\sc
gr}|} << 1$. Fig.~\ref{fig7} shows the dependence of the quantity
${\tau_{\rm\sc v} \over |\tau_{\rm\sc gr}|}$ on the stellar rotation
frequency, and also on its internal temperature. While for the
Crab ${\tau_{\rm\sc v} \over |\tau_{\rm\sc gr}|} \sim 10^{-3}$, however
its value is much less for the older pulsars, due to both their
lower $\Omega$ as well as lower $T$ values. The dependence on the
temperature is however seen to be much less than that on the
rotation frequency, in contrast to the dominant role of the
temperature in determining the value of the total damping time
$\tau$, as indicated above. As is seen in Fig.~\ref{fig7}, for the Vela
${\tau_{\rm\sc v} \over |\tau_{\rm\sc gr}|} < 10^{-7}$, which means a
maximum predicted value of \(|{\Delta\Omega \over\Omega}|_\infty
<10^{-8} \), even for the large values of $\alpha_0 \sim 1$. This
has to be contrasted with the glitch induced values of
\(|{\Delta\Omega \over\Omega}| \sim1 0^{-6} \) in Vela, which
shows the insignificance of the role of r-modes in its post-glitch
behaviour.

\newpage

\newpage

\newpage

\newpage

\setcounter{sub}{0} \setcounter{subeqn}{0}
\renewcommand{\theequation}{7.\thesub\thesubeqn}

\chapter{Concluding remarks}

In many cosmological and astrophysical situations, an idealized
fluid model of matter is inappropriate, and a self-consistent
microscopic model based on relativistic kinetic theory gives a
more detailed physical description. Kinetic theory offers a
microscopic approach to describe the macroscopic features of
matter, rather than the phenomenological fluid dynamics and its
associated thermodynamics. Starting from a microscopic approach,
to obtain an effective macroscopic description, is the most
fascinating feature of this theory. The theory is based on a
simple function which is called the distribution function which is
a solution of Boltzmann's or Liouville's equation, and describes
the dynamics of the system at the microscopic level.  At the
macroscopic level, the mass density, flow density, pressure, and
the other macroscopic quantities are obtained from the
distribution function. On the other hand,  observations at large
scales, such as stellar systems, can help us to improve our
understanding of microphysics.

In chapter 3, general relativistic Liouville's equation in the
post-Newtonian approximation was studied.  In the static case, the
equilibrium, two static solutions of the Liouville's equation in
this approximation are obtained.  These integrals are
generalizations of the classical energy, $E={1\over
2}v^2+\phi+(2\phi^2+\psi)/c^2$, and angular momentum,
$l_i=\varepsilon_{ijk}x^jv^k \exp(-\phi/c^2)$.  In this spirit,
the polytropic model, a simple model for a neutron
 star, was studied. Our results show that the post-Newtonian
corrections tend to reduce the radius of any polytrope. This is a
consequence of the fact that the post-Newtonian correction is more
significant for systems with larger density \cite{RSo00}.

Linear perturbations of phase space distribution functions was
investigated in chapter 4. We introduced the linearized
Liouville-Einstein equation in this approximation. We showed that,
if the underlying potentials are spherically symmetric, the
evolution equation is O(3) symmetric, ie. the linearized
Liouville-Einstein operator commutes with the angular momentum
operator in phase space,
$\varepsilon_{ijk}(x^j{\partial\over\partial
x^k}+v^j{\partial\over\partial v^k})$.  Then the modes can be
characterized by a pair of angular momentum eigennumbers, $(j,
m)$.  The eigenvalues $\omega_j$ are, however, $(2j+1)$ fold
degenerate.

Furthermore, we showed that the post-Newtonian gravitational
potentials may excite some of the neutral modes of the star and
that these modes are purely relativistic effects.  Using the O(3)
property of $pnl$, we proposed distribution functions for
perturbations that are functions of classical energy and classical
angular momentum, Eqs. (\ref{pertdist}) and (\ref{lambda}):
\[
f_{jm}=Re\;\Lambda_{jm}\;\;; \Lambda_{jm}=af(e,
l^2)J_+^{j+m}l_-^j=bf(e, l^2)J_-^{j-m}l_+^j,
\]
Although these functions are neutral in classical approximation,
they are not so in $pn$ order. Neutral, here, means to belong to zero
frequency modes. The weak $pn$ forces generate a sequence of low
frequency modes from such perturbations. In their hydrodynamic
behavior, they constitute a sequence of low frequency toroidal
modes.  There is an oscillatory $g_{0i}$ component of the metric
tensor associated with these modes. From a conceptual point of
view, they are similar to toroidal modes of slowly rotating fluids
generated by Coriolis forces or to the standing Alfven waves of a
weakly magnetized fluids \cite{SRe00}.

The latter perturbations are analogue of the recent quasi normal
modes in relativistic systems believed to have been originated
from the perturbations of the space-time metric, gravitational
wave modes ($w$-modes).   Kokkotas and Schutz \cite{KSc86} first
recognized the $w$-modes in a toy model of a finite string (to
mimic a fluid) coupled to a semi-infinite one (to substitute the
dynamical space-time). Such a system accommodates a family of
damped oscillations due to the emission of gravitational wave.
Different investigators have proposed different mathematical and
numerical schemes to isolate these modes
\cite{Koj88}--\cite{LMI97}.  They verified that strongly damped
($w$-) modes, due to the space-time metric perturbation, do indeed
exist in realistic stellar models.

In chapters 5 and 6, we studied the recent and interesting instability in
rotating neutron stars.  Recently it has been shown that the
instability of perturbations of rotating stars are important
during the early history of hot neutron stars. These perturbations
are driven by the Coriolis force which is always present in a
rotating star and are known as $r$-modes. The instability of these
modes cause a rotating neutron star's rotation rate to slow down,
emitting gravitational radiation in the process. This emission of
gravitational radiation is important both as a possibly detectable
source and as a mechanism to explain the observed spin rate of
neutron stars.

It is important to improve our understanding of the various
factors that go into $r$-mode stability analysis. One important
aspect which has needed further elaboration is the role of
dissipation in the fluid. Normally, the fluid's viscosity is
thought to damp out any instability if the star is relatively
cool. However, the earliest analyses only considered a model for
the viscosity based on the  Navier-Stokes theory (which is known
to have serious problems, such as faster-than-light propagation of
signals).

As a first attempt to include more realistic microphysics, we have
considered the role of vorticity on the stability of $r$-modes.
This effect is predicted by kinetic theory when the unperturbed
equilibrium state is rotating, but is absent in Navier-Stokes
theory. In standard Navier-Stokes theory, the angular velocity of
the fluid has no effect on viscous stress or heat flux. We
calculated the vorticity-shear viscosity coupling and showed that
the coupling between vorticity and shear viscous stress predicted
by kinetic theory can in principle have a significant effect on
$r$-mode instability in neutron stars. The M\"uller-Israel-Stewart
correction of Navier-Stokes theory predicts that colder stars can
remain stable at higher spin rates, so that accreting spin-up
could be protected from $r$-mode instability \cite{RMa00}.

Normally all neutron stars which have been observed are seen to be
rotating while showing a slow down of their rotation rate.
However, some pulsars, such as the Crab pulsar exhibit glitches,
which are brief periods during which their rotation rate suddenly
increases. We have studied the role of $r$-modes in the
post-glitch relaxation of radio pulsars. We have shown that
excitation of the $r$-modes at a glitch may provide a solution to
an unsolved observed effect in post-glitch relaxation of the Crab
pulsar \cite{RJa00}.

Of course, our analysis is limited by the fact that we have
followed the standard assumption in viscous stability analysis and
ignored superfluid effects that will become important at lower
temperatures (see, e.g., \cite{LMe99}). Superfluid ``friction"
effects are thought to prevent $f$-mode instability, and these
effects are likely to be relevant also for $r$-modes. These
effects may strongly alter the vorticity correction effect, and
the possibility of $r$-modes excitation in the Crab like pulsars
($ T \sim 10^8$ K).\\

\appendix

\setcounter{sub}{0}
\setcounter{subeqn}{0}
\renewcommand{\theequation}{A.\thesub\thesubeqn}

\chapter{ The post-Newtonian approximation}

The Einstein field equations are nonlinear, and therefore cannot
in general be solved exactly.     In most cases, by imposing some
symmetries such as time independence, spatial isotropy and/or
homogenity, we were able to find some exact solutions, the
Schwarzschild and the Freidmann-Robertson-Walker metrics for
examples.  But we cannot actually make use of the symmetries in
all problems. Solar system is the familiar example of non-static
and anisotropic case.

In most problems what we need is not to find the exact solutions of the
problems, but we need a systematic approximation method to extract the
solutions without any assumed symmetry properties of the problem.

The post-Newtonian approximation was historically derived
\cite{Ein,EIn40,EIn49}, to the study of the
problem
of motion.  But in the last three decades, It is used largely to study
dynamics of stellar systems like compact stars and black holes.

In this appendix we introduce the post-Newtonain approximation in
details.  We follow \cite{Wei72} to present this method.

Consider a system of particles that, like the sun and the planets,
are bound together by their mutual attraction. Let ${\bar M}$,
${\bar r}$, and ${\bar v}$ be typical values of the masses,
separations, and velocities of these particles.   The Newtonian
typical kinetic energy  ${\bar M}{\bar v}^2/2$ will be roughly of
the same order of magnitude as the typical potential energy
$G{\bar M}/{\bar r}$, so \st
\be
{\bar v}^2\approx\frac{G{\bar M}}{{\bar r}}. \ee The relation will
be exact for a particle moves with velocity ${\bar v}$ in circular
orbit of radius ${\bar r}$ about a central mass ${\bar M}$.   The
post-Newtonian approximation may be described as a method for
obtaining the motion of the system to one higher power of the
small parameters  $G{\bar M}/{\bar r}$ and ${\bar v}^2$ than given
by Newtonian mechanics.    It is also referred to as an expansion
in inverse powers of the speed of light, $c$. Here we prefer to
use ${\bar v}/c$ as the expansion parameter.

From our experience with the Schwarzschild solution, we expect
that it should be possible to find a coordinate system in which
the metric tensor is nearly equal to the Minkowski tensor
$\eta_{\mu\nu}$, the corrections being expandable in powers of
${\bar v}/c$.    In particular, we expect \st \stq \bea
&&g_{oo}=-1+ \;^2g_{oo}+ \;^4g_{oo}+\cdots,\\ \stq
&&g_{ij}=\delta_{ij}+ \;^2g_{ij}+ \;^4g_{ij}+\cdots,\\ \stq
&&g_{oi}= \;^3g_{oi}+ \;^5g_{oi}+\cdots, \eea where the symbol
$^Ng_{\mu\nu}$ denotes the term in $g_{\mu\nu}$ of order $({\bar
v}/c)^N$.    Odd powers of ${\bar v}/c$ occur in $g_{io}$ because
$g_{io}$ must change sign under time-reversal transformation
$t\rightarrow -t$. These expansion lead to a consistent solution
of Einstein field equations. The inverse of the metric tensor is
defined by the equation \st\stq \bea
&&g^{i\mu}g_{o\mu}=g^{io}g_{oo}+g^{ij}g_{oj}=0,\\ \stq
&&g^{o\mu}g_{o\mu}=g^{oo}g_{oo}+g^{oi}g_{oi}=1,\\ \stq
&&g^{i\mu}g_{j\mu}=g^{io}g_{jo}+ g^{ik}g_{jk}=\delta_{ij}. \eea We
expect that \st\stq \bea &&g^{oo}=-1+ \;^2g^{oo}+
\;^4g^{oo}+\cdots,\\ \stq &&g^{ij}=\delta_{ij}+ \;^2g^{ij}+
\;^4g^{ij}+\cdots,\\ \stq &&g_{oi}= \;^3g^{oi}+ \;^5g^{oi}+\cdots,
\eea and inserting these expansions into the Eqs. (A.3), we find
\st
\be
^2g^{oo}=- \;^2g_{oo};\;\; ^2g^{ij}=- \;^2g_{ij};\;\;  ^3g^{oi}= \;^3g_{oi}.
\ee
The affine connection may be obtained from the familiar formula
\st
\be
\Gamma^{\lambda}_{\mu\nu}=\frac{1}{2}g^{\lambda\rho}\lp\frac{\partial
g_{\mu \rho}}{\partial x^{\nu}}+\frac{\partial
g_{\nu\rho}}{\partial x^{\mu}} -\frac{\partial
g_{\mu\nu}}{\partial x^{\rho}}\rp. \ee In computing
$\Gamma^{\lambda}_{\mu\nu}$ we must note that the scales of
distance and time, ${\bar r}$ and ${\bar r}/{\bar v}$,
respectively.    So the space and time derivatives should be
regarded as being of order $$ \frac{\partial}{\partial
x^i}\approx\frac{1}{\bar r};\;\; \frac{\partial}{c\partial
t}\approx\frac{\bar{v}/c}{\bar r}. $$ Using the metric expansions
we find that the various components of $\Gamma^{\lambda}_{\mu\nu}$
have the expansions \st\stq \bea
&&\Gamma^{\lambda}_{\mu\nu}=\;^2\Gamma^{\lambda}_{\mu\nu}+\;^4\Gamma^{\lambda}_
{\mu\nu}+\cdots;\;\;\;(\rm{for}\;\Gamma^i_{oo},\;\Gamma^i_{jk},\;\Gamma^o_{oi})
, \\ \stq
&&\Gamma^{\lambda}_{\mu\nu}=\;^3\Gamma^{\lambda}_{\mu\nu}+\;^5\Gamma^{\lambda}_
{\mu\nu}+\cdots;\;\;\;(\rm{for}\;\Gamma^i_{oj},\;\Gamma^o_{oo},\;\Gamma^o_{ij})
. \eea The symbol $^N\Gamma^{\lambda}_{\mu\nu}$ denoting the term
in $\Gamma^{\lambda}_{\mu\nu}$ of order $({\bar v}/c)^N/{\bar r}$.
After some manipulating one finds \st\stq \bea
&&^2\Gamma^i_{oo}=-\frac{1}{2}\frac{\partial \;^2g_{oo}}{\partial
x^i},\\ \stq &&^4\Gamma^i_{oo}=-\frac{1}{2}\frac{\partial
\;^4g_{oo}}{\partial x^i} +\frac{\partial \;^3g_{oi}}{c\partial
t}+\frac{1}{2} \;^2g_{ij} \frac{\partial \;^2g_{oo}}{\partial
x^j},\\ \stq &&^3\Gamma^i_{oj}=\frac{1}{2}\lp\frac{\partial
\;^3g_{oi}}{\partial x^j} +\frac{\partial \;^2g_{oi}}{c\partial
t}+\frac{\partial \;^3g_{jo}} {\partial x^i}\rp ,\\ \stq
&&^2\Gamma^i_{jk}=\frac{1}{2}\lp\frac{\partial
\;^2g_{ij}}{\partial x^k} +\frac{\partial \;^2g_{ik}}{\partial
x^j}-\frac{\partial \;^2g_{jk}} {\partial x^i}\rp ,\\ \stq
&&^3\Gamma^o_{oo}=-\frac{1}{2}\lp \frac{\partial
\;^2g_{oo}}{c\partial t}\rp ,\\ \stq
&&^2\Gamma^o_{oi}=-\frac{1}{2}\lp \frac{\partial
\;^2g_{oo}}{\partial x^i}\rp ,\\ \stq &&^1\Gamma^o_{ij}=0. \eea
The Ricci tensor is defined by \st
\be
R_{\mu\nu}\equiv R^{\lambda}_{\mu\lambda\nu}=\frac{\partial \Gamma^{\lambda}
_{\mu\lambda}}{\partial x^{\nu}}-\frac{\partial \Gamma^{\lambda}
_{\mu\nu}}{\partial x^{\lambda}}-\Gamma^{\eta}_{\mu\lambda}
\Gamma^{\lambda}_{\nu\eta}-\Gamma^{\eta}_{\mu\nu}\Gamma^{\lambda}_{
\eta\lambda}.
\ee
Using Eqs. (A.6)-(A.7) we find that the components of $R_{\mu\nu}$ have the
expansions
\st\stq
\bea
&&R_{oo}= \;^2R_{oo}+ \;^4R_{oo}+\cdots,\\
\stq
&&R_{ij}= \;^2R_{ij}+ \;^4R_{ij}+\cdots,\\
\stq
&&R_{oi}= \;^3R_{oi}+ \;^5R_{oi}+\cdots,
\eea
Inserting Eqs. (A.6) in Eqs. (A.9) we obtain
\st\stq
\bea
&&^2R_{oo}= -\frac{\partial \;^2\Gamma^i_{oo}}{\partial x^i},\\
\stq
&&^4R_{oo}= \frac{\partial \;^3\Gamma^i_{oi}}{c\partial t}-
\frac{\partial \;^4\Gamma^i_{oo}}{\partial x^i}+ \;^2\Gamma^o_{oi}
{\;^2\Gamma^i_{oo}}- \;^2\Gamma^i_{oo} {\;^2\Gamma^j_{ij}},\\
\stq
&&^3R_{oi}=\frac{\partial \;^2\Gamma^j_{ij}}{c\partial t}-
\frac{\partial \;^3\Gamma^j_{oi}}{\partial x^j},\\
\stq
&&^2R_{ij}=\frac{\partial \;^2\Gamma^o_{oi}}{\partial x^j}+
\frac{\partial \;^2\Gamma^k_{ik}}{\partial x^j}-
\frac{\partial \;^2\Gamma^k_{ij}}{\partial x^k}.
\eea
Therefore in terms of metric tensor, the Ricci tensor will be
\st\stq
\bea
&&^2R_{oo}= \frac{1}{2}\nabla^2 {\;^2g_{oo}},\\
\stq
&&^4R_{oo}= \frac{1}{2}\frac{\partial^2 {\;^2g_{ii}}}{c^2\partial t^2}-
\frac{\partial^2 {\;^3g_{io}}}{c\partial x^i\partial t}+
\frac{1}{2}\nabla^2 {\;^4g_{oo}}- \frac{1}{2} \;^2g_{ij}
\frac{\partial^2 {\;^2g_{oo}}}{\partial x^i\partial x^i}\nonumber\\
&&\hspace{1cm}-\frac{1}{2}\lp \frac{\partial \;^2g_{ij}}{\partial x^j}\rp
\lp \frac{\partial \;^2g_{oo}}{\partial x^i}\rp +
\frac{1}{4}\lp \frac{\partial \;^2g_{oo}}{\partial x^i}\rp
\lp \frac{\partial \;^2g_{oo}}{\partial x^i}\rp \nonumber\\
&&\hspace{4cm}+\frac{1}{4}\lp \frac{\partial\;^2g_{jj}}{\partial x^i}\rp
\lp \frac{\partial \;^2g_{oo}}{\partial x^i}\rp ,\\
\stq
&&^3R_{oi}=\frac{1}{2}\frac{\partial^2\;^2g_{ij}}{c\partial x^j\partial t}-
\frac{1}{2}\frac{\partial^2\;^3g_{oj}}{\partial x^i\partial x^j}
-\frac{1}{2}\frac{\partial^2\;^2g_{ij}}{c\partial x^i\partial t}
+\frac{1}{2}\nabla^2\;^3g_{oi},\\
\stq
&&^2R_{ij}=-\frac{1}{2}\frac{\partial^2\;^2g_{oo}}{\partial x^i\partial x^j}+
\frac{1}{2}\frac{\partial^2\;^2g_{kk}}{\partial x^i\partial x^j}
-\frac{1}{2}\frac{\partial^2\;^2g_{kj}}{\partial x^k\partial x^i}
+\frac{1}{2}\nabla^2\; ^2g_{ij}.
\eea
By choosing a suitable coordinates system, one can simplify the above
equations.
It is always possible to define the $x^{\mu}$ so that they obey the harmonic
conditions
\st
\be
g^{\mu\nu}\Gamma^{\lambda}_{\mu\nu}=0.
\ee
Substituting Eqs. (A.3) and (A.7) in Eq. (A.12), we find that the
vanishing of the third-order term in $g^{\mu\nu}\Gamma^o_{\mu\nu}$ gives
\stq
\be
\frac{1}{2}\frac{\partial\; ^2g_{oo}}{c\partial t}
-\frac{\partial\; ^3g_{oi}}{\partial x^i}
+\frac{1}{2}\frac{\partial\; ^2g_{ii}}{c\partial t}=0,
\ee
while the vanishing of the second order term in $g^{\mu\nu}\Gamma^o_{\mu\nu}$
gives
\stq
\be
\frac{1}{2}\frac{\partial\; ^2g_{oo}}{\partial x^i}
+\frac{\partial\; ^2g_{ij}}{\partial x^j}
-\frac{1}{2}\frac{\partial\; ^2g_{jj}}{\partial x^i}=0.
\ee
But
\begin{eqnarray*}
&&\frac{\partial}{c\partial t}\lp g^{\mu\nu}\Gamma^o_{\mu\nu}\rp =
\frac{1}{2}\frac{\partial^2\; ^2g_{oo}}{c^2\partial t^2}
-\frac{\partial^2\; ^3g_{oi}}{c\partial x^i\partial t}
+\frac{1}{2}\frac{\partial^2\; ^2g_{ii}}{c^2\partial t^2}=0,\\
&&\frac{\partial}{\partial x^j}\lp g^{\mu\nu}\Gamma^o_{\mu\nu}\rp -
\frac{\partial}{c\partial t}\lp g^{\mu\nu}\Gamma^j_{\mu\nu}\rp =
\frac{\partial^2\; ^2g_{ii}}{c\partial t\partial x^j}
-\frac{\partial^2\; ^3g_{oi}}{\partial x^i\partial x^j}\nonumber\\
&&\hspace{6cm}-\frac{1}{2}\frac{\partial^2\; ^2g_{ij}}
{c\partial t\partial x^i}=0,\\
&&\frac{\partial}{\partial x^k}\lp g^{\mu\nu}\Gamma^i_{\mu\nu}\rp -
\frac{\partial}{\partial x^i}\lp g^{\mu\nu}\Gamma^k_{\mu\nu}\rp =
\frac{\partial^2\; ^2g_{ij}}{\partial x^j\partial x^k}
+\frac{\partial^2\; ^2g_{kj}}{\partial x^i\partial x^j}
-\frac{\partial^2\; ^2g_{jj}}{\partial x^i\partial x^k}\nonumber\\
&&\hspace{6cm}+\frac{\partial^2\; ^2g_{oo}}{\partial x^i\partial x^k}=0.
\end{eqnarray*}
So Eqs. (A.11) now give simplified formulas for the Ricci tensor
\st\stq
\bea
&&^2R_{oo}= \frac{1}{2}\nabla^2 {\;^2g_{oo}},~~~~\\
\stq
&&^4R_{oo}= \frac{1}{2}\nabla^2 \;^4g_{oo}
-\frac{1}{2}\frac{\partial^2 \;^2g_{oo}}{c^2\partial t^2}-
\frac{1}{2} \;^2g_{ij}\frac{\partial^2 \;^2g_{oo}}
{\partial x^i\partial x^j}
+\frac{1}{2}(\nabla \;^2g_{oo})^2,~~~~\\
\stq
&&^3R_{oi}=\frac{1}{2}\nabla^2\;^3g_{oi},~~~~\\
\stq
&&^2R_{ij}=\frac{1}{2}\nabla^2\; ^2g_{ij}.~~~~
\eea
The Einstein field equations are
\st
\be
R_{\mu\nu}=-\frac{8\pi G}{c^4}
(T_{\mu\nu}-\frac{1}{2}g_{\mu\nu}T^{\lambda}_{\lambda}).
\ee
The various components of energy momentum tensor will have the expansions
\st\stq
\bea
&&T^{oo}= \;^oT^{oo}+ \;^2T^{oo}+\cdots,\\
\stq
&&T^{ij}= \;^2T^{ij}+ \;^4T^{ij}+\cdots,\\
\stq
&&T^{oi}= \;^1T^{oi}+ \;^3T^{oi}+\cdots,
\eea
where $^NT^{\mu\nu}$ denotes the term in $T^{\mu\nu}$ of order $({\bar M}/{\bar
r}^3)({\bar v}/c)^N$.  Therefore
\st
\be
S_{\mu\nu}=T_{\mu\nu}-\frac{1}{2}g_{\mu\nu}T^{\lambda}_{\lambda},
\ee we find \stq \bea &&S_{oo}= \;^oS_{oo}+ \;^2S_{oo}+\cdots,\\
\stq &&S_{ij}= \;^oS_{ij}+ \;^2S_{ij}+\cdots,\\ \stq &&S_{oi}=
\;^1S_{oi}+ \;^3S_{oi}+\cdots. \eea In particular \st\stq \bea
&&^oS_{oo}=\frac{1}{2}\;^oT^{oo},\\ \stq &&^2S_{oo}=\frac{1}{2}\lp
\;^2T^{oo}-2\;^2g_{oo}\;^oT^{oo}+\;^2T^{ii}\rp ,\\ \stq
&&^oS_{ij}=\frac{1}{2}\;^oT^{oo}\delta_{ij},\\ \stq
&&^1S_{oi}=-\;^1T^{oi}. \eea Using Eqs. (A.14) and (A.18) in field
equations, we find that the field equations in harmonic
coordinates conditions \st\stq \bea &&\nabla^2\;^2g_{oo}=
-\frac{8\pi G}{c^4}\;^oT^{oo},\\ \stq
&&\nabla^2\;^4g_{oo}=\frac{\partial^2 \;^2g_{oo}}{c^2\partial
t^2}+ \;^2g_{ij}\frac{\partial^2 \;^2g_{oo}}{\partial x^i\partial
x^j} -\lp \nabla \;^2g_{oo}\rp ^2\nonumber\\
&&\hspace{2cm}-\frac{8\pi G}{c^4}\lp
\;^2T^{oo}-2\;^2g_{oo}\;^oT^{oo}+\;^2T^{ii}\rp ,\\ \stq
&&\nabla^2\;^3g_{oi}=\frac{16\pi G}{c^4}\;^1T^{oi},\\ \stq
&&\nabla^2\; ^2g_{ij}=-\frac{8\pi G}{c^4}\;^oT^{oo}. \eea From Eq.
(A.19a) we find \st\stq
\be
^2g_{oo}=-2\phi, \ee where $\phi$ is the Newtonian potential,
defined by Poisson's equation \stq
\be
\nabla^2\phi=\frac{4\pi G}{c^4}\;^oT^{oo}.
\ee
Also $^2g^{oo}$ must vanish at infinity, so the solution is
\stq
\be
\phi (\x, t)=-\frac{G}{c^4}\int\frac{\;^oT^{oo}(\xp, t)}{\mid {\bf
x}- \xp\mid}d^3x'.
\ee
From Eq. (A.19d) we find that the solution for $^2g_{ij}$  that vanishes at
infinity is
\st
\be
^2g_{ij}=-2\phi\delta_{ij}.
\ee
$^3g_{io}$ is a new vector potential $\xi_i$
\st\stq
\be
^3g_{oi}=\xi_i,
\ee
and the solution of Eq. (A.19c) that vanishes at infinity is
\stq
\be
\xi_i (\x, t)=-4G\int\frac{\;^1T^{oi}(\xp, t)}{\mid \x-
\xp\mid}d^3x'.
\ee
Using Eqs. (A.19b) and (A.20) and the identity
$$
\frac{\partial\phi}{\partial x^i}\frac{\partial\phi}{\partial x^i}\equiv
\frac{1}{2}\nabla^2\phi^2-\phi\nabla^2\phi,
$$
we obtain
\st\stq
\be
^4g_{oo}=-2\phi^2-2\psi.
\ee
The scalar potential $\psi$ satisfies
\stq
\be
\nabla^2\psi=\frac{\partial^2 \phi}{\partial t^2}+4\pi G\lp \;^2T^{oo}+
^2T^{ii}\rp ,
\ee
with solution
\stq
\be
\psi (\x, t)=-4\int\frac{d^3x'}{\mid \x-\xp\mid}\lp
\frac{1}{4\pi}\frac{\partial^2 \phi (\xp, t)}{c^2\partial t^2}
+ G\;^2T^{oo} (\xp, t)+ G\;^2T^{ii} (\xp, t)\rp .
\ee
The coordinates condition, Eqs. (A.13), imposes on $\phi$ and
$\xxi$ the further relation
\st
\be
\frac{\partial \phi}{c\partial t}+\nabla\cdot\xxi =0,
\ee
while the other coordinate condition, Eq. (A.12b), is automatically satisfied.

The various components of the affine connection are
\st\stq
\bea
&&^2\Gamma^i_{oo}=\frac{\partial \phi}{\partial x^i},\\
\stq
&&^4\Gamma^i_{oo}=\frac{\partial}{\partial x^i}(2\phi^2+\psi)
+\frac{\partial \xi_i}{c\partial t},\\
\stq
&&^3\Gamma^i_{oj}=\frac{1}{2}\lp \frac{\partial \xi_i}{\partial x^j}-
\frac{\partial \xi_j}{\partial x^i}\rp -\delta_{ij}
+\frac{\partial \phi}{c\partial t},\\
\stq
&&^2\Gamma^i_{jk}=-\delta_{ij}\frac{\partial \phi}{\partial x^k}
+\delta_{ik}\frac{\partial \phi}{\partial x^j}+\delta_{jk}
\frac{\partial \phi}{\partial x^i},\\
\stq
&&^3\Gamma^o_{oo}=\frac{\partial\phi}{c\partial t},\\
\stq
&&^2\Gamma^o_{oi}=\frac{\partial\phi}{\partial x^i},\\
\stq
&&^3\Gamma^o_{ij}=-\frac{1}{2}\lp \frac{\partial \xi_i}{\partial x^j}+
\frac{\partial \xi_j}{\partial x^i}\rp -\delta_{ij}\frac{\partial\phi}
{c\partial t},\\
\stq
&&^4\Gamma^o_{oi}=\frac{\partial\psi}{\partial x^i},\\
\stq
&&^5\Gamma^o_{oo}=\frac{\partial\psi}{c\partial t}
\xxi\cdot\nabla\phi.
\eea

\setcounter{sub}{0}
\setcounter{subeqn}{0}
\renewcommand{\theequation}{B.\thesub\thesubeqn}

\chapter{ Derivation of Eqs. (3.5)}
Consider a general coordinate transformation $(X, U)=(X^{\mu}, U^i)$ to
$(Y, V)=(Y^{\mu}, V^i)$.   The corresponding partial derivatives transform as
\[\left( \begin{array}{c}
\partial/\partial X\\ \partial/\partial U
          \end{array} \right)\;= M
\left( \begin{array}{c}
\partial/\partial Y\\ \partial/\partial V
          \end{array} \right)\;,\]

\st
\be
\hspace{5.9cm}=
\left( \begin{array}{cc}
 \partial Y/\partial X &\partial V/\partial
X\\
\partial Y/\partial U &\partial V/\partial U
            \end{array}\right)
\left( \begin{array}{c}
\partial/\partial Y\\\partial/\partial V
          \end{array} \right) ,
\ee
where $M$ is the $7\times 7$ Jacobian matrix of transformation.   Setting
$X=Y=x^{\mu}$, $V=v^i$ and $U=U^i$ for our problem, one finds
\st
\stq
\be
M=\left( \begin{array}{cc}
\partial x^{\mu}/\partial x^{\nu}&\partial v^i/\partial
x^{\nu}\\
\partial x^{\mu}/\partial U^j&\partial v^i/\partial U^j
\end{array}\right),
\ee
and
\stq
\be
M^{-1}=\left( \begin{array}{cc}
\partial x^{\mu}/\partial x^{\nu}&\partial U^i/\partial
x^{\nu}\\
\partial x^{\mu}/\partial v^j&\partial U^i/\partial v^j
\end{array}\right).
\ee
One easily finds
\st
\stq
\bea
&&\partial x^{\mu}/\partial x^{\nu}=\delta_{\mu\nu};\;\;\;\;\;\;\;
\partial x^{\mu}/\partial v^j=0,\\
\stq
&&\partial U^i/\partial x^{\nu}=v^i\partial U^0/\partial
x^{\nu}
=\frac{{U^0}^3v^i}{2}\frac{\partial g_{\alpha\beta}}{\partial
x^{\nu}}v^{\alpha}v^{\beta},\\
\stq
&&\partial U^i/\partial v^j=U^0\delta_{ij}+v^i \partial
U^0/\partial v^j
=U^0\delta_{ij}-{U^0}^3v^ig_{j\beta}v^{\beta}.
\eea
Inserting the latter in $M^{-1}$ and inverting the result one arrives at $M$
from which Eqs. (3.5) can be read out.


\setcounter{sub}{0}
\setcounter{subeqn}{0}
\renewcommand{\theequation}{C.\thesub\thesubeqn}

\chapter{ Post-Newtonian hydrodynamics}
Mathematical manipulations in composing of this work has been tasking.
To ensure that no error has crept in the course of calculations we try
to derive the post-Newtonian hydrodynamical equations from the post-Newtonian
Liouville equation derived earlier.        From Eq. (3.6a) one has
\st
\bea
{\Lpn}_U F&&=U^0(\Lcl+\Lpn)F\nonumber\\
&&=[(c^2+\phi+\frac{1}{2}{\bf v}^2)\Lcl +\Lpn ]F,
\eea
where $\Lcl$ and $\Lpn$ are given by Eq. (3.10).
    We integrate ${\Lpn}_U F$ over the ${\bf v}$-space:
\st
\be
\int {\Lpn}_UFd^3v=\int [(c^2+\phi+\frac{1}{2}{\bf v}^2)
\Lcl +\Lpn ]Fd^3v.
\ee
Using Eqs. (3.12) and (3.13), one finds the continuity equation
\st
\bea
&&\frac{\partial}{c\partial
t}(\;^0T^{00}+\;^2T^{00})+\frac{\partial}{\partial x^j}(\; ^1T^{0j}
+\; ^3T^{0j})-\;^0T^{00}\frac{\partial \phi}{c^3\partial t} =0,\nonumber\\
\eea
which is the $pn$ expansion of the continuity equation
\st
\be
T^{0\nu}_{\;\;\;;\nu}=0,
\ee
Next,
we multiply ${\Lpn}_UF$ by $v^i$ and integrate over the ${\bf
v}$-space:
\st
\be
\int v^i {\Lpn}_UFd^3v=\int v^i[(c^2+\phi+\frac{1}{2}{\bf v}^2)
\Lcl+\Lpn ]Fd^3v. \ee After some calculations one finds \st \bea
&&\frac{\partial}{c\partial
t}\left(\;^1T^{0i}+\;^3T^{0i}\right)+\frac{\partial}{\partial
x^j}\left(\; ^2T^{ij} +\; ^4T^{ij}\right)\nonumber\\ &&\;\;\;+\;
^0T^{00}\left(\frac{\partial}{\partial
x^i}(\phi+2\phi^2/c^2+\psi/c^2)+ \frac{\partial \xi_i}{c\partial
t}\right)/c^2+\; ^2T^{00}\frac{\partial \phi}{c^2\partial
x^i}\nonumber\\ &&\;\;\;+\; ^1T^{0j}\left(\frac{\partial
\xi_i}{\partial x^j}-\frac{\partial \xi_j}{\partial
x^i}-4\delta_{ij}\frac{\partial \phi}{c\partial t}\right)/c^3+ \;
^2T^{jk} \left(\delta_{jk}\frac{\partial \phi}{\partial
x^i}-4\delta_{ik}\frac{\partial \phi} {\partial
x^j}\right)/c^2=0.\nonumber\\ \eea The latter, the $pn$ expansion
of \st
\be
T^{i\nu}_{\;\;\;;\nu}=0;\;\;\;\;i=1,2,3, \ee is the same as that
of Weinberg \cite{Wei72}, QED.

\setcounter{sub}{0}
\setcounter{subeqn}{0}
\renewcommand{\theequation}{D.\thesub\thesubeqn}

\chapter{ Eigensolutions of  J$^2$ and J{\small z}}
As pointed out earlier, $J_i$'s of Eq. (4.15) have the angular momentum
algebra,
\stepcounter{sub}
\begin{equation}
[J_i,J_j]=i\varepsilon_{ijk}J_k.
\end{equation}
Therefore, the simultaneous eigensolutions of $J^2$ and $J_z$,
$\Lambda_{jm}({\bf x}, {\bf u})$, obey the following
\stepcounter{sub}
\begin{equation}
J^2\Lambda_{jm}=j(j+1)\Lambda_{jm},\;\;\;\;\;j=0,1,\cdots,
\end{equation}
\stepcounter{sub}
\begin{equation}
J_z\Lambda_{jm}=m\Lambda_{jm},\;\;\;\;\;-j\le m\le j.
\end{equation}
The ladder operators, $J_{\pm}=J_x\pm iJ_y$, raise and lower the $m$ values:
\stepcounter{sub}
\begin{equation}
J_{\pm}\Lambda_{jm}=\sqrt{(j\mp m)(j\pm
m+1)}\Lambda_{j m\pm 1}.
\end{equation}
In particular
\stepcounter{subeqn}
\begin{equation}
J_{\pm}\Lambda_{j,\pm j}=0.
\end{equation}
The effect of $J_i$ on classical energy integral, $e=u^2/2-\theta (r)$, and
the classical angular momentum integral, $l_i=\varepsilon_{ijk}x_ju_k$, are as
follows
\stepcounter{sub}
\stepcounter{subeqn}
\begin{eqnarray}
&&J_i e=J_i l^2=J_i f(e, l^2)=0,\\
\stepcounter{subeqn}
&&J_i l_j=i\varepsilon_{ijk} l_k.
\end{eqnarray}
{\it Theorem 1:}
\stepcounter{sub}
\begin{equation}
\Lambda_{j,\pm j}=l_{\pm}^j=(\frac{1}{2})^j(l_x\pm il_y)^j.
\end{equation}
{\it Proof:}
\stepcounter{sub}
\stepcounter{subeqn}
\begin{eqnarray}
&&J_zl_{\pm}^j=jl_{\pm}^{j-1}(J_zl_{\pm})=\pm
jl_{\pm}^j,\hspace{2.6cm}{\rm by\; (D.5b)},~~~~~~\\ \stepcounter{subeqn}
&&J^2l_+^j=(J_-J_++J_z^2+J_z)l_+^j=j(j+1)l_+^j,\;{\rm by ~(D.4a)
~and ~(D.7a)},~~~~~~\\ \stepcounter{subeqn}
&&J^2l_-^j=(J_+J_-+J_z^2-J_z)l_-^j=j(j+1)l_-^j,~~~~~~
\end{eqnarray}
QED.   Combining Eqs. (D.6), (D.4) and (D.5) one obtains
\stepcounter{sub}
\begin{equation}
\Lambda_{jm}=af(e, l^2)J_+^{j+m}l_-^j=bf(e,
l^2)J_-^{j-m}l_+^j,\label{lambda}
\end{equation}
where $f(e, l^2)$ is an arbitrary function of its arguments, and $a$ and $b$
are normalization constants.    Examples:  Aside from an arbitrary factor of
classical constants of motion, one has

\stepcounter{sub}
\stepcounter{subeqn}
\begin{eqnarray}
&&\Lambda_{1\;0}=l_z,\\
\stepcounter{subeqn}
&&\Lambda_{1\;\pm1}=l_{\pm},\\
\stepcounter{subeqn}
&&\Lambda_{2\;0}=2l_+l_--l_z^2=\frac{1}{2}(3l_z^2-l^2),\\
\stepcounter{subeqn}
&&\Lambda_{2\;\pm 1}=l_{\pm}l_z,\\
\stepcounter{subeqn}
&&\Lambda_{2\;\pm 2}=l_{\pm}^2.
\end{eqnarray}
{\it Theorem 2:}   The vector field ${\bf V}^{jm}=\int\Lambda_{jm}{\bf
u}d\Omega$ is a toroidal vector field belonging to the spherical harmonic
numbers ($j,m$), where integration is over the directions of ${\bf u}$.\\
{\it Preliminaries:}   Let ($\vartheta,\varphi$) and ($\alpha,\beta$)
denote the polar angles of ${\bf x}$, of ${\bf u}$, respectively, and $\gamma$
be  the angle
between (${\bf x}, {\bf u}$).     Also choose magnitudes of ${\bf
x}$ and ${\bf u}$ to be unity, for only integrations over the
direction angles are of concern.     One has $\cos\gamma
=\cos\vartheta\;\cos\alpha\;+\;\sin\vartheta\;\sin\alpha\;\cos(\varphi-\beta)$
\stepcounter{sub}
\stepcounter{subeqn}
\begin{eqnarray}
&&u_r=\cos\gamma ,\\
\stepcounter{subeqn}
&&u_{\vartheta}=-\sin\vartheta\;\cos\alpha\;+\;\cos\vartheta\;\sin\alpha\;
\cos(\varphi-\beta),\\
\stepcounter{subeqn}
&&u_{\varphi}=-\sin\alpha\;\sin(\varphi-\beta),\\
\stepcounter{subeqn}
&&l_+=i(\sin\vartheta\;\cos\alpha\;e^{i\varphi}-\cos\vartheta\;\sin\alpha\;e^{
 i\beta}).
\end{eqnarray}
{\it Proof:}   By induction, we show that a) ${\bf V}^{jj}$ is a toroidal field
and b) if ${\bf V}^{jm}$ is a toroidal field, so is ${\bf V}^{j\;m-1}$.\\
a) Direct integrations over $\alpha$ and $\beta$ gives
\stepcounter{sub}
\stepcounter{subeqn}
\begin{eqnarray}
&&V_r^{jj}=\int l_+^j u_r d\Omega=0,\;\;\;\;\;\;d\Omega
=\sin\alpha\;d\alpha\;d\beta,\\ \stepcounter{subeqn}
&&V_{\vartheta}^{jj}=\int l_+^j u_{\vartheta}
d\Omega=-\frac{1}{\sin\vartheta}\frac{\partial}{\partial\varphi}Y_{jj}(
\vartheta, \varphi), \\ \stepcounter{subeqn}
&&V_{\varphi}^{jj}=\int l_+^j u_{\varphi}
d\Omega=\frac{\partial}{\partial\vartheta}Y_{jj}(\vartheta,
\varphi).\;\;{\rm QED.}
\end{eqnarray}
b)  Suppose ${\bf V}^{jm}$ is a toroidal vector field and calculate ${\bf
V}^{j\;m-1}=
\int(J_-\Lambda_{jm}){\bf u}d\Omega$, where $J_{\pm}=L_{\pm}+K{\pm}$,
$L_{\pm}=
\pm e^{\pm i\varphi}(\frac{\partial}{\partial\vartheta}\pm i{\rm cotg}\vartheta
\frac{\partial}{\partial\varphi})$,
$K_{\pm}=
\pm e^{\pm i\beta}(\frac{\partial}{\partial\alpha}\pm i{\rm cotg}\alpha
\frac{\partial}{\partial\beta})$.     Again  direct integrations gives
\stepcounter{sub}
\stepcounter{subeqn}
\begin{eqnarray}
&&\hspace{-1cm}V_r^{j\;m-1}=L_-V_r^{jm}=0,\hspace{2cm} {\rm
if}\;\; V_r^{jm}=0,~~~~~~\\ \stepcounter{subeqn}
&&\hspace{-1cm}V_{\vartheta}^{j\;m-1}=
-\frac{1}{\sin\vartheta}\frac{\partial}{\partial\varphi}Y_{j\;m-1}(
\vartheta, \varphi), \hspace{.15cm} {\rm if}\;
V_{\vartheta}^{j\;m}=
-\frac{1}{\sin\vartheta}\frac{\partial}{\partial\varphi}Y_{j\;m}(
\vartheta, \varphi),~~~~~~\\ \stepcounter{subeqn}
&&\hspace{-1cm}V_{\varphi}^{j\;m-1}=
\frac{\partial}{\partial\vartheta}Y_{j\;m-1}(\vartheta,
\varphi),\hspace{1.4cm} {\rm if}\;\; V_{\varphi}^{j\;m}=
\frac{\partial}{\partial\vartheta}Y_{j\;m}( \vartheta, \varphi).~~~~~~
\end{eqnarray}
QED.

\end{document}